\documentclass[submission,Phys]{SciPost}

\usepackage[utf8]{inputenc} 
\usepackage[T1]{fontenc} 	
\usepackage[english]{babel} 

\usepackage[bitstream-charter]{mathdesign}
\usepackage{geometry} 		
\usepackage{amsmath} 		
\usepackage{mathtools} 		
\usepackage{float} 			
\usepackage{graphicx} 		
\usepackage{tabularx} 		
\usepackage{booktabs} 		
\usepackage{color, xcolor} 	
\usepackage{pdfpages} 		
\usepackage{extarrows} 		
\usepackage{multirow} 		
\usepackage{multicol} 		
\usepackage[subrefformat=parens]{subcaption}
\usepackage{enumitem} 		
\usepackage{xspace} 		
\usepackage{stackrel} 		
\usepackage{tikz} 			
\usetikzlibrary{calc}
\usepackage{braket} 		
\usepackage{bm} 			
\usepackage{tensor} 		
\usepackage{slashed} 		
\usepackage{siunitx} 
\usepackage{lastpage} 		
\usepackage{cite} 			
\usepackage[normalem]{ulem} 
\usepackage{fontawesome} 	
\usepackage{tocloft} 		
\usepackage{titlesec} 		
\usepackage{doi} 			
\usepackage{hyperref} 		
\usepackage[most]{tcolorbox} 					
\usepackage[nameinlink, capitalize]{cleveref} 	
\usepackage[nottoc, notlot, notlof]{tocbibind} 	
\usepackage[ruled, vlined]{algorithm2e} 		
\usepackage{makecell}
\usepackage{setspace}

\makeatletter
\def\BState{\State\hskip-\ALG@thistlm}
\makeatother

\makeatletter
\@ifundefined{pdfoutput}{}{\DeclareGraphicsRule{*}{mps}{*}{}}
\makeatother

\makeatletter
\DeclareRobustCommand*{\bfseries}{%
   \not@math@alphabet\bfseries\mathbf
   \fontseries\bfdefault\selectfont
   \boldmath
}
\makeatother

\hypersetup{
	pdftitle={sfitter_likely},
	pdfauthor={Elmer et al.},
	colorlinks=true, 			
	linkcolor={red!50!black}, 	
	citecolor={blue!50!black}, 	
	urlcolor={blue!80!black} 	
} 

\DeclareSymbolFont{usualmathcal}{OMS}{cmsy}{m}{n}
\DeclareSymbolFontAlphabet{\mathcal}{usualmathcal}

\SetArgSty{textnormal}
\SetKwComment{Comment}{{\small\#}~}{}
\SetCommentSty{mycommfont}

\setitemize{itemsep=0pt, parsep=0pt} 				
\setenumerate{itemsep=0pt, parsep=0pt} 				
\setitemize{itemsep=2pt,topsep=2pt,parsep=0pt,partopsep=0pt,leftmargin=*}
\setenumerate{itemsep=0pt,topsep=2pt,parsep=0pt,partopsep=0pt,labelindent=3pt,leftmargin=*}
\usepackage{placeins}

\newcommand{\sfitter}{\textsc{SFitter}\xspace}
\newcommand{\pois}{\text{Poiss}}

\newcommand{\qqquad}{\qquad\quad}

\newcommand{\lag}{\mathscr{L}}
\newcommand{\ope}{\mathcal{O}}

\newcommand{\arXiv}[2][]{%
	\ifthenelse{\equal{#1}{}}%
	{\href{http://arxiv.org/abs/#2}{arXiv:#2}}%
	{\href{http://arxiv.org/abs/#2}{arXiv:#2~[#1]}}}

\newcommand{\gev}{\text{GeV}}

\def\slashchar#1{\setbox0=\hbox{$#1$}           
   \dimen0=\wd0                                 
   \setbox1=\hbox{/} \dimen1=\wd1               
   \ifdim\dimen0>\dimen1                        
      \rlap{\hbox to \dimen0{\hfil/\hfil}}      
      #1                                        
   \else                                        
      \rlap{\hbox to \dimen1{\hfil$#1$\hfil}}   
      /                                         
   \fi}

\newcommand{\tikznode}[2]{%
\ifmmode%
\tikz[remember picture,baseline=(#1.base),inner sep=0pt] \node (#1) {$#2$};%
\else
\tikz[remember picture,baseline=(#1.base),inner sep=0pt] \node (#1) {#2};%
\fi}

\def\mathswitchr#1{\relax\ifmmode{\text{#1}}\else$\text{#1}$\xspace\fi}
\def\mathswitch#1{\relax\ifmmode#1\else$#1$\xspace\fi}

\graphicspath{{./figs/}}
\begin{document}

\vspace*{-2.5em}
\hfill{}
\vspace*{0.5em}

\begin{center}{\Large \textbf{
Staying on Top of SMEFT-Likelihood Analyses
}}\end{center}

\begin{center}
Nina Elmer,
Maeve Madigan,
Tilman Plehn,
and Nikita Schmal
\end{center}

\begin{center}
Institut f\"ur Theoretische Physik, Universit\"at Heidelberg, Germany
\end{center}


\vspace{-1cm}
\section*{Abstract}
{\bf We present a new global SMEFT analysis of LHC data in the top sector. 
After updating our set of measurements, we show how public 
ATLAS likelihoods can be incorporated into an external global analysis 
and how our analysis benefits from the additional information. We find 
that, unlike for the Higgs and electroweak sector, the SMEFT analysis of 
the top sector is mostly limited by the theory uncertainties.
Finally,
we present the first global \sfitter analysis combining the top and
electroweak-Higgs sectors.}

\vspace{10pt}
\noindent\rule{\textwidth}{1pt}
\tableofcontents\thispagestyle{fancy}
\noindent\rule{\textwidth}{1pt}
\vspace{10pt}

\clearpage
\section{Introduction}
\label{sec:intro}

Over the last decade, LHC physics has seen a paradigm shift, from
testing models for physics beyond the Standard Model (BSM) to
precision measurements and a complete understanding of LHC physics in
terms of fundamental quantum field theory. This change represents
impressive progress in experiment and theory in a holistic approach
to the huge LHC dataset. On the theory side, a driving force is the
development of an effective field theory version of the Standard Model
(SMEFT), which allows us to ask and answer the question: \textsl{Does 
the LHC data agree with the Standard Model altogether?}

SMEFT is a perturbative quantum field theory that respects the gauge
symmetries and covers all sectors of the Standard Model. It is
renormalizable and allows for QCD and electroweak precision
predictions.  It is built on the idea that BSM particles affecting LHC
measurements might be too heavy to be produced on-shell. Assuming that
the Higgs and Goldstone fields form the usual doublet, SMEFT 
embeds the Standard Model in an effective field theory (EFT). The
SMEFT
idea~\cite{Buchmuller:1985jz,Leung:1984ni,Degrande:2012wf,Brivio:2017vri}
was developed for a gauge-invariant description of anomalous gauge
interactions at LEP~\cite{Hagiwara:1986vm,GonzalezGarcia:1999fq}. Its
big success is the unified global analysis of the Higgs and
electroweak sector, including electroweak precision
data~\cite{Butter:2016cvz,deBlas:2016ojx,daSilvaAlmeida:2018iqo},
a major part of the legacy of the LHC
Run~2~\cite{Ellis:2018gqa,Almeida:2018cld,Biekotter:2018rhp,Kraml:2019sis,Almeida:2021asy}.
Obviously, the same
approach~\cite{Degrande:2010kt,Zhang:2010dr,Greiner:2011tt} can be
used to systematically test the top quark
sector~\cite{Hioki:2013hva,Buckley:2015nca,Buckley:2015lku,Hartland:2019bjb,Brivio:2019ius,Aoude:2022imd,Aguilar-Saavedra:2018ksv,Maltoni:2019aot,Aoude:2022deh,Degrande:2018fog,Faham:2021zet}.
This sector is especially interesting because it can be combined with
the bottom
sector~\cite{Bissmann:2019gfc,Aoude:2020dwv,Bruggisser:2021duo,Bruggisser:2022rhb,Bartocci:2023nvp,Durieux:2019rbz}
and a much broader set of precision
measurements~\cite{Garosi:2023yxg}, eventually testing the impact of
flavor symmetries. Recently, several groups have provided combined
SMEFT analyses of the electroweak and top
sectors~\cite{Ellis:2020unq,Ethier:2021bye}, SMEFT analyses combined
with parton density extraction~\cite{Gao:2022srd,Kassabov:2023hbm,Carrazza:2019sec,Greljo:2021kvv},
and even SMEFT analyses with lighter new
particles~\cite{Galda:2021hbr,Biekotter:2023mpd,Lessa:2023tqc}.

The systematic search for BSM effects in the top sector has some
unique aspects. Experimentally, precision measurements at the LHC go
far beyond simple kinematic distributions of top pair
production. Associated top pair production with gauge bosons, single
top production, and top decay kinematics are also probed with
increasing precision. This effort is beautifully matched by precision
predictions~\cite{Czakon:2023hls}. Theoretically, the top sector is
still closely related to the hierarchy problem or the dynamic origin
of the Higgs VEV, a problem which should be understood at the
LHC~\cite{Morrissey:2009tf}. Phenomenologically, the ATLAS and CMS top
groups are providing experimental results in a way that can be
implemented in an external global analysis easily and optimally. This
includes unfolded rate measurements, unfolded kinematic distributions,
and most recently published
likelihoods~\cite{ATLAS:2020aln,ATLAS:2021fzm,ATLAS:2022wfk}.

The publication of experimental data in the format of public likelihoods 
is a major step in the way experimental results can be 
re-interpreted~\cite{Kraml:2012sg,Boudjema:2013qla,Cranmer:2013hia,Feickert:2020wrx,LHCReinterpretationForum:2020xtr,Cranmer:2021urp}.
The HistFactory format~\cite{Cranmer:2012sba}, software such as \texttt{pyhf}~\cite{pyhf,pyhf_joss}, \texttt{Spey}~\cite{Araz:2023bwx}, and simplified likelihood 
frameworks~\cite{Buckley:2018vdr} allow for an efficient use of likelihoods.
In the classic BSM sector, many likelihoods have already been made public and 
analyzed~\cite{MahdiAltakach:2023bdn}. In the top sector, public ATLAS 
likelihoods~\cite{ATLAS:2020aln,ATLAS:2021fzm,ATLAS:2022wfk} still have to be 
used outside the collaborations. 
We aim to fill this gap by using them as the basis of a 
global SMEFT analysis of the top sector using \sfitter.

In this paper, we update an earlier NLO-SMEFT analysis of the top
sector in the \sfitter framework~\cite{Brivio:2019ius}. \sfitter is
unique in the sense, that it does not rely on pre-processed experimental
measurements and includes its own comprehensive uncertainty
treatment~\cite{Lafaye:2004cn,Lafaye:2007vs,Lafaye:2009vr}. This
makes it a promising candidate to, for the first time, include
public likelihoods in a global analysis and determine their
impact. After a general introduction to the \sfitter methodology and
our dataset in Sec.~\ref{sec:setup}, we will discuss three public
likelihoods in the top sector in detail in Sec.~\ref{sec:like}.  
We will then include these likelihoods in the first \sfitter analysis of
the electroweak and top sectors in Sec.~\ref{sec:global}. 
While the physics behind combining these two sectors is largely
understood~\cite{Ellis:2020unq,Ethier:2021bye}, 
in our global \sfitter analysis, we will focus on the impact of theory uncertainties. In addition, we will 
probe the impact of a profile likelihood vs Bayesian
marginalization when extracting limits on single Wilson
coefficients, where we saw significant effects on the Higgs and
electroweak sector~\cite{Brivio:2022hrb}. Finally, we will provide a
short comparison between the Markov chains used in this analysis and
the Monte Carlo experiment method used in earlier \sfitter analyses in
the Appendix.

\section{Setup}
\label{sec:setup}

\subsection{SMEFT Lagrangian}
\label{sec:eft}

By fundamental theory arguments, the SMEFT Lagrangian is the
appropriate interpretation framework to interpret LHC searches for
effects of particles which are too heavy to be produced
on-shell~\cite{Brivio:2017vri}.  While in the Higgs sector one can
argue about the proper way to implement electroweak symmetry-breaking
and the doublet nature of the Higgs and Goldstone fields, the SMEFT
description of the top sector is fixed.  An open question is how to
combine it with the light-flavor sector and its range of potential
global symmetries. This renders the impact of flavor measurements on
the top sector somewhat unclear, so we will not exploit this link and
instead refer to dedicated
analyses~\cite{Bissmann:2019gfc,Aoude:2020dwv,Bruggisser:2021duo,Bruggisser:2022rhb,Bartocci:2023nvp}.

The goal of our analysis is to probe effective higher-dimensional
interactions in the top sector using an increasing set of LHC
measurements~\cite{AguilarSaavedra:2018nen,Brivio:2019ius}. Because 
at dimension six the set of allowed operators already exceeds
the power of the available measurements, we truncate the effective
Lagrangian
\begin{align}
\lag_\text{eff} = \sum_j \left(\frac{C_j}{\Lambda^2}\,^\ddagger \ope_j + \text{h.c.} \right) 
               + \sum_k \frac{C_k}{\Lambda^2}\, \ope_k \; .
\label{eq:d6lag}
\end{align}
This means the sum runs over all operators at mass dimension six,
involving top quarks. Non-hermitian operators are denoted as
$^\ddagger \ope$. We neglect the Weinberg operator at dimension 5, as well as all operators of mass dimension seven and
higher in the EFT expansion, assuming that their $\Lambda$-suppression
translates into a suppression of their effects on LHC
observables. This assumption is formally well-motivated but given the
rather modest scale separation between the LHC and the accessible
$\Lambda$-values, it has to be checked for a given dataset and a given
UV-completion matched to the SMEFT
Lagrangian~\cite{Dawson:2020oco,Dawson:2021xei,Brivio:2021alv,Carmona:2021xtq,terHoeve:2023pvs}.

Because the underlying symmetry structure is an input to an
EFT construction, and we are hesitant to leave the test of fundamental
symmetries to a numerically tricky and hardly conclusive global
analysis~\cite{Brehmer:2017lrt}, we ignore CP-violating operators.  
Finally, the fact that the
top-sector measurements included in our analysis are blind to the
light-quark flavor we assume separate $U(2)$ symmetries in the first
and second generation~\cite{Durieux:2014xla,Degrande:2014tta},
\begin{alignat}{5}
q_i & = (u^i_L,d^i_L) & \qqquad 
u_i & = u^i_R,\,d_i=d^i_R 
\quad \text{for} \quad i=1,2 \notag  \\
Q & = (t_L,b_L) & \qqquad 
t &= t_R,\,b=b_R \; .
\label{eq:symm}
\end{alignat}
All quark masses except for the top mass are assumed to be zero.

Our assumptions leave us with 22 independent operators in the top
sector.  Eight operators come with a chiral $LL$ or $RR$ structure of
interacting fermion currents
\begin{align}
  \ope_{Qq}^{1,8} & = (\bar{Q}\gamma_\mu T^A Q) \; (\bar{q}_i\gamma^\mu T^A q_i) 
& \ope_{Qq}^{1,1} & = (\bar{Q}\gamma_\mu Q) \; (\bar{q}_i\gamma^\mu q_i) \notag \\
  \ope_{Qq}^{3,8} & = (\bar{Q}\gamma_\mu T^A\tau^I Q) \; (\bar{q}_i\gamma^\mu T^A \tau^I q_i)
& \ope_{Qq}^{3,1} & = (\bar{Q}\gamma_\mu\tau^I Q) \; (\bar{q}_i\gamma^\mu\tau^I q_i) \notag \\
  \ope_{tu}^8 & = (\bar{t}\gamma_\mu T^A t) \; (\bar{u}_i\gamma^\mu T^A u_i) 
& \ope_{tu}^1 & = (\bar{t}\gamma_\mu t) \; (\bar{u}_i\gamma^\mu u_i) \notag \\
  \ope_{td}^8 & = (\bar{t}\gamma^\mu T^A t) \; (\bar{d}_i\gamma_\mu T^A d_i) 
& \ope_{td}^1 & = (\bar{t}\gamma^\mu t) \; (\bar{d}_i\gamma_\mu d_i) \; .
\label{eq:ops_llrr}
\end{align}
Six operators show a $LR$ or $RL$ chirality in the
current-current interaction,
\begin{align}
  \ope_{Qu}^8 & = (\bar{Q}\gamma^\mu T^A Q) \; (\bar{u}_i\gamma_\mu T^A u_i)
& \ope_{Qu}^1 & = (\bar{Q}\gamma^\mu Q) \; (\bar{u}_i\gamma_\mu u_i) \notag \\
  \ope_{Qd}^8 & = (\bar{Q}\gamma^\mu T^A Q) \; (\bar{d}_i\gamma_\mu T^A d_i)
& \ope_{Qd}^1 & = (\bar{Q}\gamma^\mu Q) \; (\bar{d}_i\gamma_\mu d_i) \notag \\
  \ope_{tq}^8 & = (\bar{q}_i\gamma^\mu T^A q_i) \; (\bar{t}\gamma_\mu T^A t)
& \ope_{tq}^1 & = (\bar{q}_i\gamma^\mu q_i) \; (\bar{t}\gamma_\mu t) \; .
\label{eq:ops_lr}
\end{align}
Finally, there are eight operators, which couple two heavy quarks to
the gauge bosons~\cite{Franzosi:2015osa},
\begin{align}
\ope_{\phi Q}^{1} & = (\phi^\dagger\,i \stackrel{\longleftrightarrow}{D_\mu} \phi) \; (\bar{Q}\gamma^{\mu}Q) & ^\ddagger \ope_{tB} & = (\bar{Q}\sigma^{\mu\nu} t)\,\widetilde{\phi}\,B_{\mu\nu} \notag \\
\ope_{\phi Q}^{3} & = (\phi^\dagger\,i \stackrel{\longleftrightarrow}{D_\mu^I} \phi) \; (\bar{Q}\gamma^{\mu}\tau^I Q) & ^\ddagger \ope_{tW} & = (\bar{Q}\sigma^{\mu\nu} t)\,\tau^I\widetilde{\phi}\,W_{\mu\nu}^I \notag \\
\ope_{\phi t} & = (\phi^\dagger\,i \stackrel{\longleftrightarrow}{D_\mu} \phi) \; (\bar{t}\gamma^{\mu}t) & ^\ddagger \ope_{bW} & = (\bar{Q}\sigma^{\mu\nu} b)\,\tau^I\phi \,W_{\mu\nu}^I \notag \\
^\ddagger \ope_{\phi tb} & = (\widetilde{\phi}^\dagger iD_\mu \phi) \; (\bar{t}\gamma^{\mu}b) & ^\ddagger \ope_{tG} & = (\bar{Q}\sigma^{\mu\nu} T^A t)\,\widetilde{\phi}\,G_{\mu\nu}^A \,. 
\label{eq:tboson}
\end{align}
The relation of these operators with the Warsaw
basis~\cite{Grzadkowski:2010es} is worked out in the appendix of
Ref.~\cite{Brivio:2019ius}.

\begin{table}[b!]
  \centering
  \begin{small} \begin{tabular}{lc|ccccccc}
  \toprule
Wilson coeff && $t\bar{t}$ & single $t$ & $tW$ & $tZ$ & $t$-decay  &$t\bar{t}Z$ & $t\bar{t}W$ \\
  \midrule
$C_{Qq}^{1,8}$ & \multirow{6}*{Eq.\eqref{eq:ops_llrr} \;} & $\ \Lambda^{-2}$ & -- & -- & -- & -- & $\Lambda^{-2}$ & $\Lambda^{-2}$ \\
$C_{Qq}^{3,8}$ && $\ \Lambda^{-2}$ & $\Lambda^{-4}\ [\Lambda^{-2}]$ & -- & $\Lambda^{-4}\ [\Lambda^{-2}]$ & $\Lambda^{-4}\ [\Lambda^{-2}]$ & $\Lambda^{-2}$ & $\Lambda^{-2}$ \\
$C_{tu}^8,\, C_{td}^8$ && $\ \Lambda^{-2}$ & -- & -- & -- & -- & $\Lambda^{-2}$ & -- \\
$C_{Qq}^{1,1}$ && $\ \Lambda^{-4}\ [\Lambda^{-2}]$ & -- & -- & -- & -- & $\Lambda^{-4}\ [\Lambda^{-2}]$ & $\Lambda^{-4}\ [\Lambda^{-2}]$ \\
$C_{Qq}^{3,1}$ && $\ \Lambda^{-4}\ [\Lambda^{-2}]$ & $\Lambda^{-2}$ & -- & $\Lambda^{-2}$ & $\Lambda^{-2}$ & $\Lambda^{-4}\ [\Lambda^{-2}]$ & $\Lambda^{-4}\ [\Lambda^{-2}]$ \\
$C_{tu}^1,\, C_{td}^1$ && $\ \Lambda^{-4}\ [\Lambda^{-2}]$ & -- & -- & -- & -- & $\Lambda^{-4}\ [\Lambda^{-2}]$ & -- \\
  \midrule
$C_{Qu}^{8}, C_{Qd}^{8}$ & \multirow{4}*{Eq.\eqref{eq:ops_lr}} &$\ \Lambda^{-2}$ & -- & -- & -- & -- & $\Lambda^{-2}$ & -- \\
$C_{tq}^{8}$  && $\ \Lambda^{-2}$ & -- & -- & -- & -- & $\Lambda^{-2}$ & $\Lambda^{-2}$ \\
$C_{Qu}^{1}, C_{Qd}^{1}$ && $\ \Lambda^{-4}\ [\Lambda^{-2}]$ & -- & -- & -- & -- & $\Lambda^{-4}\ [\Lambda^{-2}]$ & -- \\
$C_{tq}^{1}$ && $\ \Lambda^{-4}\ [\Lambda^{-2}]$ & -- & -- & -- & -- & $\Lambda^{-4}\ [\Lambda^{-2}]$ & $\Lambda^{-4}\ [\Lambda^{-2}]$ \\
  \midrule
 
$C_{\phi Q}^-$ & \multirow{8}*{Eq.\eqref{eq:tboson}} &-- & -- & -- & $\Lambda^{-2}$ & -- & $\Lambda^{-2}$ & -- \\
$C_{\phi Q}^3$ && -- & $\Lambda^{-2}$ & $\Lambda^{-2}$  & $\Lambda^{-2}$ & $\Lambda^{-2}$  & $\Lambda^{-2}$ & -- \\
$C_{\phi t}$ && -- & -- & -- & $\Lambda^{-2}$  & -- & $\Lambda^{-2}$ & -- \\
$C_{\phi tb}$ && -- & $\Lambda^{-4}$ & $\Lambda^{-4}$ & $\Lambda^{-4}$ &  $\Lambda^{-4}$  & -- & -- \\
$C_{tZ}$ && -- & -- & -- & $\Lambda^{-2}$ & -- & $\Lambda^{-2}$ & -- \\
$C_{tW}$ && -- & $\Lambda^{-2}$ & $\Lambda^{-2}$ & $\Lambda^{-2}$ & $\Lambda^{-2}$  & -- & -- \\
$C_{bW}$ && -- & $\Lambda^{-4}$ & $\Lambda^{-4}$ & $\Lambda^{-4}$ & $\Lambda^{-4}$  & -- & -- \\
$C_{tG}$ && $\Lambda^{-2}$ & $[\Lambda^{-2}]$ & $\Lambda^{-2}$ &  -- &  $[\Lambda^{-2}]$  & $\Lambda^{-2}$ & $\Lambda^{-2}$\\
  \bottomrule
  \end{tabular} \end{small}
  \caption{Wilson coefficients and their contributions to top
    observables via SM-interference ($\Lambda^{-2}$) and via
    dimension-6 squared terms only ($\Lambda^{-4}$). A square bracket
    indicates that the Wilson coefficient contributes to the
    interference at NLO in QCD. Table adapted from
    Ref.~\cite{Brivio:2022hrb}.}
\label{tab:contribs}
\end{table}

The interactions with the physical states are given by the gauge
structure of the electroweak SM, so we use the combinations
\begin{align}
C_{\phi Q}^\pm = C_{\phi Q}^1 \pm C_{\phi Q}^3
\qquad \text{and} \qquad
C_{tZ} = c_w C_{tW} - s_w C_{tB}. 
\label{eq:operator-relations}
\end{align}
This way, $C_{\phi Q}^-$ and $C_{tZ}$ describe a $t\bar{t} Z$
interaction,  
$C_{tW}$ a $tbW$ interaction, and $C_{\phi Q}^3$ both
$tbW$ and $b\bar{b}Z$ interactions. The effect of our operators on
the different LHC observables are summarized in
Tab.~\ref{tab:contribs}. Here the main question is which operators
modify the LHC rate and kinematic predictions through interference
with the SM-matrix element which only contributes at dimension-6
squared order.

Further operators, which in principle affect top observables at tree
level or at higher perturbative orders are strongly constrained by other
observables and not included in this analysis. For instance, the ubiquitous triple-gluon
coupling is strongly constrained by multi-jet production or black hole
searches~\cite{Krauss:2016ely}. Operators with four heavy quarks are
starting to be constrained by LHC measurements, but these modest
constraints are not expected to feed back into the standard top
observables: weak correlations between these operators and the remainder of
the top sector of the SMEFT were observed in Ref.~\cite{Kassabov:2023hbm}.  
The effect of normalization group evolution on the Wilson coefficients of 
the SMEFT is also neglected here, but would be an interesting question~\cite{Aoude:2022aro}.

\subsection{Data, predictions, and uncertainties}
\label{sec:data}

\begin{table}[b!]
\centering
\begin{small} \begin{tabular}{lcccccccc}
        \toprule
        Experiment & Energy [TeV] & $\mathcal{L}$ [fb$^{-1}$] & Channel & Observable & \# Bins & New & Likelihood & QCD k-factor \\
        \midrule
        CMS\hfill~\cite{CMS:2016yys} & 8 & 19.7 & $e \mu$ & $\sigma_{t \bar{t}}$ &  &  &  &\cite{Czakon:2011xx} \\
        ATLAS\hfill~\cite{ATLAS:2017wvi} & 8 & 20.2 & $lj$ & $\sigma_{t \bar{t}}$ &  &  & & \cite{Czakon:2011xx} \\
        \midrule
        CMS\hfill~\cite{CMS:2021vhb} & 13 & 137 & $lj$ & $\sigma_{t \bar{t}}$ &  & \checkmark & & \cite{Czakon:2011xx} \\
        CMS\hfill~\cite{CMS:2018fks} & 13 & 35.9 & $ll$ & $\sigma_{t \bar{t}}$ &  & & & \cite{Czakon:2011xx} \\
        ATLAS\hfill~\cite{ATLAS:2019hau} & 13 & 36.1& $ll$ & $\sigma_{t \bar{t}}$ &  & \checkmark & & \cite{Czakon:2011xx} \\
        ATLAS\hfill~\cite{ATLAS:2020ccu} & 13 & 36.1 & $aj$ & $\sigma_{t \bar{t}}$ &  & \checkmark & & \cite{Czakon:2011xx} \\
        ATLAS\hfill~\cite{ATLAS:2020aln} & 13 & 139 & $lj$ & $\sigma_{t \bar{t}}$ &  & \checkmark & \checkmark & \cite{Czakon:2011xx} \\
        \midrule
        CMS\hfill~\cite{CMS:2023qyl} & 13.6 & 1.21 & $ll$, $lj$ & $\sigma_{t \bar{t}}$ &  & \checkmark & & \cite{CMS:2023qyl} \\
        \midrule
        CMS\hfill~\cite{CMS:2015rld} & 8 & 19.7 & $lj$ & $\frac{1}{\sigma} \frac{d \sigma}{d p_{T}^{t}}$ & 7 &  & &\cite{Czakon:2017dip,Czakon:2015owf,Czakon:2016dgf} \\
    CMS\hfill~\cite{CMS:2015rld} & 8 & 19.7 & $l l$ & $\frac{1}{\sigma} \frac{d \sigma}{d p_{T}^{t}}$ & 5 &  &  &\cite{Czakon:2017dip,Czakon:2015owf,Czakon:2016dgf} \\
        ATLAS\hfill~\cite{ATLAS:2015lsn} & 8 & 20.3 & $lj$ & $\frac{1}{\sigma} \frac{d \sigma}{d m_{t \bar{t}}}$ & 7 &  &  &\cite{Czakon:2017dip,Czakon:2015owf,Czakon:2016dgf} \\
        \midrule
        CMS\hfill~\cite{CMS:2021vhb} & 13 & 137 & $lj$ & $\frac{1}{\sigma} \frac{d \sigma}{d m_{t \bar{t}}}$  & 15 & \checkmark  & &\cite{Czakon:2023hls} \\
        CMS\hfill~\cite{CMS:2018adi} & 13 & 35.9 & $ll$ & $\frac{1}{\sigma} \frac{d \sigma}{d \Delta y_{t \bar{t}}}$ &  8 &  & &\cite{Czakon:2017dip,Czakon:2015owf,Czakon:2016dgf} \\
        ATLAS\hfill~\cite{ATLAS:2019hxz} & 13 & 36 & $lj$ & $\frac{1}{\sigma} \frac{d \sigma}{d m_{t \bar{t}}}$ & 9  & \checkmark & &\cite{Czakon:2023hls} \\
        ATLAS\hfill~\cite{ATLAS:2022mlu} & 13 & 139 & $aj$, high-$p_{T}$& $\frac{1}{\sigma} \frac{d \sigma}{d m_{t \bar{t}}}$ & 13 & \checkmark & &  \\
        \midrule
        CMS\hfill~\cite{CMS:2015pob} & 8 & 19.7 & $lj$ & $A_{C}$ &  &  & &\cite{Czakon:2017lgo} \\
    CMS\hfill~\cite{CMS:2016ypc} & 8 & 19.5 & $ll$ & $A_{C}$ &  &  & &\cite{Czakon:2017lgo} \\
        ATLAS\hfill~\cite{ATLAS:2015jgj} & 8 & 20.3 & $lj$ & $A_{C}$ &  &  & &\cite{Czakon:2017lgo} \\
        ATLAS\hfill~\cite{ATLAS:2016ykb} & 8 & 20.3 & $ll$ & $A_{C}$ &  &  & &\cite{Czakon:2017lgo} \\
        \midrule
	CMS\hfill~\cite{CMS:2022ged} & 13 & 138 & $lj$ & $A_{C}$ &  & \checkmark & &\cite{Czakon:2017lgo} \\
	ATLAS\hfill~\cite{ATLAS:2022waa} & 13 & 139 & $lj$ & $A_{C}$ &   & \checkmark  & &\cite{Czakon:2017lgo} \\
        \midrule
    ATLAS\hfill~\cite{ATLAS:2021fzm} & 13 & 139 & & $\sigma_{t \bar{t} Z}$ &  & \checkmark & \checkmark  &\cite{Kulesza:2018tqz} \\
CMS\hfill~\cite{CMS:2019too} & 13 & 77.5 & & $\sigma_{t \bar{t} Z}$  &   &   &  &\cite{Kulesza:2018tqz} \\
	\midrule
	CMS\hfill~\cite{CMS:2017ugv} & 13 & 35.9 & & $\sigma_{t \bar{t} W}$ &  &   &  &\cite{Kulesza:2018tqz} \\
	ATLAS\hfill~\cite{ATLAS:2019fwo} & 13 & 36.1 & & $\sigma_{t \bar{t} W}$ &  & \checkmark   &  &\cite{Kulesza:2018tqz} \\
    \midrule
	CMS\hfill~\cite{CMS:2017tzb} & 8 & 19.7 & & $\sigma_{t \bar{t} \gamma}$ &  &  \checkmark &  & \\
ATLAS\hfill~\cite{ATLAS:2017yax} & 8 & 20.2 & & $\sigma_{t \bar{t} \gamma}$ &  & \checkmark   &  & \\
        \bottomrule
\end{tabular} \end{small}
\caption{Top pair observables included in our global analysis.  `New'
  is defined relative to the previous \sfitter
  analysis~\cite{Brivio:2019ius}.  `Likelihood' indicates a dataset
  for which a public likelihood is available --- further details of
  these datasets are provided in
  Sec.~\ref{sec:like}.}
\label{tab:datasets_ttbar}
\end{table}

\begin{table}[b!]
\centering
	\begin{small} \begin{tabular}{lcccccccc}
        \toprule
		Exp. & $\sqrt{s}$ [TeV] & $\mathcal{L}$ [fb$^{-1}$] & Channel & Observable & \# Bins & New  & Likelihood & QCD k-factor \\
        \midrule
		ATLAS\hfill~\cite{ATLAS:2014sxe} & 7 & 4.59 &$t$-ch & $\sigma_{tq + \bar{t} q}$  &    &  & & \\
    CMS\hfill~\cite{CMS:2012xhh} & 7 & \footnotesize{1.17 ($e$), 1.56 ($\mu$)} &$t$-ch & $\sigma_{tq + \bar{t} q}$  &   &  & & \\
		ATLAS\hfill~\cite{ATLAS:2017rso} & 8 & 20.2 &$t$-ch & $\sigma_{tq}$, $\sigma_{\bar{t}q}$  &   &  & & \\
		CMS\hfill~\cite{CMS:2014mgj} & 8 & 19.7 &$t$-ch & $\sigma_{tq}$, $\sigma_{\bar{t}q}$  &   &  & & \\
		ATLAS\hfill~\cite{ATLAS:2016qhd} & 13 & 3.2 &$t$-ch & $\sigma_{tq}$, $\sigma_{\bar{t}q}$ &    &  & &\cite{Berger:2016oht} \\
		CMS\hfill~\cite{CMS:2016lel} & 13 & 2.2 &$t$-ch & $\sigma_{tq}$, $\sigma_{\bar{t}q}$   &  & &  &\cite{Berger:2016oht} \\
		CMS\hfill~\cite{CMS:2019jjp} & 13 & 35.9 &$t$-ch & $\frac{1}{\sigma} \frac{d \sigma}{d |p_{T,t}|}$  &  5  &  \checkmark & &  \\
	\midrule 
		CMS\hfill~\cite{CMS:2016xoq} & 7 & 5.1 & $s$-ch & $\sigma_{t \bar{b} + \bar{t} b}$   &  &   & &  \\
		CMS\hfill~\cite{CMS:2016xoq} & 8 & 19.7 & $s$-ch & $\sigma_{t \bar{b} + \bar{t} b}$   &  &  & & \\
		ATLAS\hfill~\cite{ATLAS:2015jmq} & 8 & 20.3 & $s$-ch & $\sigma_{t \bar{b} + \bar{t} b}$  &   &  & & \\
		ATLAS\hfill~\cite{ATLAS:2022wfk} & 13 & 139 & $s$-ch & $\sigma_{t \bar{b} + \bar{t} b}$ & & \checkmark & \checkmark &   \\
	\midrule
		ATLAS\hfill~\cite{ATLAS:2012bqt} & 7 & 2.05 & $tW$ $(2l)$ & $\sigma_{t W + \bar{t} W}$ & &      & & \\
		CMS\hfill~\cite{CMS:2012pxd} & 7 & 4.9 & $tW$ $(2l)$ & $\sigma_{t W + \bar{t} W}$   &  &  & &  \\
		ATLAS\hfill~\cite{ATLAS:2015igu} & 8 & 20.3 & $tW$ $(2l)$ & $\sigma_{t W + \bar{t} W}$   &  &  & &  \\
		ATLAS\hfill~\cite{ATLAS:2020cwj} & 8 & 20.2 & $tW$ $(1l)$ & $\sigma_{t W + \bar{t} W}$ & & \checkmark   & &  \\
		CMS\hfill~\cite{CMS:2014fut} & 8 & 12.2 & $tW$ $(2l)$ & $\sigma_{t W + \bar{t} W}$   &  &  & &  \\
		ATLAS\hfill~\cite{ATLAS:2016ofl} & 13 & 3.2 & $tW$ $(1l)$ & $\sigma_{t W + \bar{t} W}$ &  &  & &  \\
		CMS\hfill~\cite{CMS:2018amb} & 13 & 35.9 & $tW$ $(e \mu j)$  & $\sigma_{t W + \bar{t} W}$ &   &   & &  \\
		CMS\hfill~\cite{CMS:2021vqm} & 13 & 36 & $tW$ ($2l$) & $\sigma_{t W + \bar{t} W}$ & &  \checkmark    & &  \\
    \midrule
ATLAS\hfill~\cite{ATLAS:2017dsm} & 13 & 36.1 & $tZ$ & $\sigma_{t Zq}$ & &     & &  \\
	\midrule
        ATLAS\hfill~\cite{ATLAS:2012nhi} & 7 & 1.04 &  & $F_0$,$F_L$ &   &  & &  \\
	CMS\hfill~\cite{CMS:2013xxb} & 7 & 5 & & $F_0$,$F_L$ & & & &  \\
    ATLAS\hfill~\cite{ATLAS:2016fbc} & 8 & 20.2 &  & $F_0$,$F_L$  &  & &  &  \\
	CMS\hfill~\cite{CMS:2016asd} & 8 & 19.8 &  & $F_0$,$F_L$  &  & &  &  \\
	ATLAS\hfill~\cite{ATLAS:2022rms} & 13 & 139 & & $F_0$,$F_L$  & & \checkmark   & & \\
        \bottomrule
\end{tabular} \end{small}
\caption{Single top and top decay observables included in our global
  analysis.  `New' is defined relative to the previous \sfitter
  analysis~\cite{Brivio:2019ius}.  `Likelihood' indicates a dataset
  for which a public likelihood is available --- further details of
  these datasets are provided in Sec.~\ref{sec:like}.}
\label{tab:datasets_singletop}
\end{table}

The technical goal of our study is to integrate, for the first time, published
experimental statistical models into an analysis of the SMEFT.
For the purpose of this study, we analyze three measurements for which
likelihoods are available in the
\texttt{HistFactory}~\cite{Cranmer:2012sba} format on
\texttt{HEPData}: an ATLAS measurement of the total inclusive $t
\bar{t}$ cross section~\cite{ATLAS:2020aln}, an ATLAS measurement of
the total inclusive $t \bar{t} Z$ cross section~\cite{ATLAS:2021fzm}
and an ATLAS measurement of the total inclusive single-top cross
section in the $s$-channel~\cite{ATLAS:2022wfk}. The implementation
of these likelihoods into the \sfitter framework will be discussed in
more detail in Section~\ref{sec:like}. To obtain a realistic
assessment of the effect of these likelihoods on the SMEFT, we incorporate them into a global analysis.

With this in mind, our analysis in the top sector will consider all measurements listed in
Tables~\ref{tab:datasets_ttbar} and~\ref{tab:datasets_singletop}.
The analysis is an update to a previous global top analysis performed by \sfitter, in
Ref.~\cite{Brivio:2019ius}.  We highlight in Tables~\ref{tab:datasets_ttbar}
and~\ref{tab:datasets_singletop} the measurements that are new
relative to those included in Ref.~\cite{Brivio:2019ius}, as well as
those for which a public likelihood is available.  
Where possible, we make use of measurements
encompassing the full Run II LHC luminosity and choose measurements
in the boosted regime in which sensitivity to energy-growing SMEFT
operators is maximized; see, for example, the top pair
production invariant mass distribution of Ref.~\cite{ATLAS:2022mlu}.
The dataset consists of a total of 122 data points spanning the 
$t \bar{t}$, $t \bar{t} + X (Z, W, \gamma)$ and single top ($s$, $t$-channel, $tW$ 
and $tZ$) sectors, including measurements of top-pair production 
charge asymmetries $A_{C}$ and $W$ boson polarization in top decays ($F_0, F_L$).


A key ingredient to all global analyses are precision predictions from 
perturbative quantum field theory. 
Most observables considered in this analysis are unfolded to parton level,
assuming stable top quarks. This allows us to use fixed-order
calculations to determine the SM predictions at NLO in QCD using
\texttt{MadGraph5 aMC@NLO}~\cite{Frederix:2018nkq,Alwall:2014hca} and
\texttt{NNPDF 4.0}~\cite{NNPDF:2021njg} interfaced with
\texttt{LHAPDF}~\cite{Buckley:2014ana}.
Alongside the observables listed in Tables~\ref{tab:datasets_ttbar}
and~\ref{tab:datasets_singletop}, we note whether the SM
predictions for these observables are approximated at NNLO in QCD
using a $K$-factor approximation and referencing the source of these
QCD $K$-factors. In the case of new top quark pair production
observables these QCD $K$-factors
are calculated using \texttt{HighTea}~\cite{Czakon:2023hls}.

Calculations of the effect of the SMEFT on all updated measurements in the top sector are
performed at NLO in QCD using the FeynRules~\cite{Alloul:2013bka}
model \texttt{SMEFTatNLO}~\cite{Degrande:2020evl} up to quadratic
order in the EFT expansion. The exceptions are the measurements of 
the $t \bar{t} \gamma$ total cross sections at 8 TeV by ATLAS~\cite{ATLAS:2017yax} 
and CMS~\cite{CMS:2017tzb}, for which the SMEFT predictions at LO in QCD 
are taken from Ref.~\cite{Kassabov:2023hbm}.

Theory uncertainties appear whenever we compare a measurement to a
first-principle description. In principle, they cover a wide range of
approximations which we make to be able to calculate, for example, an
LHC cross section from a fundamental renormalized Lagrangian. For the
LHC, they are dominated by the truncation of the perturbative series,
in QCD and the electroweak gauge coupling. Because these
perturbative series converge very slowly for LHC rates, theory
predictions have become limiting factors for the interpretation of
many LHC measurements in terms of actual physics. Aside from the size
of the theory uncertainties, it is problematic that they do not follow any
statistical pattern or model~\cite{Ghosh:2022lrf}, and assuming a 
Gaussian distribution is neither justified nor conservative.

Because of their impact on global analyses of effective Lagrangians,
\sfitter puts an emphasis on the proper description of these
uncertainties, including their correlations between different
observables. We will describe this treatment in more detail in
Sec.~\ref{sec:sfitter}. In the top sector, the theory uncertainties are
critical for the precisely measured top pair production
rates~\cite{Brivio:2019ius} and are correlated between different final
states for rate measurements. We typically use the theory
uncertainties reported in the respective publications, with the 
exception that we enforce a minimum scale uncertainty of
10\% for total rates in single top production and 2\% for bin-wise kinematic
distributions.

\subsubsection*{Boosted top pair production}
\label{subsec:boost}

\begin{figure}[t]
\includegraphics[width=0.49\textwidth]{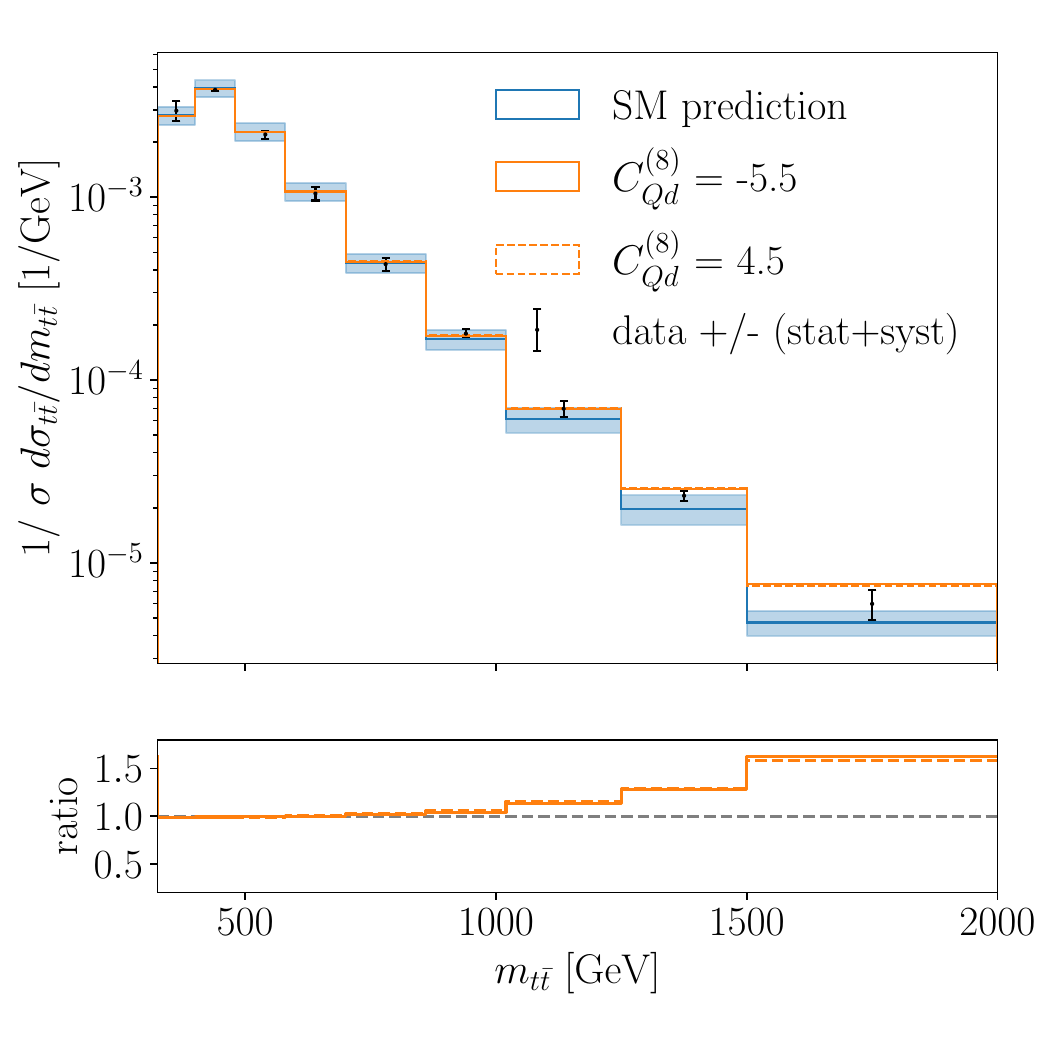}
\includegraphics[width=0.49\textwidth]{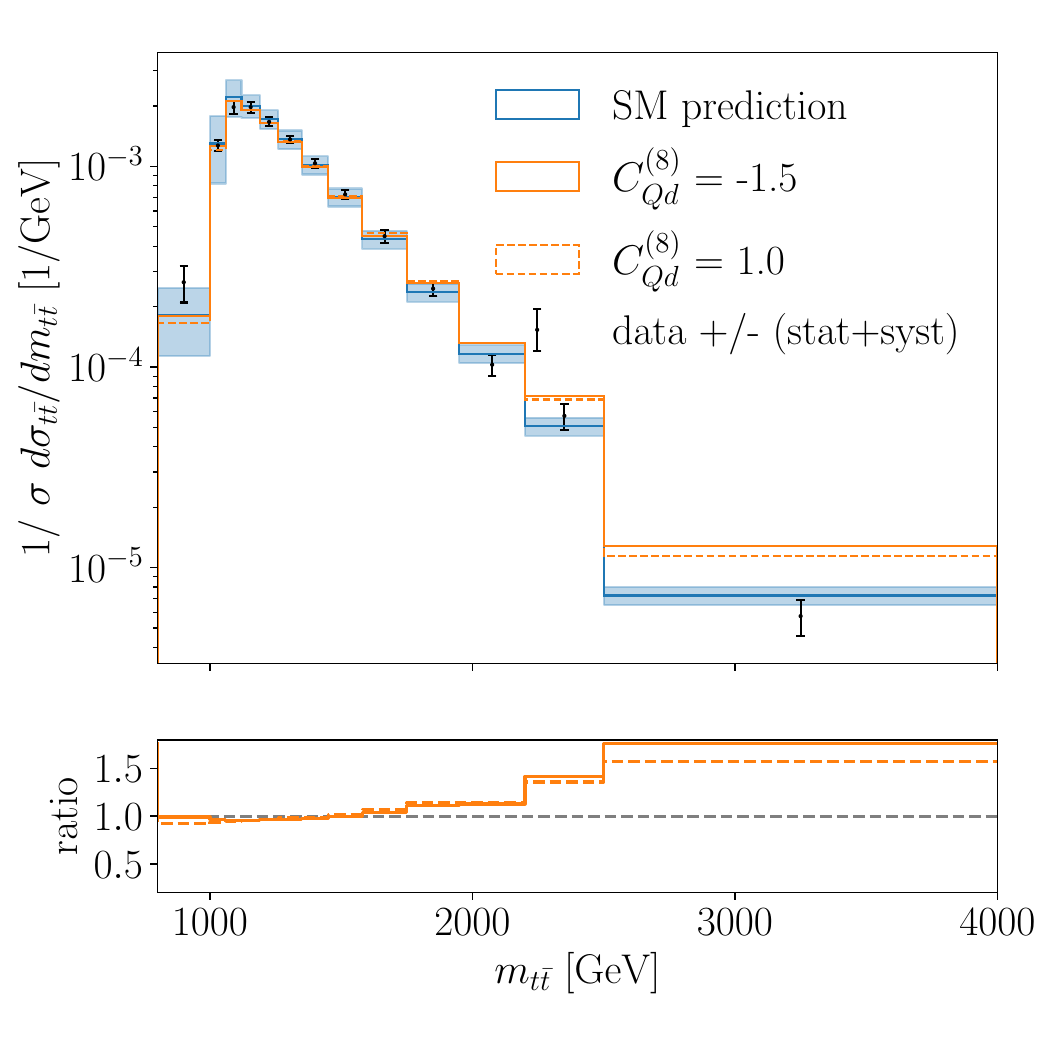}
\caption{Left: impact of $\ope_{Qd}^{(8)}$ on the unfolded ATLAS $m_{t\bar{t}}$ distribution in the 
lepton+jets channel~\cite{ATLAS:2019hxz}.  Right: impact of this operator on the unfolded ATLAS $m_{t\bar{t}}$ distribution in the 
all-hadronic channel measured with boosted top quarks~\cite{ATLAS:2022mlu}.}
\label{fig:boosted_SFitter_results}
\end{figure}

As part of our dataset, we highlight the 
reinterpretation of the ATLAS measurement of $t \bar{t}$ 
production in the lepton+jets channel~\cite{ATLAS:2019hxz} and the ATLAS 
measurement of
$t \bar{t}$ production using boosted top quarks in the all-hadronic 
channel~\cite{ATLAS:2022mlu}. Both are differential in the top-pair invariant 
mass, as shown in Fig.\ref{fig:boosted_SFitter_results}.  The measurement using 
boosted top quarks is unfolded to a fiducial parton-level phase space, 
defined by 
\begin{align}
 p_{T,t_1} > 500~\gev
 \qquad \text{and} \qquad 
 p_{T,t_2} > 350~\gev \; ,
\end{align}
allowing for an easy comparison with fixed-order calculations. 
This, alongside the high-$m_{t \bar{t}}$ reach of this 
distribution makes it an excellent candidate for constraining the energy-
growing SMEFT four-fermion operators of the top sector. We display the impact of 
one of these operators, $\ope_{Qd}^{8}$, in 
Fig.\ref{fig:boosted_SFitter_results}.  

The theory uncertainty is shown in blue in both figures and compared to the 
statistical and systematic uncertainties in the experimental data.  In both 
cases, the values of $\mathcal{C}_{Qd}^{8}$ chosen are those which would produce 
a $3 \sigma$ deviation in a one-parameter analysis.
We observe that, while the measurement unfolded to the full phase space is 
sensitive to the energy-growing effects of $\ope_{Qd}^{8}$, this 
sensitivity is significantly enhanced by the measurement of boosted top 
quarks. 

\subsection{SFitter}
\label{sec:sfitter}

The \sfitter
framework~\cite{Lafaye:2004cn,Lafaye:2007vs,Lafaye:2009vr} has been
developed for global analyses of LHC measurements in the context of
BSM physics and Higgs
properties~\cite{Klute:2012pu,Corbett:2015ksa,Butter:2016cvz,Biekotter:2018rhp,Brivio:2019ius},
including comprehensive studies on Higgs and electroweak properties
induced by actual UV-completion of the
SM~\cite{Lopez-Val:2013yba,Brivio:2021alv}, an extrapolation for
the HL-LHC~\cite{Brivio:2022hrb}, as well as for future electron-positron colliders~\cite{Klute:2013cx}.

The relation to full models and a proper treatment of
uncertainties in precision matching is crucial for the LHC because
the typical scale separation between directly probed energies and
indirectly accessible energies is not very large. On the other hand,
the consistency of the EFT description is not a universal property of
the EFT Lagrangian, but only defined by possible on-shell propagators
in the observables and relative to the UV-completion and its typical
coupling strengths. Without additional information on the underlying
model the Lagrangian in Eq.\eqref{eq:d6lag} is degenerate along $C_k
\sim \Lambda^2$, which means the EFT assumption of large $\Lambda$
improves for larger postulated couplings. This is the reason why
\sfitter SMEFT analyses start with the truncated dimension-6
Lagrangian at face value.

From Tab.~\ref{tab:contribs}, we know that some Wilson coefficients do
not interfere with the SM matrix elements at leading order, so we
include dimension-6 squared contributions to the LHC observables. This
means we truncate the Lagrangian rather than the LHC rate
prediction. We emphasize that all our assumptions are neither inherently
right nor wrong, and need to be validated for a given dataset
and a given UV-completion~\cite{Lopez-Val:2013yba,Dawson:2020oco,Dawson:2021xei,Brivio:2021alv}.
However, our assumptions ensure that the
\sfitter analysis makes optimal use of the kinematic information,
especially in the tails of momentum or energy distributions.

At the heart of \sfitter is the extraction of the fully exclusive
likelihood, given a rate measurement $d$ from 
Sec.~\ref{sec:data}, evaluated over the combined space of Wilson
coefficients $c$ and nuisance parameters $\theta$,
\begin{align}
  p(d|c,\theta) 
  = \text{Pois}(d|m(c,\theta,b)) \; 
    \text{Pois}(b_\text{CR}|b\,k) \; 
    \prod_{i} \mathcal{C}_i(\theta_{i},\sigma_{i}) \; .
    \label{eq:sfitter_like}
\end{align}
It incorporates the effects of  the statistical, systematic, and theory uncertainties.
The first Poisson distribution gives the probability to observe $d$
events given the corresponding theory prediction $m(c,\theta,b)$,
which in turn depends on the predicted background count $b$. The
background rate is, itself, constrained by measurements $b_\text{CR}$
in the control region, implemented as a scaled prediction $kb$ with a
suitable factor $k$. The constraint function $\mathcal{C}$ gives the
distribution of the nuisance parameter $\theta_i$, given a width
measure $\sigma_i$. Depending on the source of the uncertainty, it can
be chosen as follows:
\begin{itemize}
    \item Gaussian, for systematic uncertainties related to
      independent measurements in other channels. Examples are other
      LHC rates, but also calibration. We take $\sigma_i$ from the
      respective experimental publications. As we will discuss in
      Sec.~\ref{sec:like}, public likelihoods will help here;
    \item Flat, for theory uncertainties which do not have a
      well-defined maximum and could be thought of as a
      range~\cite{Ghosh:2022lrf}. Examples are scale uncertainties for
      QCD predictions and PDF uncertainties. They are usually taken
      from the experimental publications, but we increase them
      whenever the standard choice appears not conservative.
\end{itemize}
The flat scale uncertainty is not parametrization invariant, as one
would expect from a fixed range, but without a preferred central value
we consider it conservative. Scale uncertainties are obtained by 
varying the renormalization and factorization scales $\mu_{R}$ and $\mu_{F}$ 
by a factor of 2 around their respective central value. These are process 
dependent and chosen to be $\mu_{R} = \mu_{F} = m_{t} + \frac{1}{2}m_{V}$ 
for associated $t\bar{t}$ production with $V=W,Z$. For $t\bar{t}$ production, 
the sum of the transverse masses of the top and anti-top is used, while for 
single top production they are set to the top mass $m_{t}$.

By ansatz, \sfitter treats all measurements $d$ as uncorrelated, 
constructing an individual likelihood for each measurement, 
as defined in Eq.\eqref{eq:sfitter_like}. This assumption is justified 
by the individual statistical uncertainties described by the Poisson distributions 
in the likelihood. The full likelihood can thus be constructed as the product 
of these individual contributions. As a consequence, the Gaussian constraint 
terms describing the systematics of each measurement can be generalized to a 
single higher-dimensional Gaussian which allows correlations between 
uncertainties to be introduced. In the case of Gaussian systematics, 
the correlations are described by a correlation matrix~\cite{Bissmann:2019qcd}

\begin{align}
    C_{ij} = \frac{\sum_{\text{syst}} \rho_{ij} \sigma_{i,\text{syst}}\sigma_{j,\text{syst}} }{\sigma_{i,\text{exp}}\sigma_{j,\text{exp}}} 
    \qquad  \text{with} \qquad 
    \sigma_{i,\text{exp}}^{2} = \sum_{\text{syst}}\sigma_{i,\text{syst}}^{2} + \sum_{\pois}\sigma_{i,\pois}^{2} \, .
\end{align}
Here $i,j$ run over all measurements and $\sigma_{\text{syst}}, \sigma_{\pois}$
are the systematic and Poisson uncertainties. We then choose $\rho_{ij}$ 
to be either uncorrelated or (essentially) fully correlated for systematics of the same type. 

Theory uncertainties are correlated for all measurements with identical
predictions. They are also correlated within one measurement across all bins 
but not across several different measurements. 
This is done by averaging them, weighted such that the final
standard deviation is minimized. Using the prediction only once for this 
weighted average, instead of each individual measurement, ensures the 
proper correlations of the corresponding theory uncertainties.
The implementation of
flat theory uncertainties allows for a shift of the prediction within 
their bounds without affecting the likelihood value.

To construct the exclusive likelihood, \sfitter uses cross section
predictions over the entire model parameter space and extracts the
quadratic behavior analytically, which guarantees sufficient precision
even for small Wilson coefficients. We then use a Markov chain to
evaluate Eq.\eqref{eq:sfitter_like} numerically and to encode the
likelihood in the distribution of points covering the combined space
of Wilson coefficients and nuisance parameters. The setup of the
Markov chain depends on the structure of the model space; for BSM
analyses an efficient search for local structures in the global model
space is important, while for SMEFT analyses we know that the global
likelihood maximum will be close to the SM-point. Adjusting the Markov
chain accordingly leads to a significant speed
improvement~\cite{Brivio:2021alv}. Our motivation to use a Markov
chain rather than so-called toys is described in the Appendix.

Finally, to combine uncertainties by removing nuisance parameters or
to reduce the space of physical Wilson coefficients, \sfitter can
employ a profile likelihood or a Bayesian
marginalization~\cite{Lafaye:2007vs,Brivio:2022hrb}.  Obviously, these
two methods give different results. Only for uncorrelated Gaussians do 
the profile likelihood and Bayesian marginalization lead to the common
result of errors added in quadrature.  For a flat likelihood, the
uncorrelated profile likelihood adds the two uncertainties linearly,
which happens for the scale uncertainty and the PDF uncertainty in
\sfitter.  The profile likelihood combination of a flat and a Gaussian
uncertainty gives the well-known RFit
prescription~\cite{Hocker:2001xe}. In contrast, when applying marginalization 
on the combination of Gaussian and flat uncertainties, the central limit 
theorem ensures that the final posterior will be Gaussian again.

\section{Public likelihoods}
\label{sec:like}

\begin{table}[b!]
\centering
    \begin{tabular}{lll}
     \toprule
     Description & Modification & Constraint $\mathcal{C}$ \\
     \midrule
     Luminosity ('lumi') & $\kappa_{sb} = \lambda$ & $\mathcal{N}(l=\lambda_0|\lambda,\sigma_{\lambda})$\\
     Normalization unc. ('normsys') & $\kappa_{sb}=g_p(\alpha|\kappa_{sb,\alpha=\pm 1})$ & $\mathcal{N}(a=0|\alpha,\sigma=1)$ \\
     Correlated Shape ('histosys') & $\Delta_{sb}= f_p(\alpha|\Delta_{sb,\alpha=\pm 1})$ & $\mathcal{N}(a=0|\alpha,\sigma=1)$\\
     MC Stat. ('staterror') & $\kappa_{sb} = \gamma_b$ & $\prod_b \mathcal{N}(a_{\gamma_b}=1|\gamma_b,\delta_b)$\\
     Uncorrelated Shape ('shapesys') & $\kappa_{sb} = \gamma_b$ & $\prod_b\text{Pois}(\sigma_b^{-2}|\sigma_b^{-2}\gamma_b)$\\
     Normalization ('normfactor') & $\kappa_{sb} = \mu$ & \\
     \bottomrule
    \end{tabular}
    \caption{List of modifiers in the construction of the HistFactory
      likelihoods, adapted from Ref.~\cite{ATLAS:2019oik}. Per-bin modifiers are 
      denoted as $\gamma_{b}$, while interpolated modifiers are denoted as $\alpha$. 
      Here $g_p$ and $f_p$ describe different interpolation strategies used to compute 
      these from the values $\kappa_{sb,\alpha=\pm 1},\Delta_{sb,\alpha=\pm 1}$ provided 
      in the likelihood. Luminosity and scale factors affect all bins equally and are 
      denoted as $\lambda$ and $\mu$, respectively.}
    \label{tab:PyhfNP}
\end{table}

For a standard \sfitter analysis, we extract systematic uncertainties for each measurement
from the respective experimental publications.  Systematics of the same type
are fully correlated between measurements of the same experiment. This approach has  
drawbacks, for instance, we can only use the uncertainty
categories reported in the experimental publications or on HEPData, and this 
information often needs to be extracted by
hand. Public likelihoods include the full information on a large
number of systematic uncertainties in a
documented manner, making their implementation more accurate and efficient.

Likelihoods are published in the HistFactory format~\cite{Cranmer:2012sba}, similar 
to the \sfitter
likelihood in Eq.\eqref{eq:sfitter_like}. For each bin $b$ measured in 
a kinematic distribution of a given channel or final state, it provides
\begin{align}
  p(d_b | \mu, \theta) 
  = \text{Pois}(d_b|m_b(\mu, \theta)) \; 
  \prod_i
  \mathcal{C}_i(a_i|\theta_i) \; ,
\label{eq:likelihoodHistfactory}
\end{align}
where $d_b$ and $m_b$ are the measured and expected number of
events in bin $b$. The nuisance parameters $\theta_i$
are constrained by $\mathcal{C}_i(a_i|\theta_i)$ with the auxiliary
data $a_i$. The parameter of interest $\mu$ describes, for instance, 
a signal strength. It corresponds to the Wilson coefficient in 
Eq.\eqref{eq:sfitter_like}.

The analysis of these likelihoods is performed using \texttt{pyhf}~\cite{pyhf,pyhf_joss}, a python module allowing for easy construction of HistFactory
likelihoods and their subsequent statistical analysis. It uses
data published in the JSON format to compute the predicted number of
events using
\begin{align}
  m_b = \sum_s
  \left( \prod_\kappa \kappa_{sb} \right)
    \left( \bar{m}_{sb} + \sum_\Delta \Delta_{sb} \right) \; ,
\end{align}
with the nominal expected rate $\bar{m}_{sb}$ and multiplicative
$(\kappa_{sb})$ and additive $(\Delta_{sb})$ modifiers for each physics process
$s$. These modifiers correspond to the nuisance parameters affecting
the event rate $m_b$. The type of modifier and the constraints on
its corresponding nuisance parameter depend on the type of
uncertainty. The most common are given in Tab.~\ref{tab:PyhfNP}.
Using the public likelihoods in terms of modifiers and nominal
rates $\bar{m}_{sb}$, we can reproduce the experimental results. 
For visualization, we use \texttt{cabinetry}~\cite{Cranmer:2021oxr}, 
a python library making use of \texttt{pyhf} for statistical analyses.

Starting from the public likelihoods, we organize the full set of 
nuisance parameters corresponding
to systematic uncertainties in a small number of categories. This allows for an easier
numerical treatment at, essentially, no cost. To compute the ranges of 
nuisance parameters for
these categories, we first use a profile likelihood to determine the
central values and, in a second step, an analysis of the
distribution of the nuisance parameter.
In this section, we will show how we implement and
test three public ATLAS likelihoods.

\begin{figure}[b!]
    \includegraphics[width=0.495\textwidth]{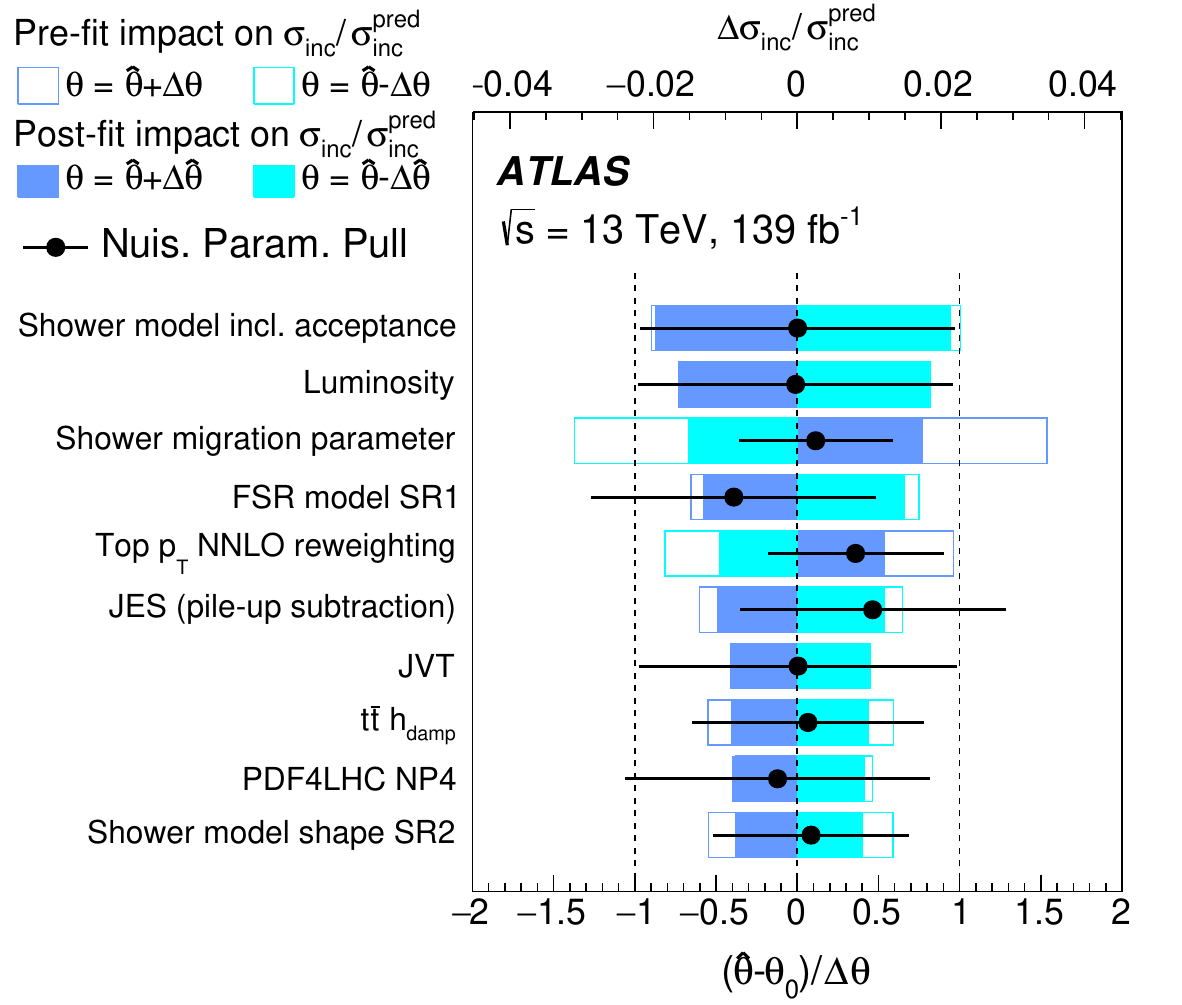}
    \includegraphics[width=0.495\textwidth]{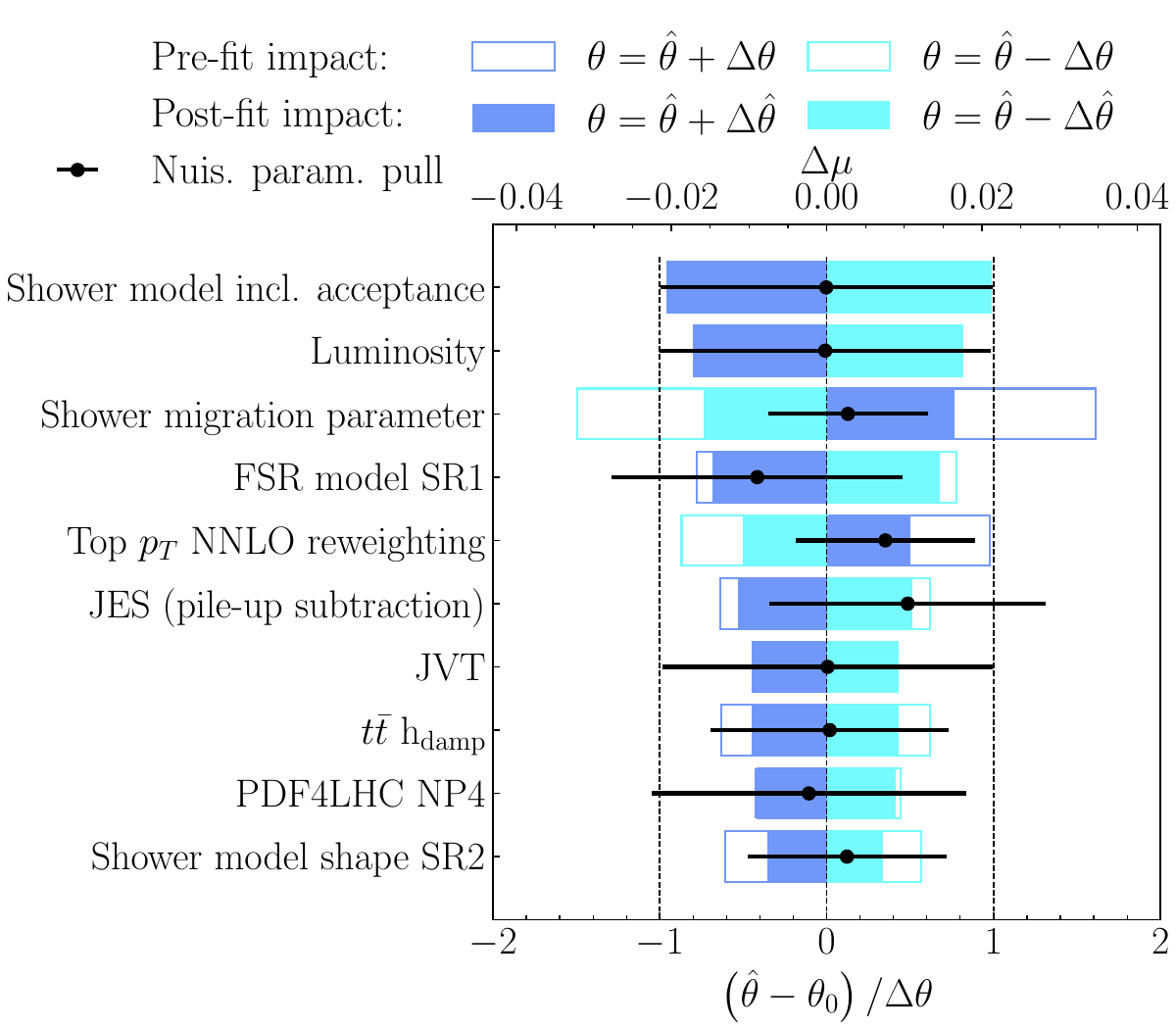}
    \caption{Impact of nuisance parameters on the $t \bar{t}$ total
      rate fit. We compare the ATLAS result~\cite{ATLAS:2020aln}
      (left) and our evaluation of the public likelihood (right).}
    \label{fig:atlas_ttbar_ranking}
\end{figure}

\subsection{ATLAS $t \bar{t}$ likelihood}

\begin{figure}[t]
    \includegraphics[width=0.45\textwidth]{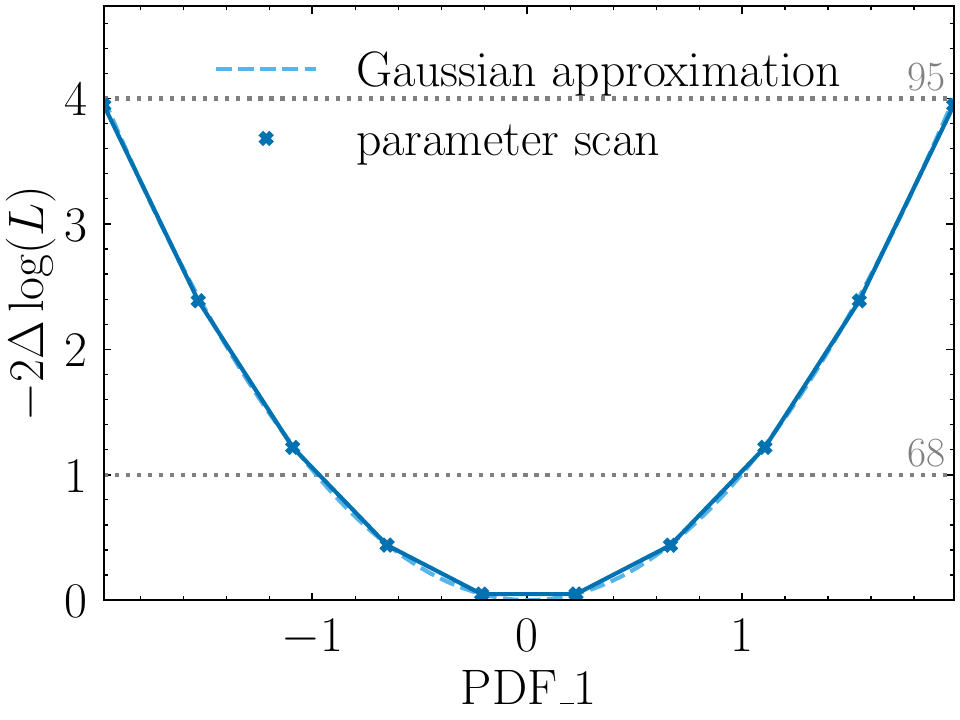}
    \hspace*{0.09\textwidth}
    \includegraphics[width=0.45\textwidth]{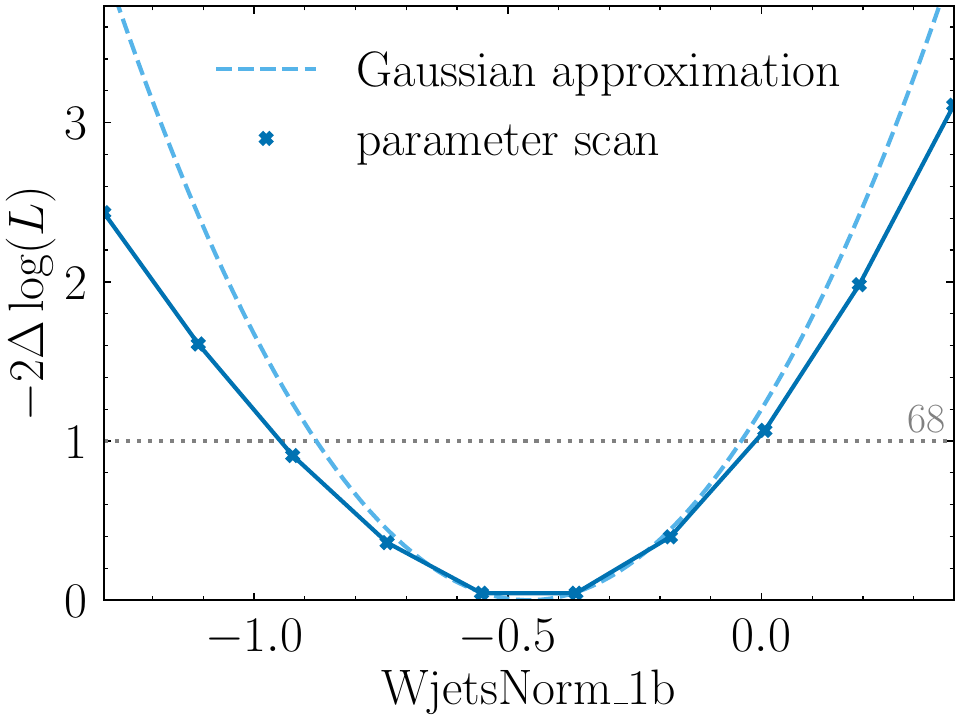}
    \caption{Dependence of the log-likelihood for the $t \bar{t}$ rate
      on two nuisance parameters, one describing the PDF uncertainty and one describing the $W +$ jets background normalization uncertainty, compared with the Gaussian
      approximation.}
    \label{fig:ttbar_likelihood_nuisance}
\end{figure}

\begin{figure}[b!]
    \begin{tabular}{p{0.5\textwidth} p{0.45\textwidth}}
        \vspace{0pt} \includegraphics[width=0.5\textwidth]{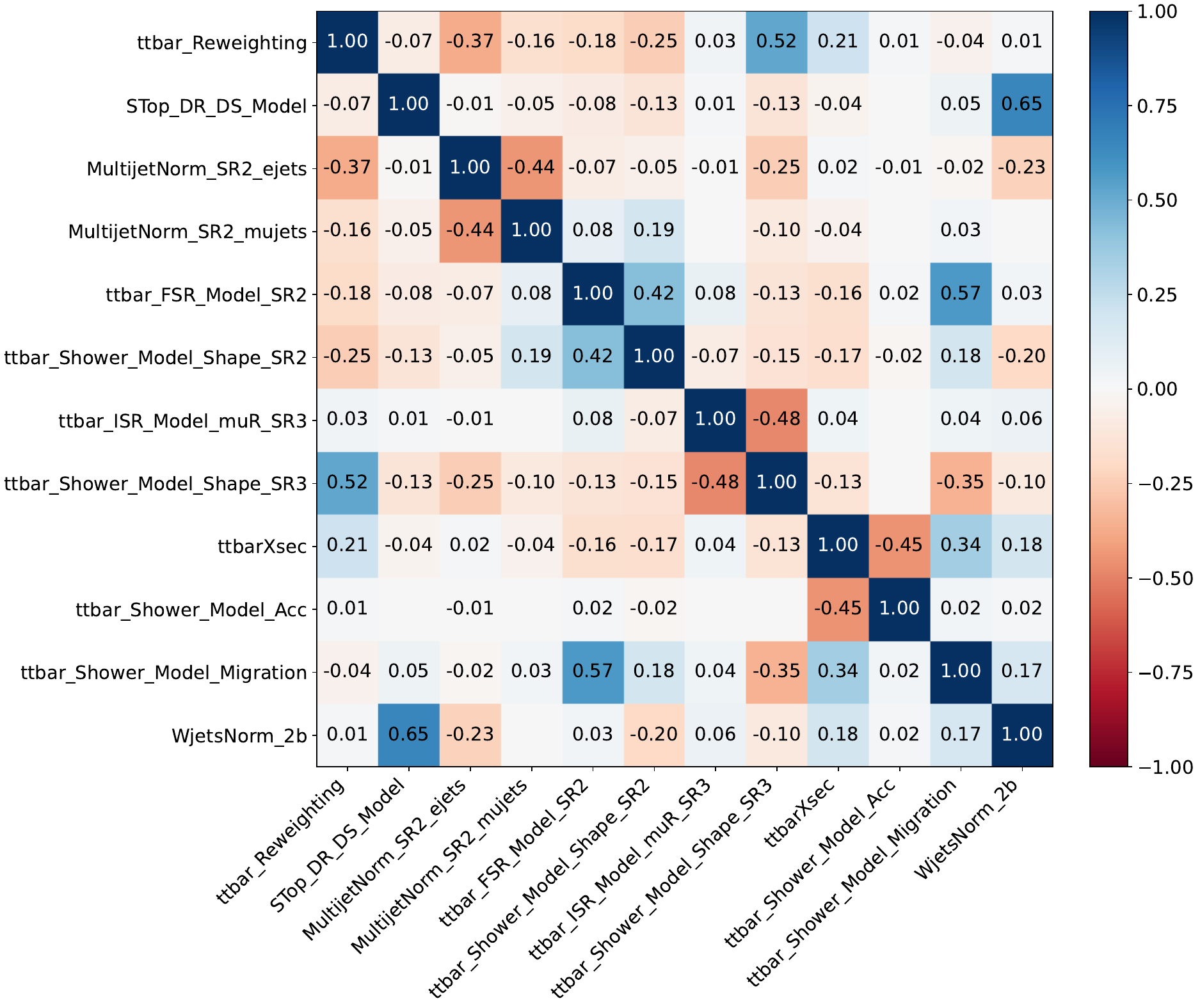} &
        \vspace{0pt} \includegraphics[width=0.475\textwidth]{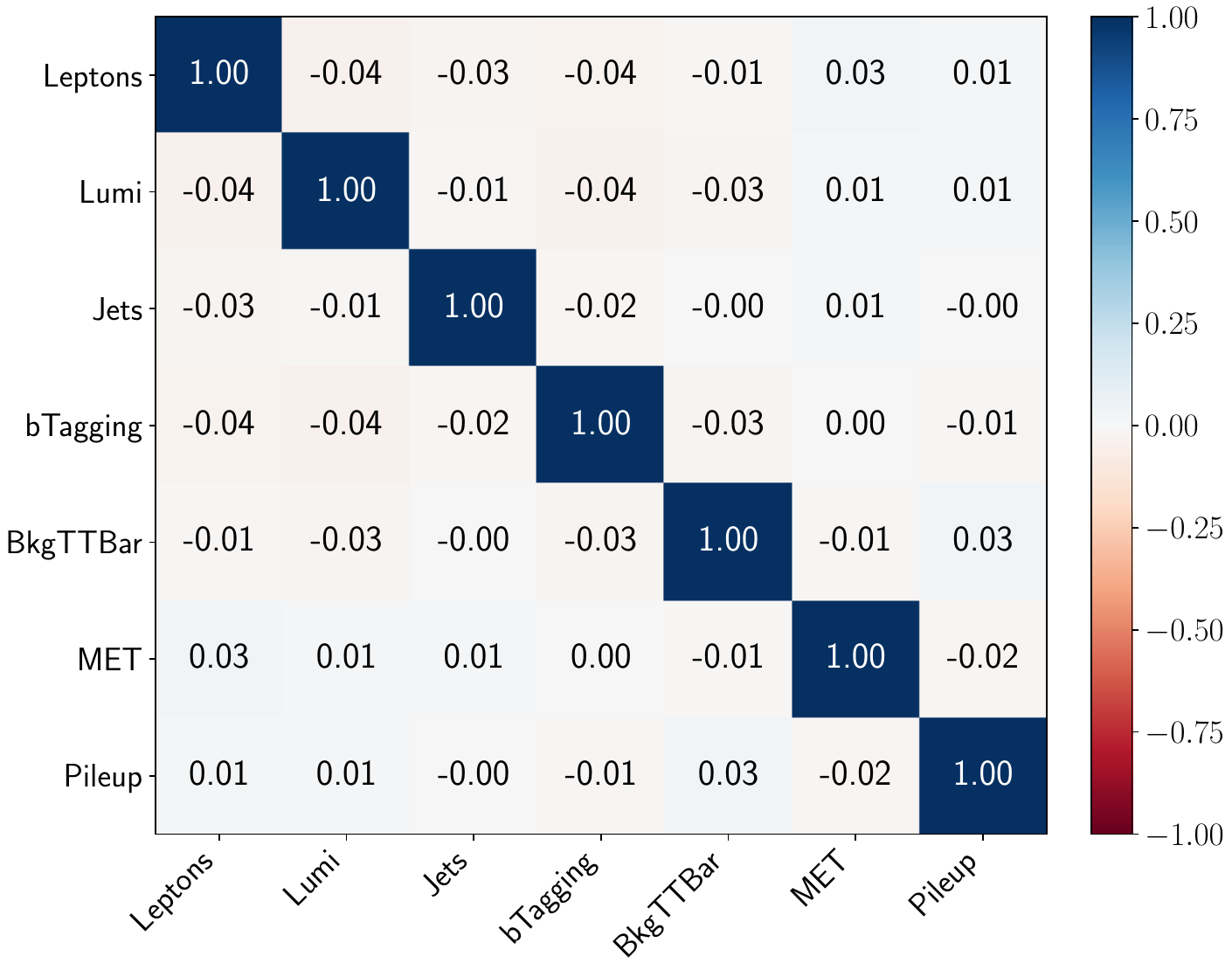}
    \end{tabular}
    \caption{Left: Correlations between individual nuisance parameters
    affecting the $t \bar{t}$ rate with at least one correlation greater than 0.4. Right:
    Correlations between categories of systematic uncertainties extracted 
    from the $t \bar{t}$ likelihood as implemented in \sfitter.}
    \label{fig:ttbar_likelihood_corr}
\end{figure}

The first public likelihood we analyze covers the $t \bar{t}$ rate measurement in 
the leptons+jets final state~\cite{ATLAS:2020aln}. It consists of three channels or 
signal regions, using the aplanarity, minimum lepton-jet mass and average angular distance 
between jets. The parameter of interest $\mu$ is the $t \bar{t}$ signal strength, with 
a total of 177 nuisance parameters covering the systematic uncertainties.

To test our implementation and evaluation of the public likelihood, we first
reproduce some key ATLAS results in Fig.~\ref{fig:atlas_ttbar_ranking}. We
show the values for each nuisance parameter that maximizes the
likelihood and the pulls,
\begin{align}
\text{pull} = \frac{\hat{\theta} - \theta_0}{\Delta\theta} \; .
\end{align}
Here, $\hat{\theta}$ describes the maximum likelihood values and
$\theta_0$ is the value before the fit, normalized to the
pre-fit uncertainty $\Delta\theta$. We also show the impact of the
individual nuisance parameters on the signal strength $\mu$. It is
determined by repeating the fit after fixing the nuisance
parameter to its maximum-likelihood value $\hat{\theta}$, shifted by its prefit
(postfit) uncertainties $\pm \Delta\theta (\pm \Delta\hat{\theta})$. The left
panel of Fig.~\ref{fig:atlas_ttbar_ranking} is taken from
Ref.~\cite{ATLAS:2020aln}, while the right panel shows our reproduced
results. Both sets show
excellent agreement, with negligible differences for a few select nuisance
parameters.

Next, we analyze the full likelihood as a function of a single
nuisance parameter. This allows us to check the validity of a Gaussian likelihood,
as assumed for systematic uncertainties in \sfitter. For each nuisance 
parameter, we generally find excellent agreement with the Gaussian assumption, as
shown on the left, with only a few exceptions. 
Figure~\ref{fig:ttbar_likelihood_nuisance} shows two such cases,
one with excellent agreement and one with poor agreement. Even the
larger deviations are under control, showing good agreement with the Gaussian approximation when we translate them into one
standard deviation. Our combination of nuisance parameters into categories
washes out non-Gaussian shapes in these exceptions.

Finally, we test the correlations between individual nuisance parameters and between 
nuisance parameters assigned to the categories implemented in \sfitter. The
left panel of Fig.~\ref{fig:ttbar_likelihood_corr} shows the
correlations of all individual nuisance parameters with at least one correlation
greater than 0.4. Since the public likelihoods do not provide additional metadata 
on all nuisance parameters, their labels do not necessarily match those used in the impact plots.
We find that out of the many nuisance parameters
included in the public likelihood, only very few are significantly correlated.  
We mainly see strong correlations, for instance, for modeling choices or
jets. 

In the standard \sfitter approach, we group these individual nuisance
parameters into uncorrelated categories and implement these categories
with a single nuisance parameter each. Standard categories cover
leptons, jets, tagging, and luminosity. Additional categories 
are process-specific, such as certain backgrounds or missing transverse energy. 
For all processes in our dataset, we use 
21 nuisance parameters describing the systematic uncertainties assumed to be uncorrelated.
Using the full likelihood, we show the 
correlations between these categories in the right panel of
Fig.~\ref{fig:ttbar_likelihood_corr}. The fact that the correlations
between categories essentially vanish 
validates this \sfitter approach.

\begin{figure}[b!]
    \includegraphics[width=0.495\textwidth]{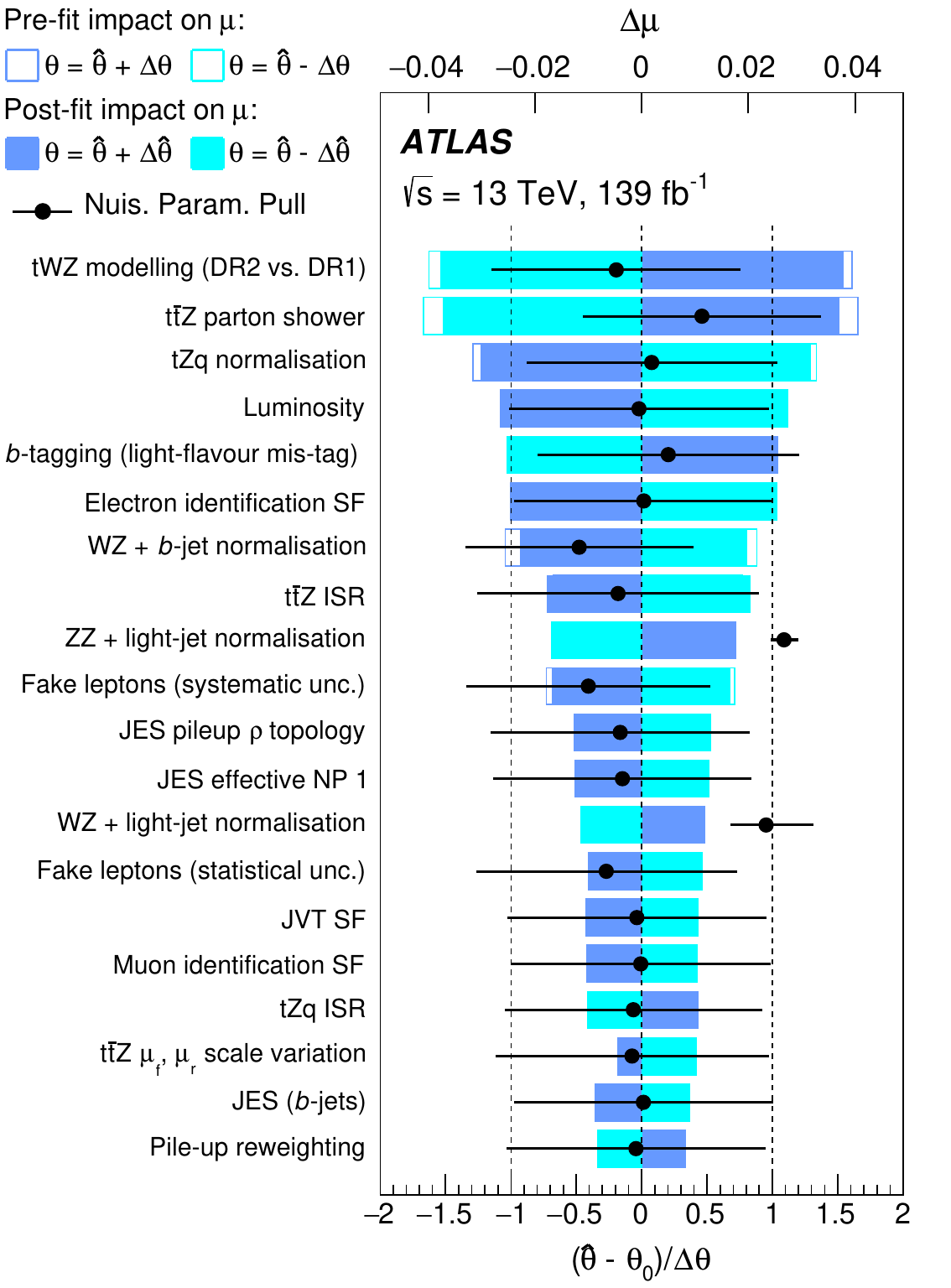}
    \includegraphics[width=0.495\textwidth]{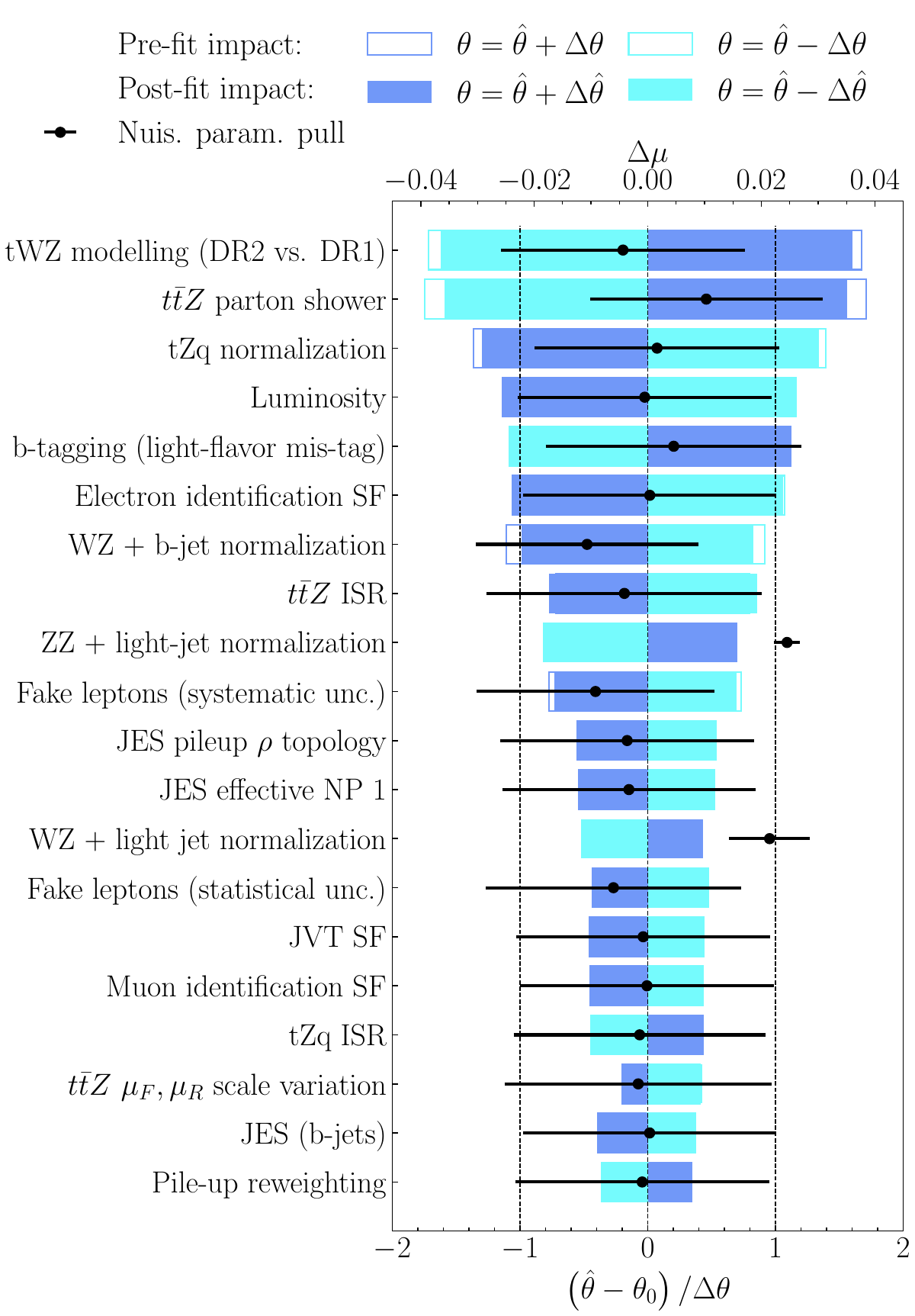}
    \caption{Impact of nuisance parameters on the $t \bar{t} Z$ total
      rate fit. We compare the ATLAS result~\cite{ATLAS:2021fzm}
      (left) and our evaluation of the public likelihood (right).}
    \label{fig:atlas_ttZ_ranking}
\end{figure}

\subsection{ATLAS $t \bar{t} Z$ likelihood}

\begin{figure}[t]
    \begin{tabular}{p{0.5\textwidth} p{0.5\textwidth}}
        \vspace{0pt} \includegraphics[width=0.485\textwidth]{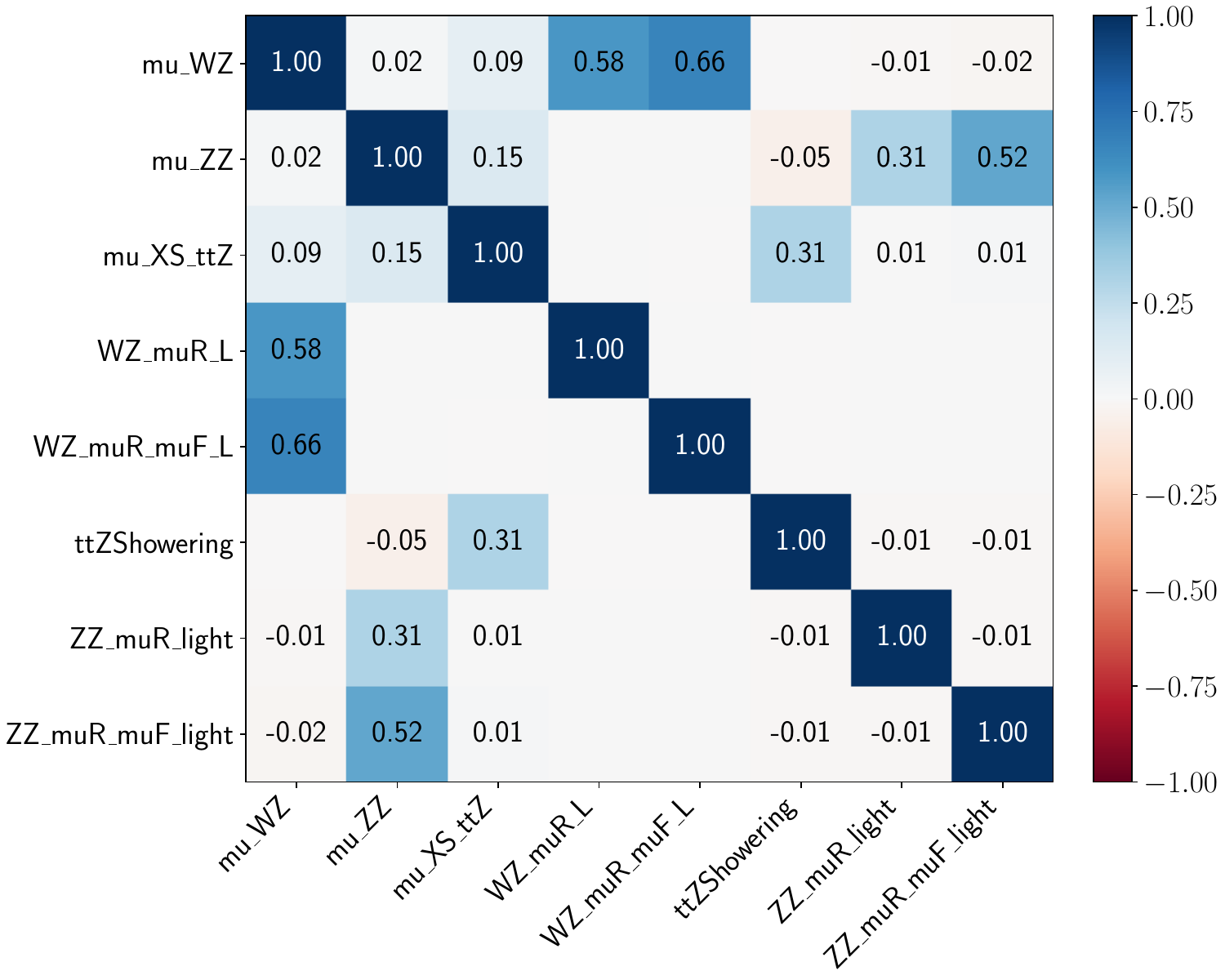} & 
        \vspace{0pt} \includegraphics[width=0.485\textwidth]{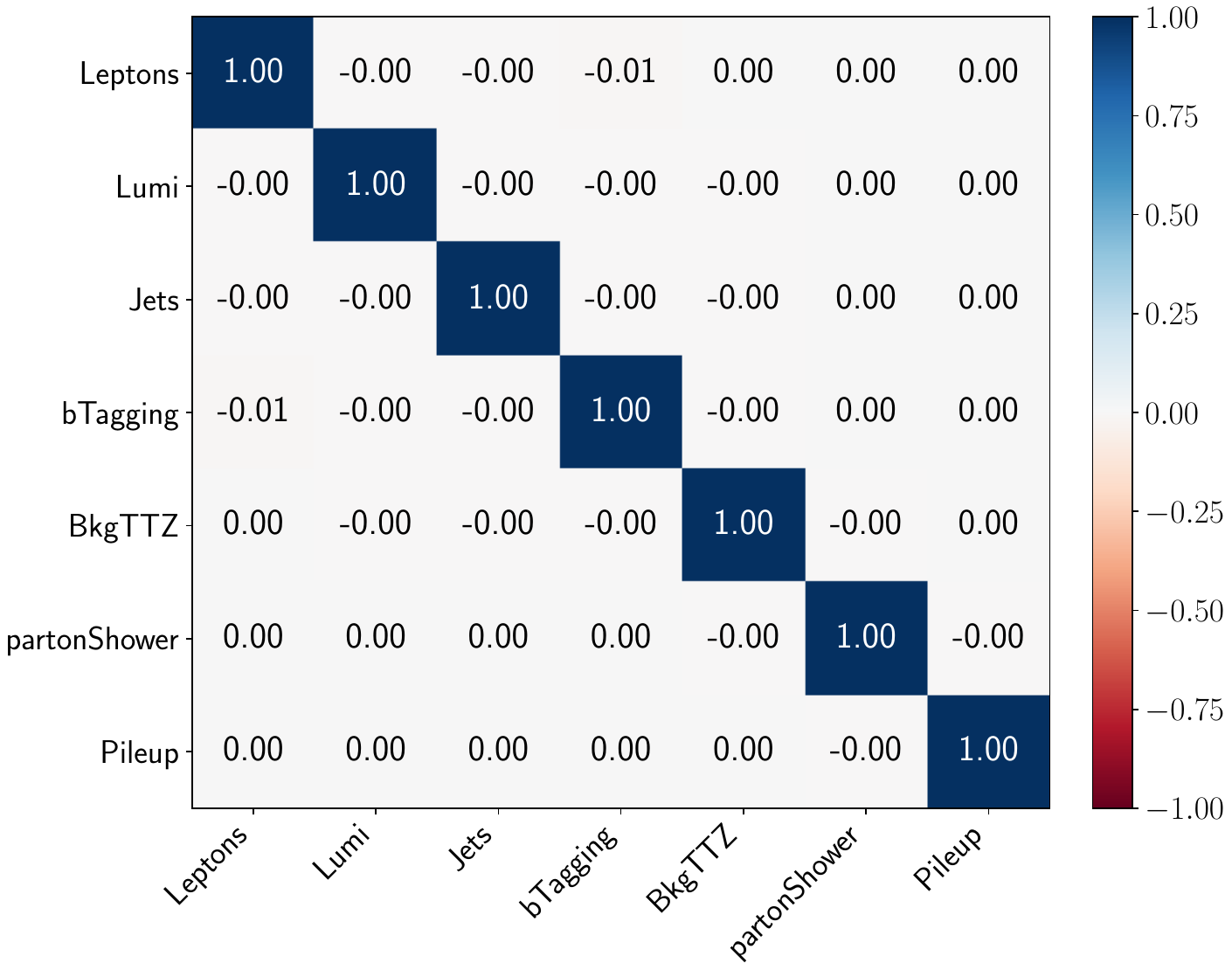}
    \end{tabular}
    \caption{Left: Correlations between individual nuisance parameters affecting the $t \bar{t} Z$ rate with at least one correlation greater than 0.3. Right:
      Correlations between categories of systematic uncertainties extracted from the
      $t \bar{t} Z$ likelihood as implemented in \sfitter.}
    \label{fig:ttZ_likelihood_corr}
\end{figure}

The second likelihood we implement is for the $t \bar{t} Z$ rate 
measurement~\cite{ATLAS:2021fzm}. It simultaneously fits both 3-lepton and 4-lepton 
signal regions and the corresponding control regions. The parameter of interest 
is the $t \bar{t} Z$ signal strength. A total of 230 nuisance parameters 
describe the systematic uncertainties. Unlike for the $t \bar{t}$ likelihood, 
there are no uncertainties on the shape of the signal since each signal region is 
described by a single bin.

Following the method described for the $t \bar{t}$ analysis, we also test 
the $t \bar{t} Z$ likelihood and our implementation.
Figure~\ref{fig:atlas_ttZ_ranking} compares the impact 
and pulls taken from Ref.~\cite{ATLAS:2021fzm} with those
reproduced by us. We see excellent
agreement for all nuisance parameters, with, at most, very minor differences.

As before, we then show the correlations between nuisance parameters
with at least one correlation greater than 0.3 in the left panel of
Fig.~\ref{fig:ttZ_likelihood_corr}. We can compare them to our
\sfitter implementation on the right.
We find that the correlations between individual nuisance 
parameters are already much smaller for this likelihood. 
The only strong correlations appear between 
scale uncertainties and the signal strength of the 
corresponding background. Consequently, the results, after combining 
all nuisance parameters into the \sfitter categories, display 
negligible correlations between categories. 

\begin{figure}[t]
    \includegraphics[width=0.495\textwidth]{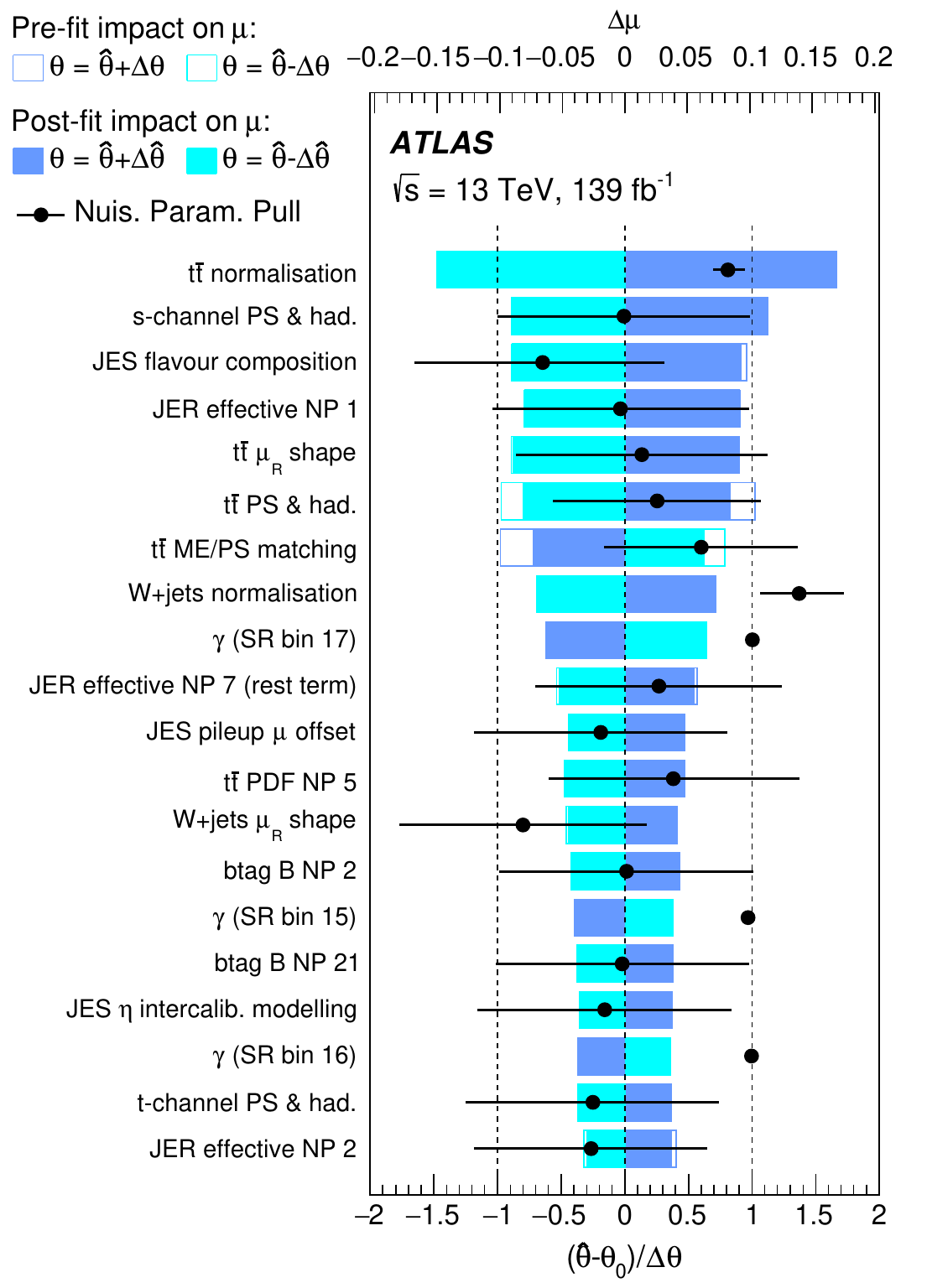}
    \includegraphics[width=0.495\textwidth]{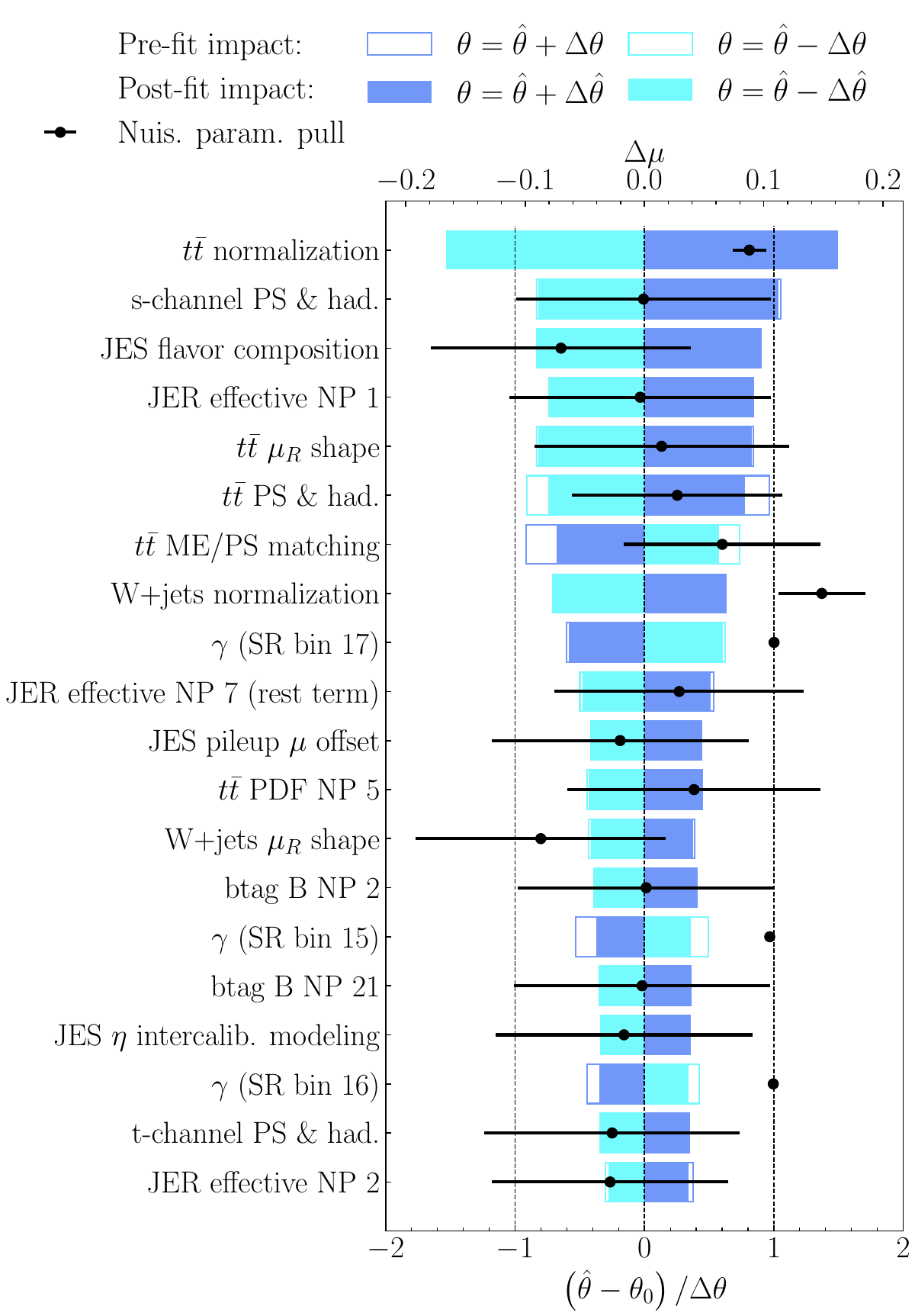}
    \caption{Impact of nuisance parameters on the $s$-channel single
      top rate fit. We compare the ATLAS result~\cite{ATLAS:2022wfk}
      (left) and our evaluation of the public likelihood (right).}
    \label{fig:atlas_STop_ranking}
\end{figure}

\subsection{ATLAS $s$-channel single top likelihood}

The third likelihood we implement is for the signal strength of $s$-channel single top 
production~\cite{ATLAS:2022wfk}. Unlike the previous measurements, it consists of 
a single channel, making use of the matrix element method (MEM) to determine the 
probability that an event is a signal event. The discriminant defined using 
the MEM gives a distribution with 171 nuisance parameters affecting the rate 
and shape of the signal.

Once again, we validate our implementation of this likelihood
in Fig.~\ref{fig:atlas_STop_ranking}, showing the impact and pulls from 
Ref.~\cite{ATLAS:2022wfk} in the left panel and our reproduction on the right.
We find perfect agreement, showing that regardless of the process
considered, the public likelihoods allow for an easy and precise
reproduction of the experimental results in more detail than most global 
analyses will ever need or want to use.

\begin{figure}[t]
    \begin{tabular}{p{0.525\textwidth} p{0.475\textwidth}}
        \vspace{0pt} \includegraphics[width=0.525\textwidth]{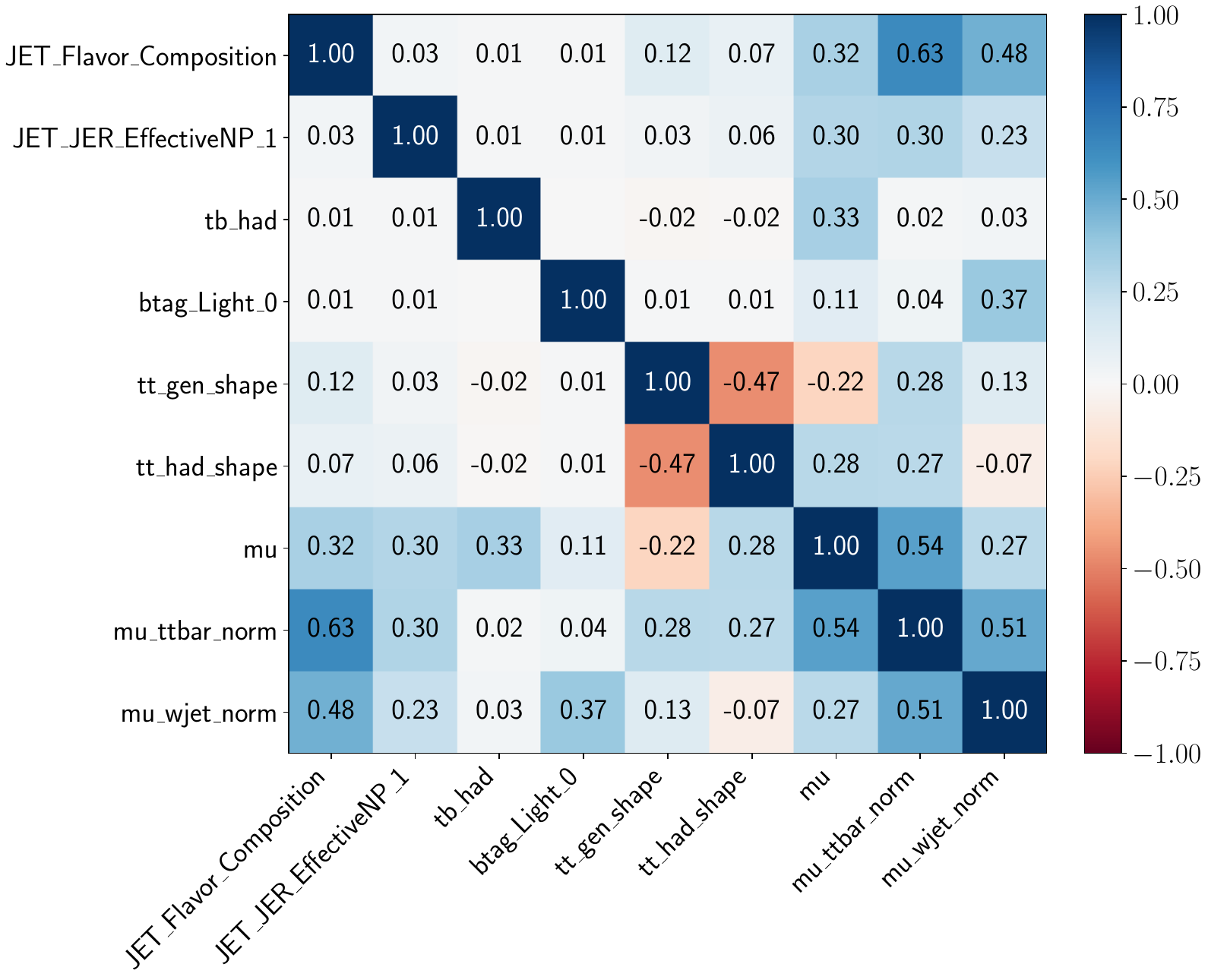} &
        \vspace{0pt} \includegraphics[width=0.475\textwidth]{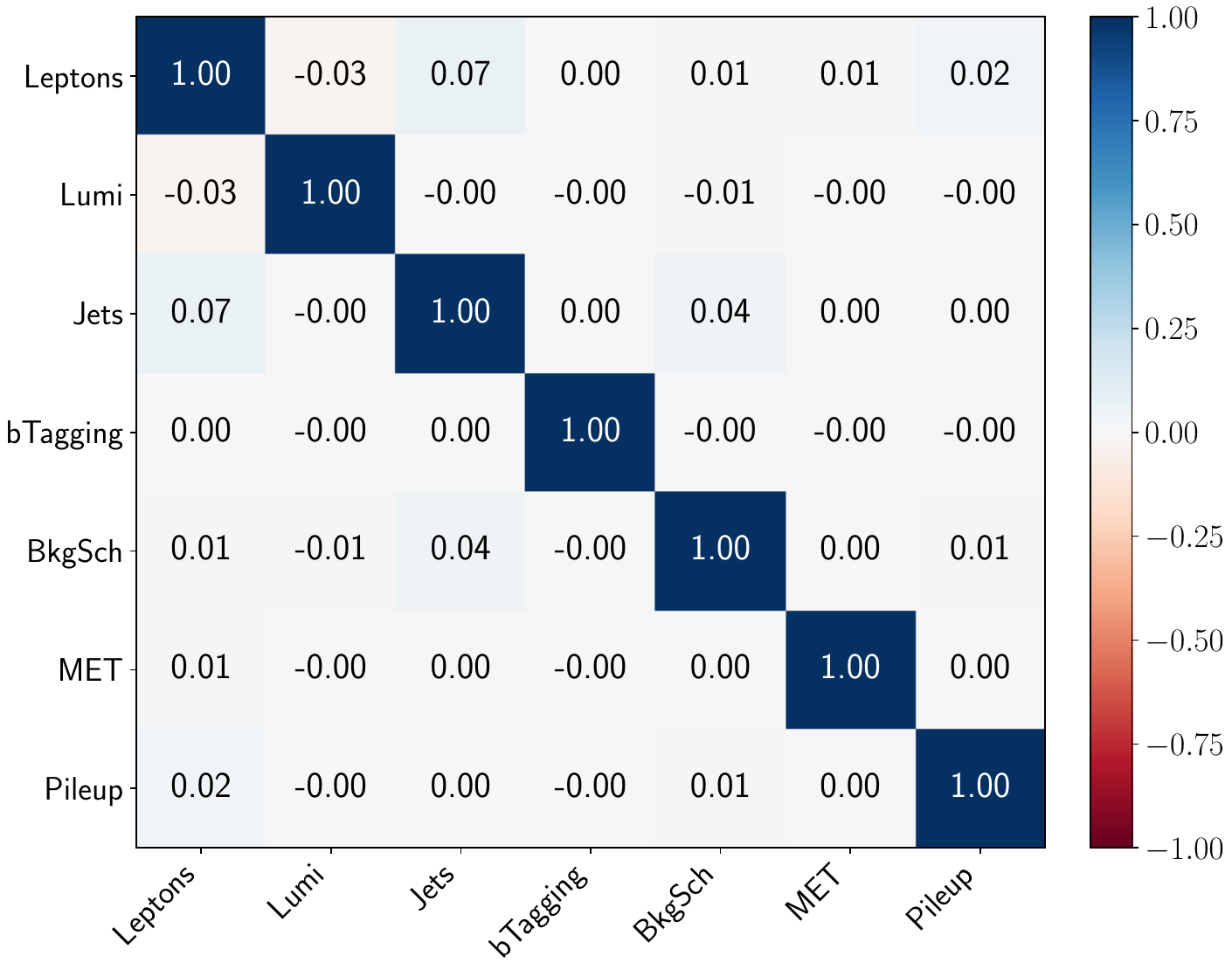}
    \end{tabular}
    \caption{Left: Correlations between individual nuisance parameters affecting the $s$-channel single top rate with
      at least one correlation greater than 0.3. Right: 
      Correlations between categories of systematic uncertainties
      extracted from the single top likelihood as 
      implemented in \sfitter.}
    \label{fig:STop_likelihood_corr}
\end{figure}

The correlations in the left panel of 
Fig.~\ref{fig:STop_likelihood_corr} show strong correlations only 
between select nuisance parameters. The strongest correlations appear between 
jet-related uncertainties and the signal strengths of the two dominant backgrounds, 
$t \bar{t}$ and $W$+jets. For \sfitter, these nuisance parameters are put 
into the background uncertainty category. These strong correlations are therefore 
implicitly included in this larger category, and the final implementation into
\sfitter is essentially uncorrelated, as one can see in the right of Fig.~\ref{fig:STop_likelihood_corr}. 
While one still finds a nonzero correlation between the jet and background uncertainties and 
between the jet and lepton uncertainties, these are all negligibly small.

\section{Global analysis}
\label{sec:global}

Using, for the first time, public likelihoods in a global SMEFT analysis 
allows us to look at different relevant questions. From the data 
included in our analysis, we know that our global analysis is 
somewhat unlikely to uncover a fundamental and statistically significant 
break-down of the SM. We first look at the impact on the constraining power from new 
measurements, especially boosted top kinematics, relative to Ref.~\cite{Brivio:2019ius}. 
We then study the impact of correlated 
uncertainties encoded in the public likelihoods. 
From a pure statistics perspective, we also check if lower-dimensional limits 
extracted by profiling and by marginalization differ. 
Finally, we provide SMEFT limits 
combining the updated top sector analysis with the 
electroweak and Higgs sector from Ref.~\cite{Brivio:2022hrb}.

\subsection{Better and boosted measurements}
\label{sec:global_new}

\begin{figure}[t]
\includegraphics[width=0.40\textwidth]{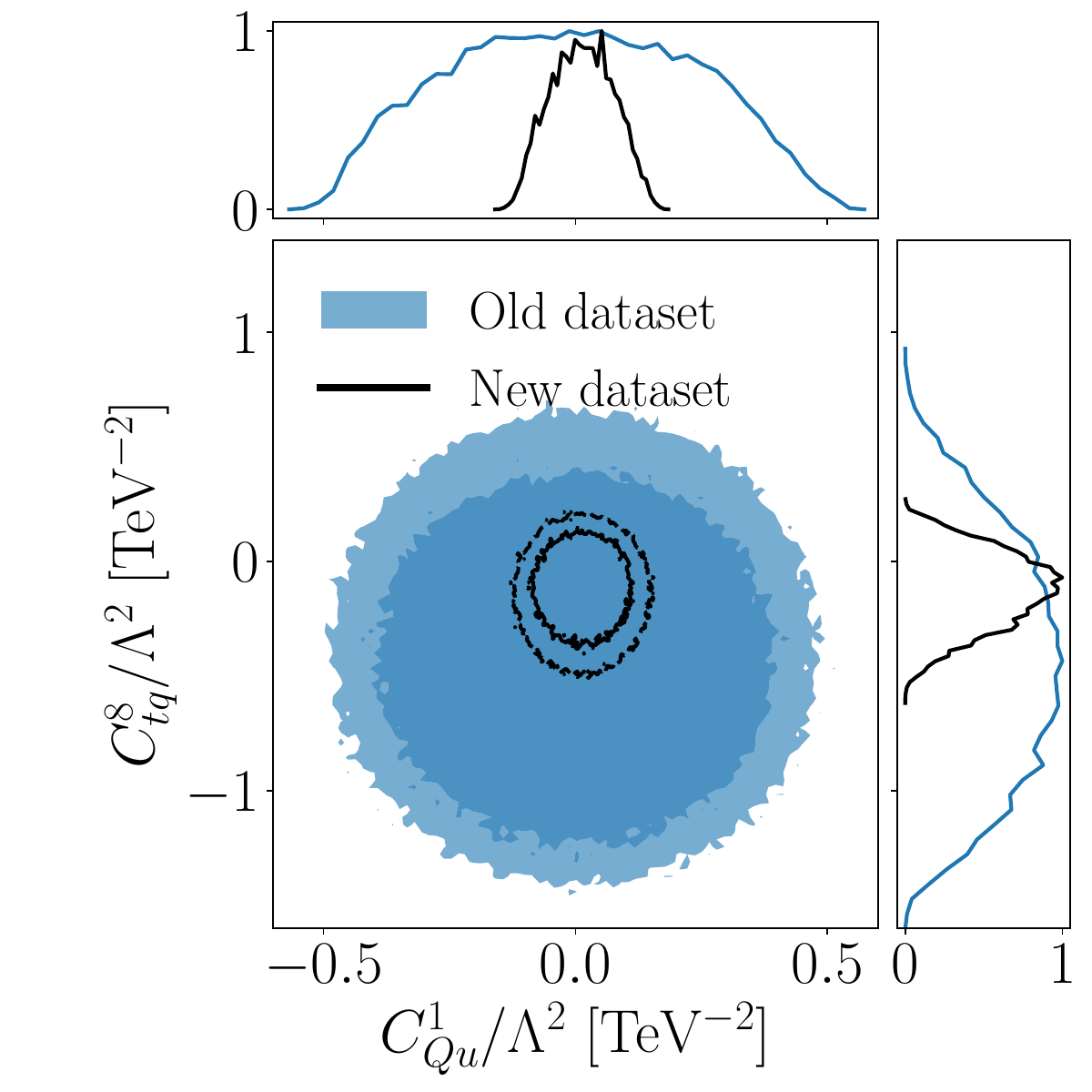}
\includegraphics[width=0.40\textwidth]{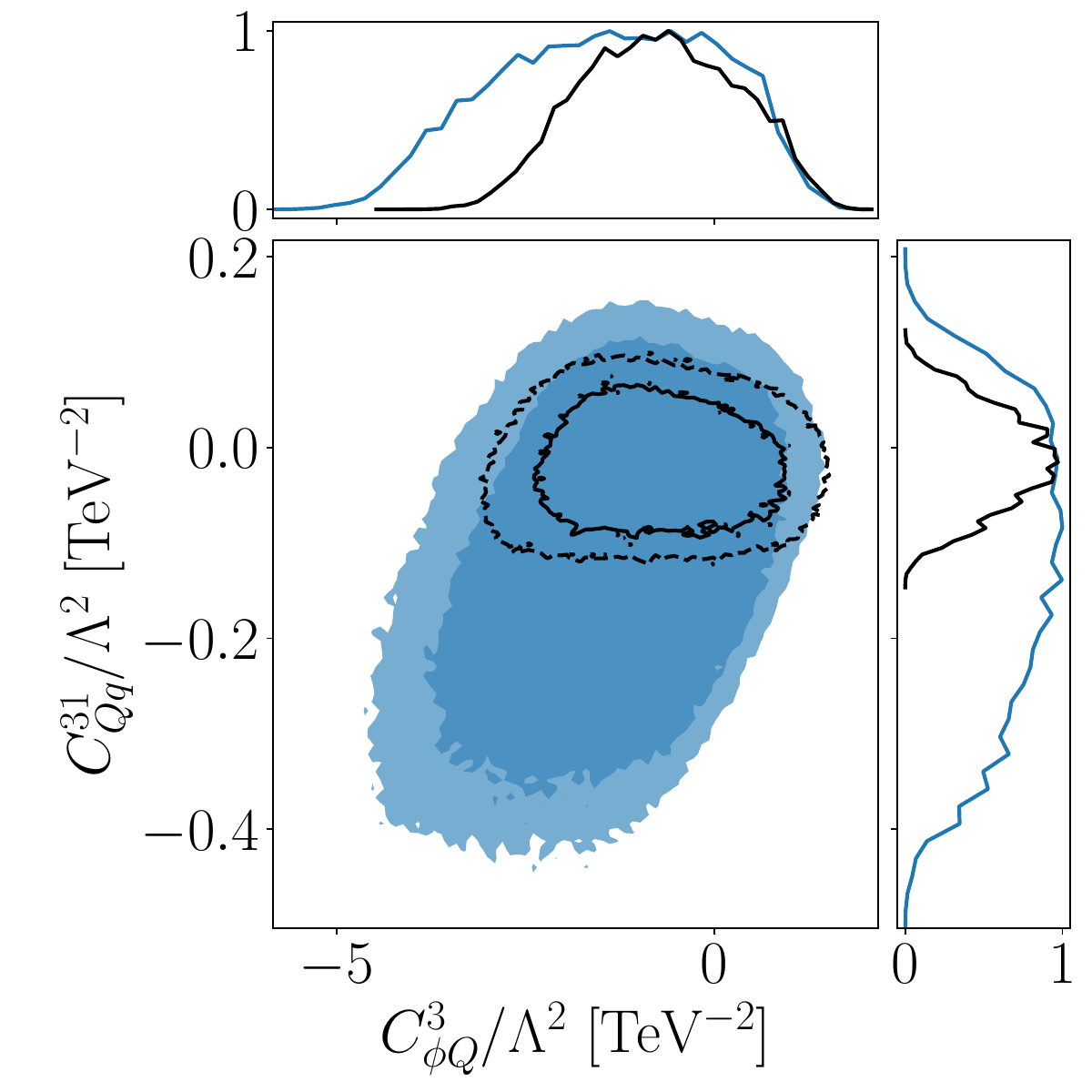}
    \centering
\includegraphics[width=0.40\textwidth]{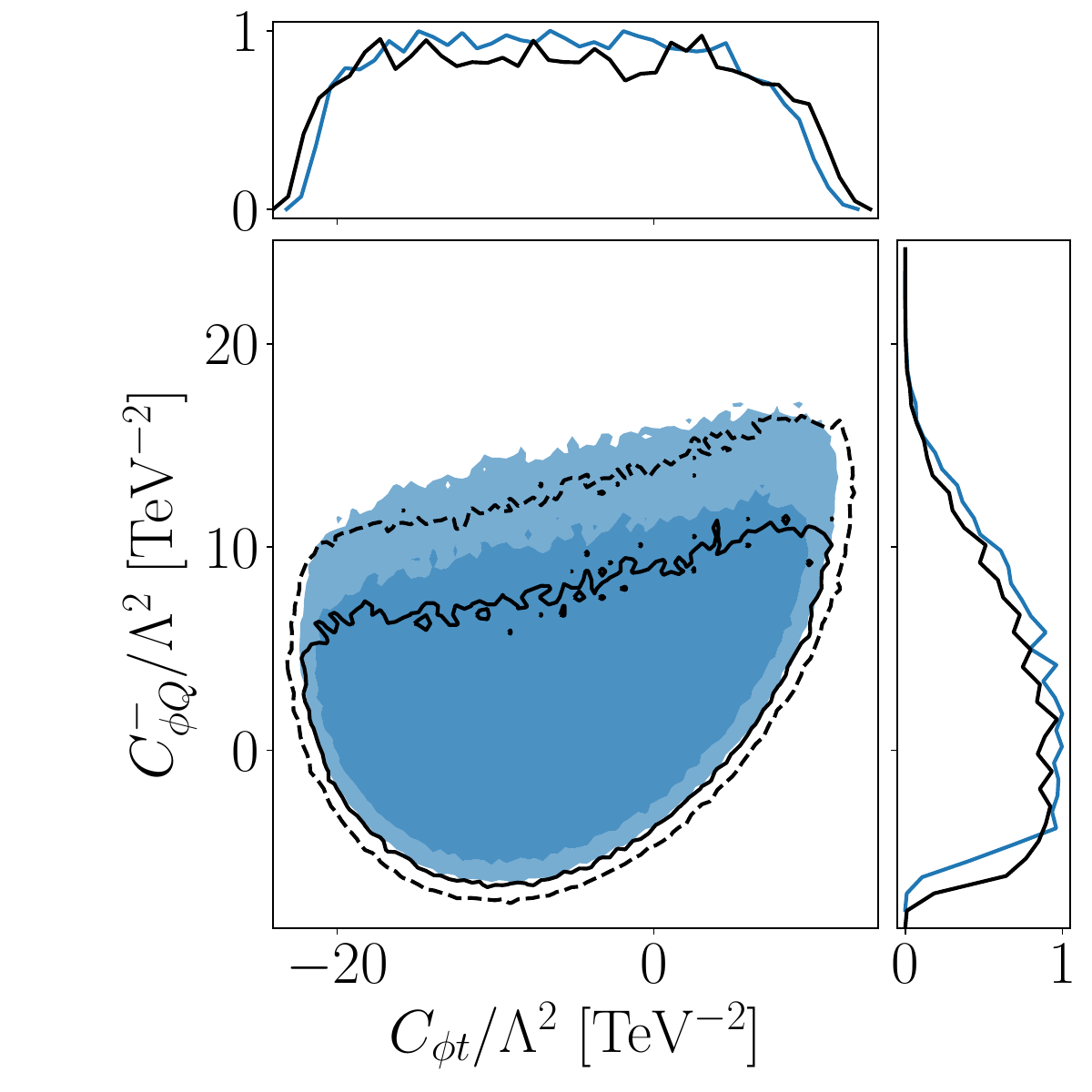}
  \caption{Profile likelihood correlations for three pairs out of the 22 Wilson coefficients, illustrating the impact of the new data listed in Tabs.~\ref{tab:datasets_ttbar} and~\ref{tab:datasets_singletop} (black) compared to the previous analysis~\cite{Brivio:2019ius} (blue).}
    \label{fig:global_newdata}
\end{figure}

Before we study more conceptual questions of global SMEFT analyses, 
we update our dataset with new measurements, as marked in 
Tabs.~\ref{tab:datasets_ttbar} and~\ref{tab:datasets_singletop}.
In Fig.~\ref{fig:global_newdata}, we show the constraints on a selection of 
2-dimensional correlations of Wilson coefficients, using all top 
data compared to the previous \sfitter top analysis~\cite{Brivio:2019ius}.  
All constraints are the result of an analysis of all 22 Wilson coefficients.  
To extract limits on pairs of coefficients, we use a profile likelihood. 
Potential differences in marginalization will be discussed in 
Sec.~\ref{sec:global_marg}.

The left panel in Fig.~\ref{fig:global_newdata} shows the impact 
on the four-fermion operators $\ope_{Qu}^{1}$ and $\ope_{tq}^{8}$.
Both operators receive constraints from top pair production, 
now with a public likelihood~\cite{ATLAS:2020aln}, as well as 
new data in the boosted regime~\cite{ATLAS:2022mlu} and at 13.6 TeV~\cite{CMS:2023qyl}.
The dominant constraining power comes from measurements of the boosted 
kinematics~\cite{ATLAS:2022mlu} and will be discussed below.

The right panel shows the improvement in constraints on 
$\ope_{\phi Q}^{3}$ and $\ope_{Qq}^{31}$. Single top production provides constraints 
on them, and we again benefit from the public 
likelihood~\cite{ATLAS:2022wfk}, a new $p_{T,t}$ distribution in $t$-channel 
single top production~\cite{CMS:2019jjp}, and new measurements of the $tW$ production 
cross section~\cite{ATLAS:2020cwj,CMS:2021vqm}.  We observe an improvement in the
individual constraints and their correlation.
In particular, $\ope_{Qq}^{31}$ receives some constraining power from 
boosted top pair production, which in turn allows single top measurements 
to constrain $\ope_{\phi Q}^{3}$.

Finally, in the lower panel of Fig.~\ref{fig:global_newdata} we highlight the 
improvement in probing $\ope_{\phi t}$  and $\ope_{\phi Q}^{-}$.  As before, these 
operators are constrained by measurements of $t \bar{t} 
Z$ production, for which we use a public likelihood. However, in this case, 
we only find a small change in the correlated likelihood. 

Altogether, we find that the public likelihoods do not have a significant 
effect on our SMEFT limits. As discussed in Sec.~\ref{sec:like}, the likelihoods 
available and included in our analysis all describe total cross sections,
limiting their impact. On the positive side, public likelihoods 
allow for an accurate modeling of correlated systematics, an aspect we will discuss
in Sec~\ref{sec:global_corr}.\medskip

\begin{figure}[t]
\centering
\includegraphics[width=0.40\textwidth]{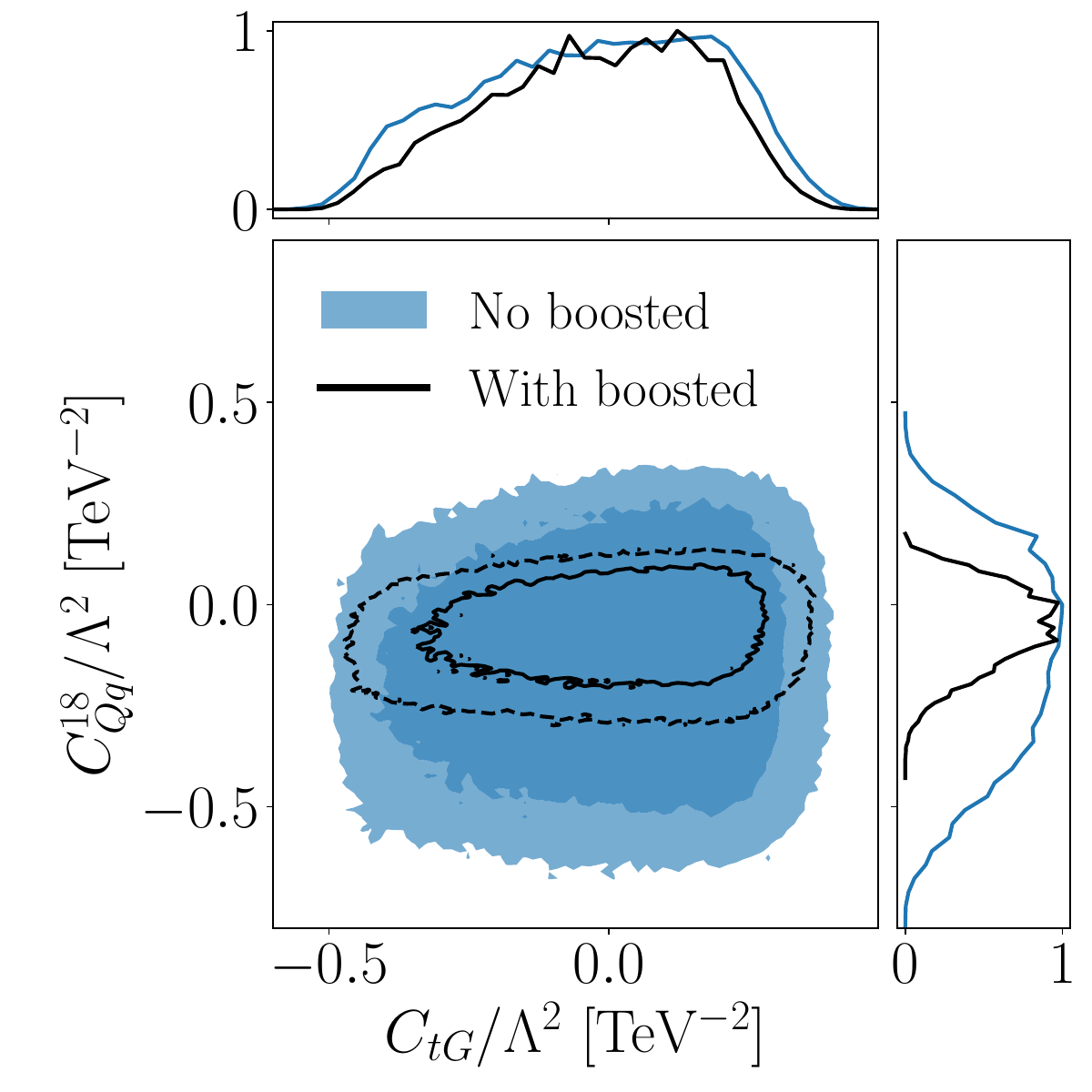}
\hspace*{0.1\textwidth}
\includegraphics[width=0.40\textwidth]{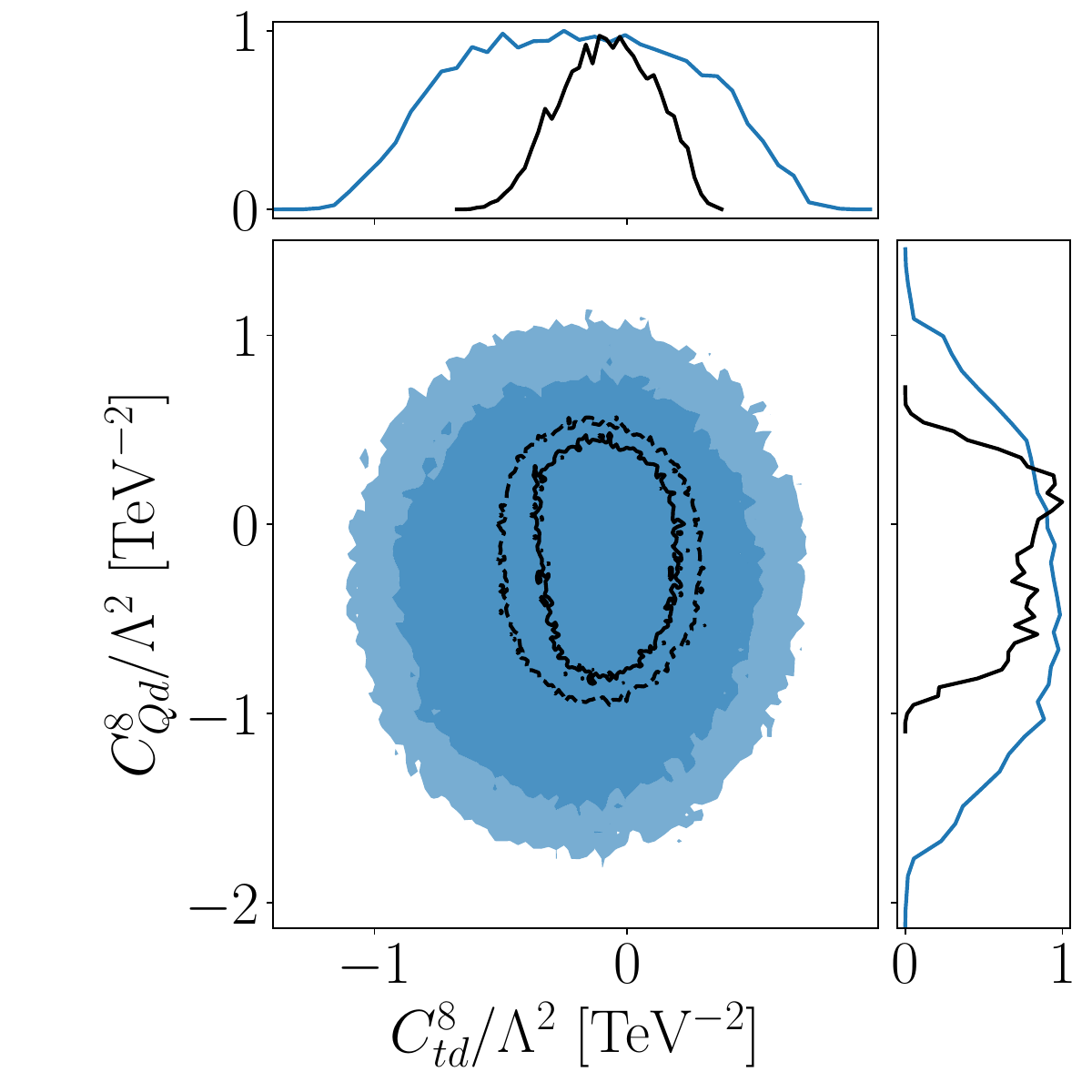}
  \caption{Profile likelihood correlations showing the impact of boosted 
  top pair kinematics~\cite{ATLAS:2022mlu} (black), compared to the same dataset
  without this one measurement (blue).}
    \label{fig:global_boosted}
\end{figure}

Much of the improvement we see from our new dataset is due to 
the boosted regime. For SMEFT analyses, such measurements are extremely helpful
to constrain operators which include momentum scaling. As discussed in Sec.~\ref{sec:data}, we add
the unfolded ATLAS measurement of boosted top pair production~\cite{ATLAS:2022mlu}.  
In Fig.~\ref{fig:boosted_SFitter_results}, we already showed the impact of a single SMEFT operator 
$\ope_{Qd}^{8}$ on the normalized $m_{t \bar{t}}$ distribution of this measurement. 

Here, we study its effect on the global analysis. 
Figure~\ref{fig:global_boosted} demonstrates the impact 
on a selection of two-operator correlations. The 
complete analysis including all data in Tabs.~\ref{tab:datasets_ttbar}, 
Tabs.~\ref{tab:datasets_singletop} is compared to the case 
where the measurement of boosted tops from Ref.~\cite{ATLAS:2022mlu} is excluded. 
In the first panel, we observe an increase in constraining power on $\ope_{Qq}^{18}$, 
while the constraints on $\ope_{tG}$ are stable.  This follows from the fact that  
this operator is instead constrained by the $t \bar{t}$ total cross section.  In 
contrast, the limits on the Wilson coefficients for energy-growing 4-fermion operators 
improve by a factor of two, as shown in the right-hand panel for $\ope_{Qq}^8$ and 
$\ope_{td}^{8}$.

\subsection{Correlated systematics}
\label{sec:global_corr}

Public likelihoods, as discussed in Sec.~\ref{sec:like}, allow us to model 
and study correlated systematic uncertainties across measurements by the same 
experiment.
For measurements without public likelihoods, the approximate treatment 
of correlations is discussed in Sec.~\ref{sec:sfitter}.
For the Higgs sector, we already know that the correlations of systematic 
uncertainties had a highly visible impact on the SMEFT analysis~\cite{Brivio:2022hrb}.
In particular, they lead to a marked shift in the most likely values of 
Wilson coefficients while leaving the width of the limits unchanged.  

\begin{figure}[t]
    \centering
    \includegraphics[width=0.40\textwidth]{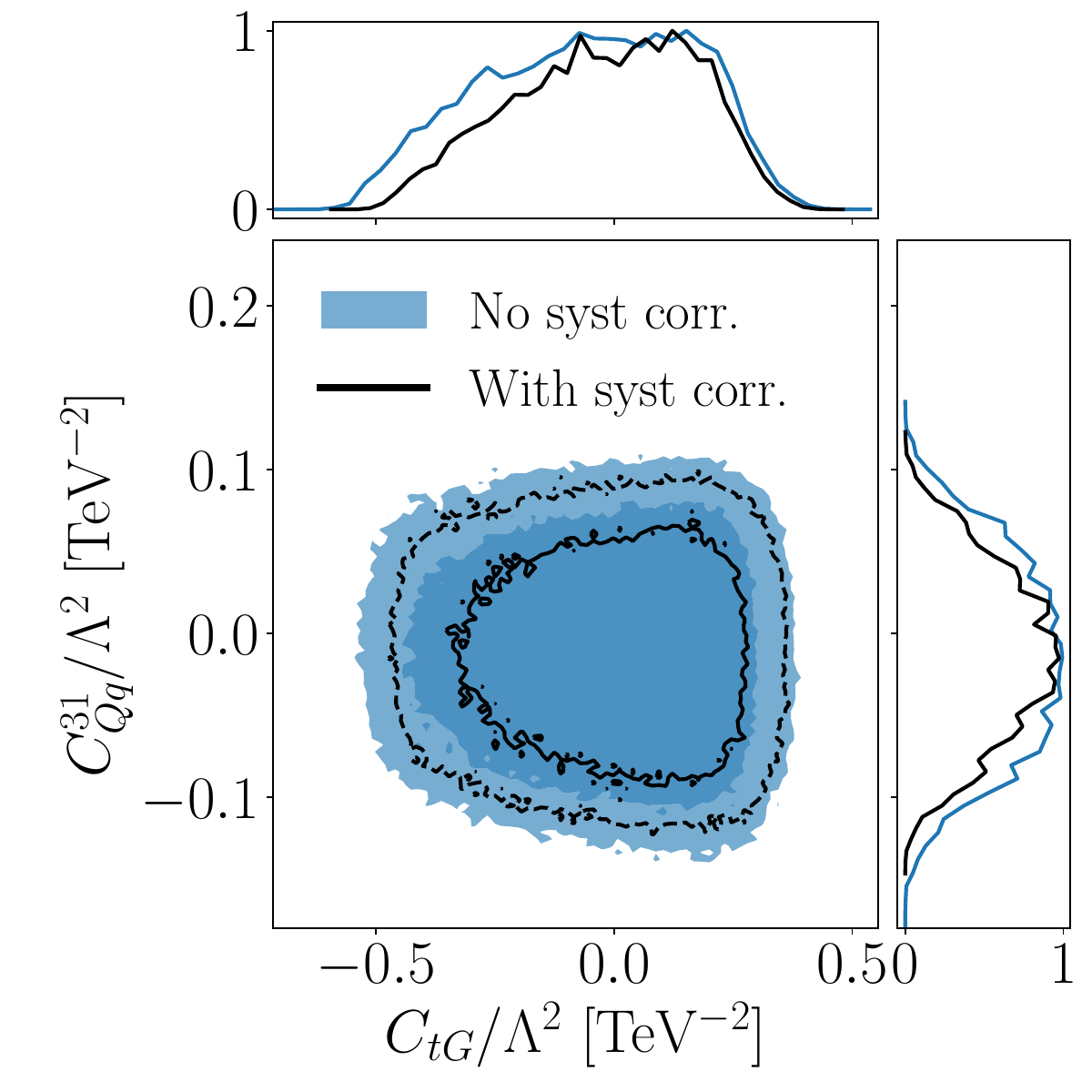}
    \hspace*{0.1\textwidth} 
    \includegraphics[width=0.40\textwidth]{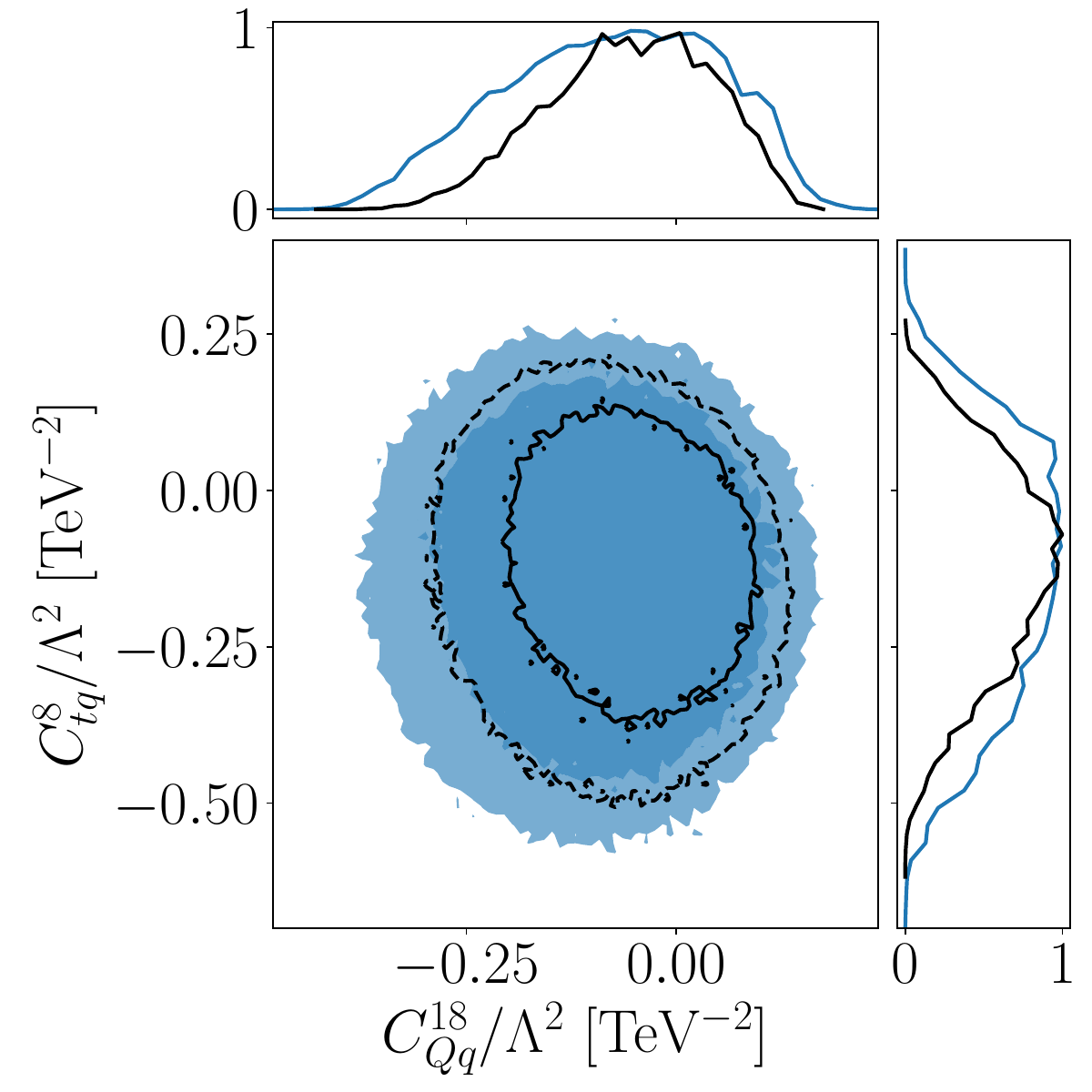}
    \caption{Profile likelihood correlations including correlated systematic and theory uncertainties (blue) 
    versus ignoring correlations between experimental systematics (black).}
    \label{fig:corrsys}
\end{figure}

\begin{figure}[t]
\centering
\includegraphics[width=0.40\textwidth]{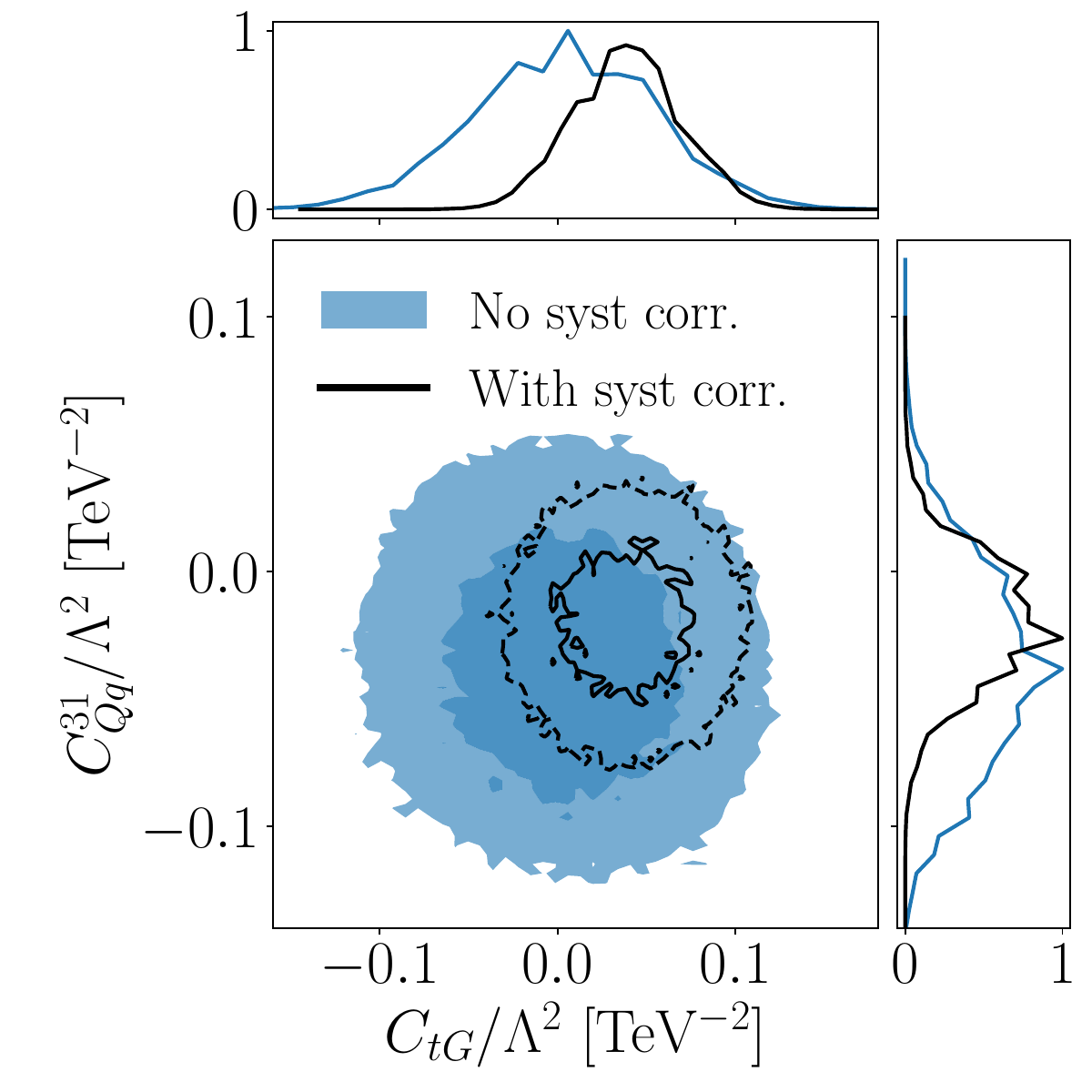}
\hspace*{0.1\textwidth}
\includegraphics[width=0.40\textwidth]{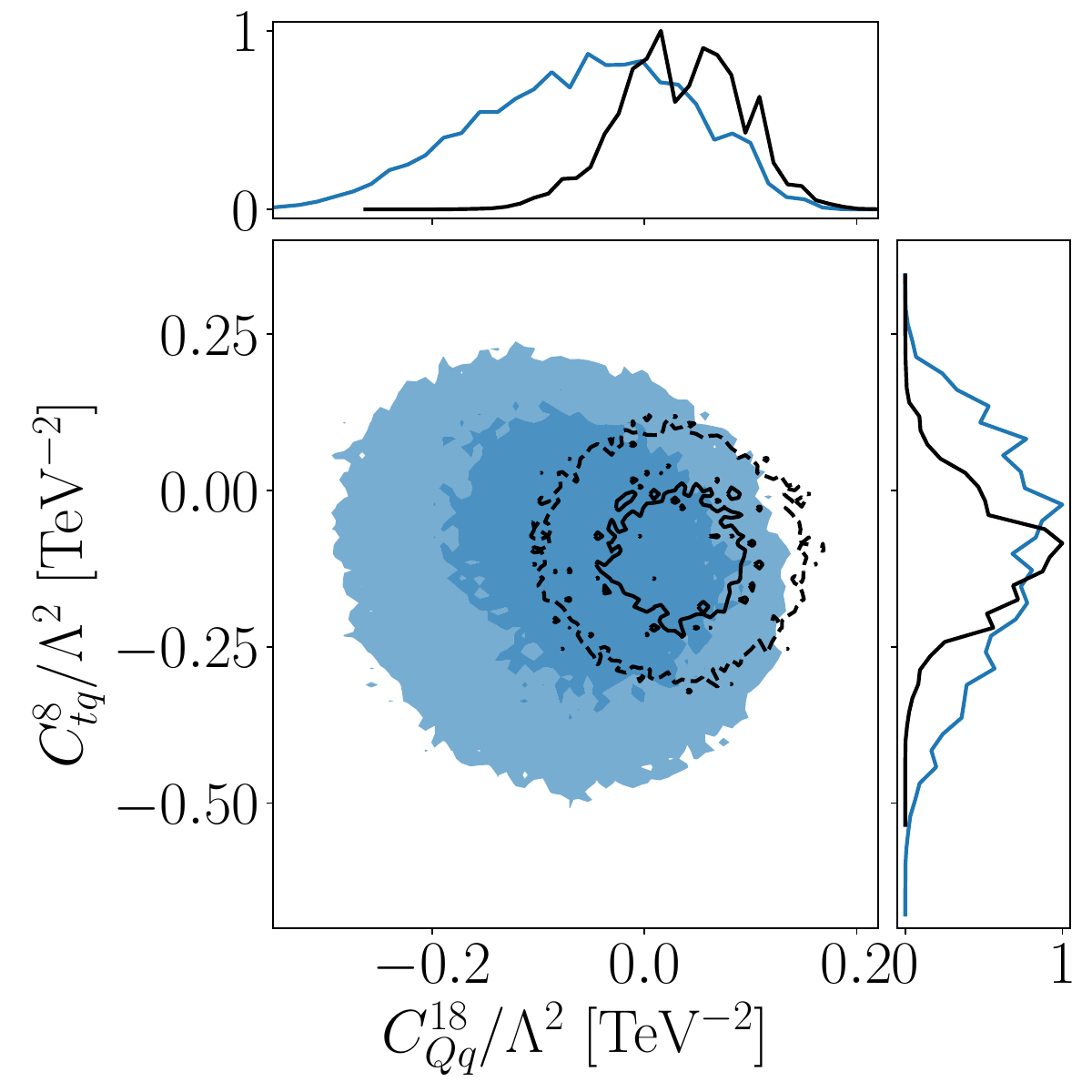}    
       \caption{Profile likelihood correlations, ignoring theory uncertainties altogether, 
       and either including correlated systematic uncertainties (blue) 
       or ignoring correlations between experimental systematics (black).}
    \label{fig:corrsysnotheo}
\end{figure}

Here, we assess the impact of correlating systematic uncertainties on the top 
sector. In Fig.~\ref{fig:corrsys}, we show two sets of constraints on a selection 
of Wilson coefficients.  In blue, we show the constraints from a global 
analysis where all correlations between experimental systematics and 
between theory uncertainties are included. In black we show the same 
results, but treating all experimental systematics as uncorrelated.
For all Wilson coefficients, we find good agreement, which 
indicates that in the top sector, the correlations of systematic uncertainties 
cannot be ignored but have a limited effect on the final SMEFT limits.

We know that statistical uncertainties are not the leading challenge for 
global SMEFT analyses. So if the correlations between
experimental systematics are not really relevant either, 
which uncertainties actually dominate the SMEFT analysis? While for the Higgs 
sector, the modeling of theory uncertainties has 
surprisingly little effect on the SMEFT limits~\cite{Brivio:2022hrb}, the 
QCD nature of top pair production suggests that the situation will be different here.
As a test, in Fig.~\ref{fig:corrsysnotheo}, we repeat the  
comparison of Fig.~\ref{fig:corrsys}, neglecting all theory uncertainties.  
As before, we show the global analysis with 
correlated systematics in blue, while in black these correlations are removed.
Now we see a significant difference. When neglecting the correlations, 
we observe an increase in the size of the constraints as well as a sizeable shift 
in the most likely point.  This is particularly marked in the 2-dimensional 
constraints on $C_{Qq}^{18}$ and $C_{tq}^{8}$.

Comparing Figs.~\ref{fig:corrsys} and~\ref{fig:corrsysnotheo}
we learn the importance of theory uncertainties in the top sector.
If we neglect the theory uncertainties, the effect of correlated 
systematics is non-negligible. While theory uncertainties currently 
wash out these effects, we expect them to become more important as 
SM calculations become more precise, and theory uncertainties are reduced. 
Moreover, we cannot make any statement about the potential impact of public
likelihoods for those kinematic measurements that drive the SMEFT sensitivity.

\subsection{Marginalization vs profiling}
\label{sec:global_marg}

\begin{figure}[t]
\centering
\includegraphics[width=0.40\textwidth]{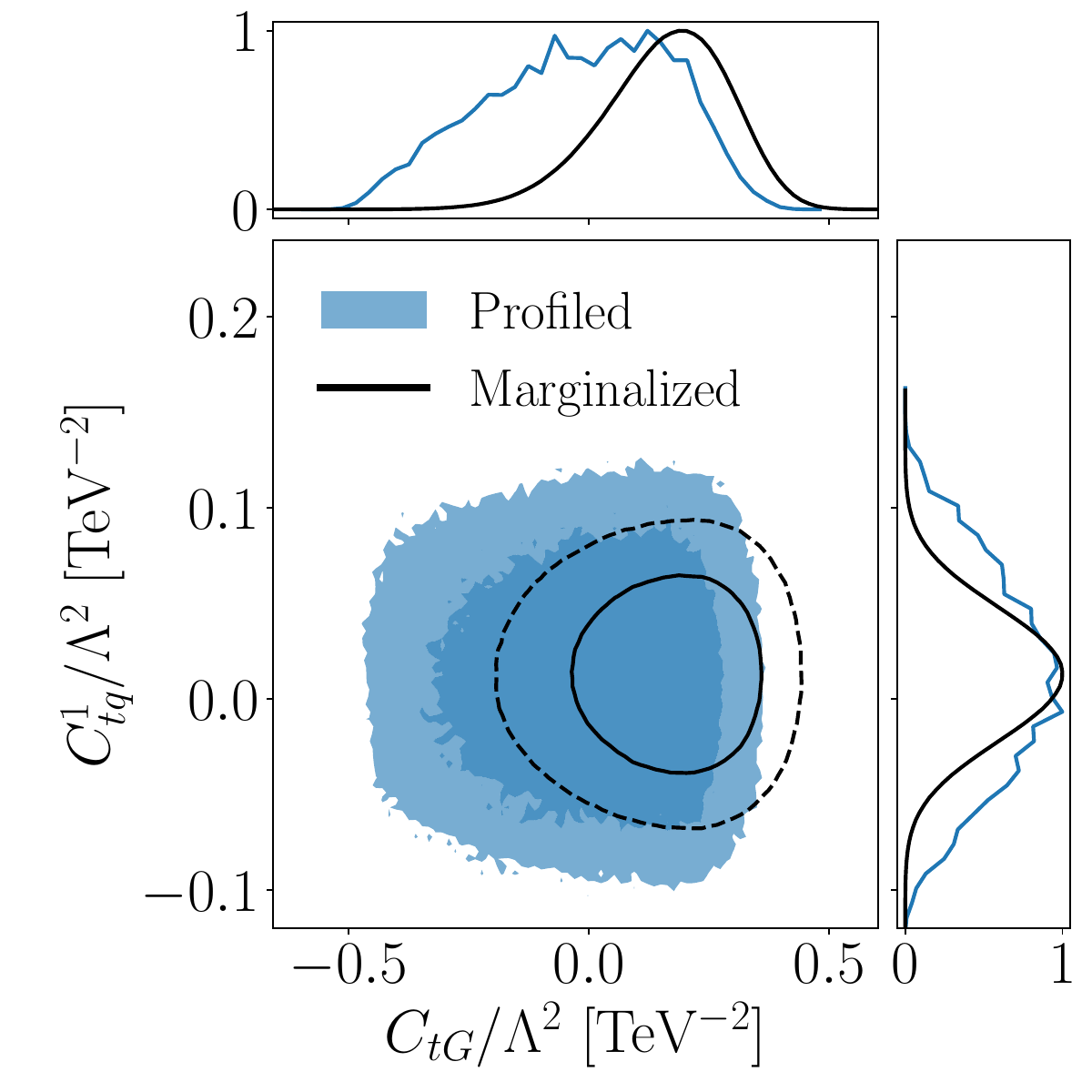}
\hspace*{0.1\textwidth}
\includegraphics[width=0.40\textwidth]{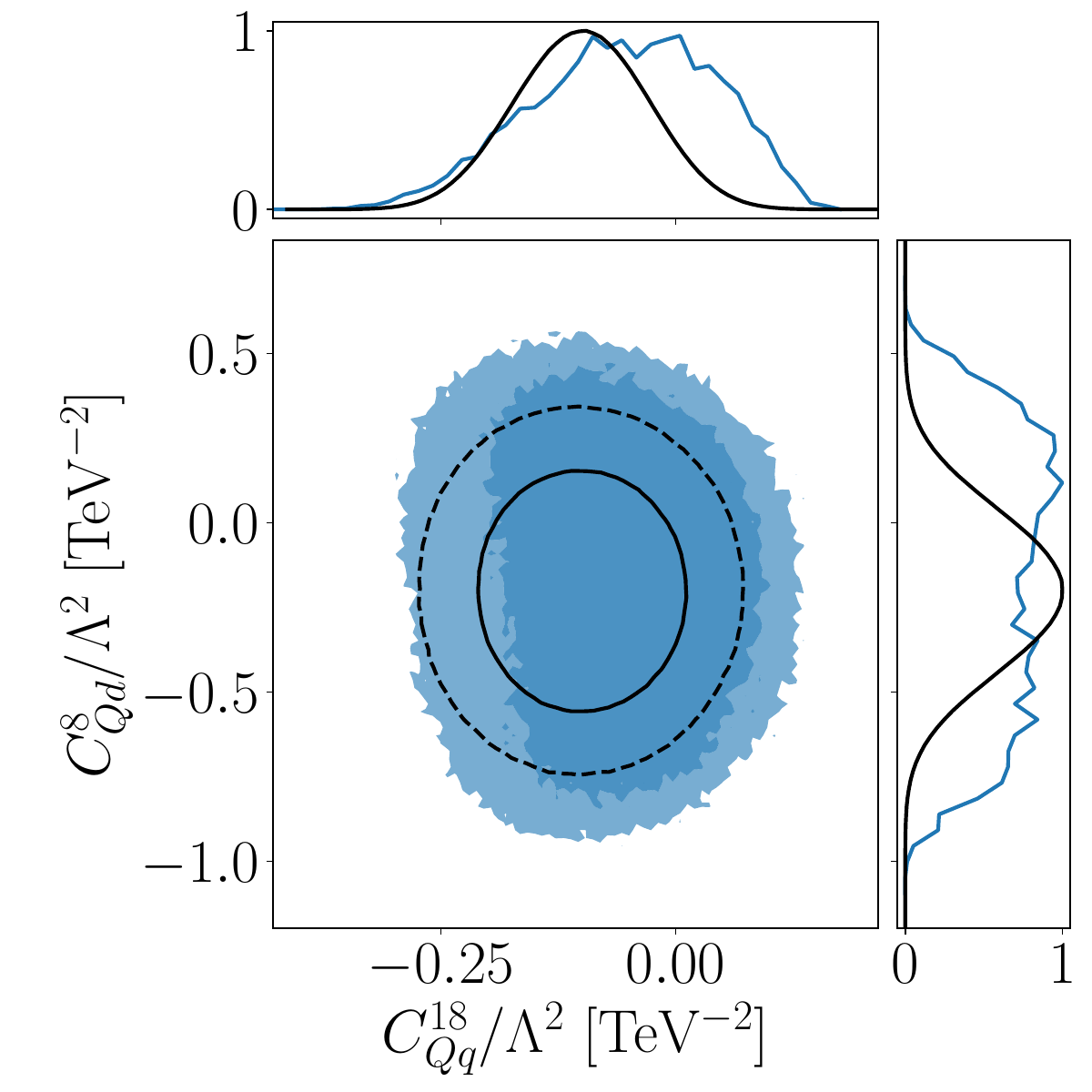}
    \caption{Comparison between marginalization (black) and profiling (blue) 
    in a global analysis of the top sector.}
    \label{fig:margvsprof}
\end{figure}

As defined in Eq.\eqref{eq:sfitter_like}, the central object of any \sfitter 
analysis is a fully exclusive likelihood. It is evaluated over the combined
space of Wilson coefficients and nuisance parameters. Obviously, the nuisance
parameters are irrelevant to the physics interpretation of the global SMEFT
analysis. In addition, we are usually not interested
in showing all 22 Wilson coefficients at the same time and instead reduce this 
space to one or two 
dimensions. Statistically, this can be done by profiling or marginalizing the likelihood.
Only for perfect Gaussian distributions do the two methods give the same results,
as discussed in Sec.~\ref{sec:sfitter}.
In the Higgs-electroweak sector, significant deviations between the two methods 
appear through a large under-fluctuation in one bin of a kinematic 
distribution~\cite{Brivio:2022hrb}.  

Fig.~\ref{fig:margvsprof} displays a selection of correlations from a 
marginalization (black) and profiling (blue) of the fully exclusive 
likelihood from all top sector measurements and Wilson coefficients.  
We show constraints for $\ope_{tG}$ vs $\ope_{tq}^{1}$ 
and $\ope_{Qq}^{18}$ vs $\ope_{Qd}^{8}$, but similar effects can be seen
in many operator pairs. In general, marginalization leads to narrower 
constraints than profiling.  This is particularly evident in the left panel 
of Fig.~\ref{fig:margvsprof}, and it is due to theory uncertainties and 
their flat likelihood distribution. With this choice, the profile likelihood 
can force a perfect agreement between data and predictions over a wide range
of values for critical Wilson coefficients. When we marginalize over the 
exclusive likelihood, the difference between Gaussian and flat uncertainties
is less pronounced, leading to more Gaussian and narrower one-dimensional 
distributions, as discussed in detail in Ref.~\cite{Brivio:2022hrb}. This 
effect is especially visible in the top sector, where theory uncertainties
are not only poorly defined~\cite{Ghosh:2022lrf}, but also large.

\subsection{Top-Higgs-electroweak combination}
\label{sec:global_combo}

\begin{figure}[t]
\centering\includegraphics[width=0.99\textwidth]{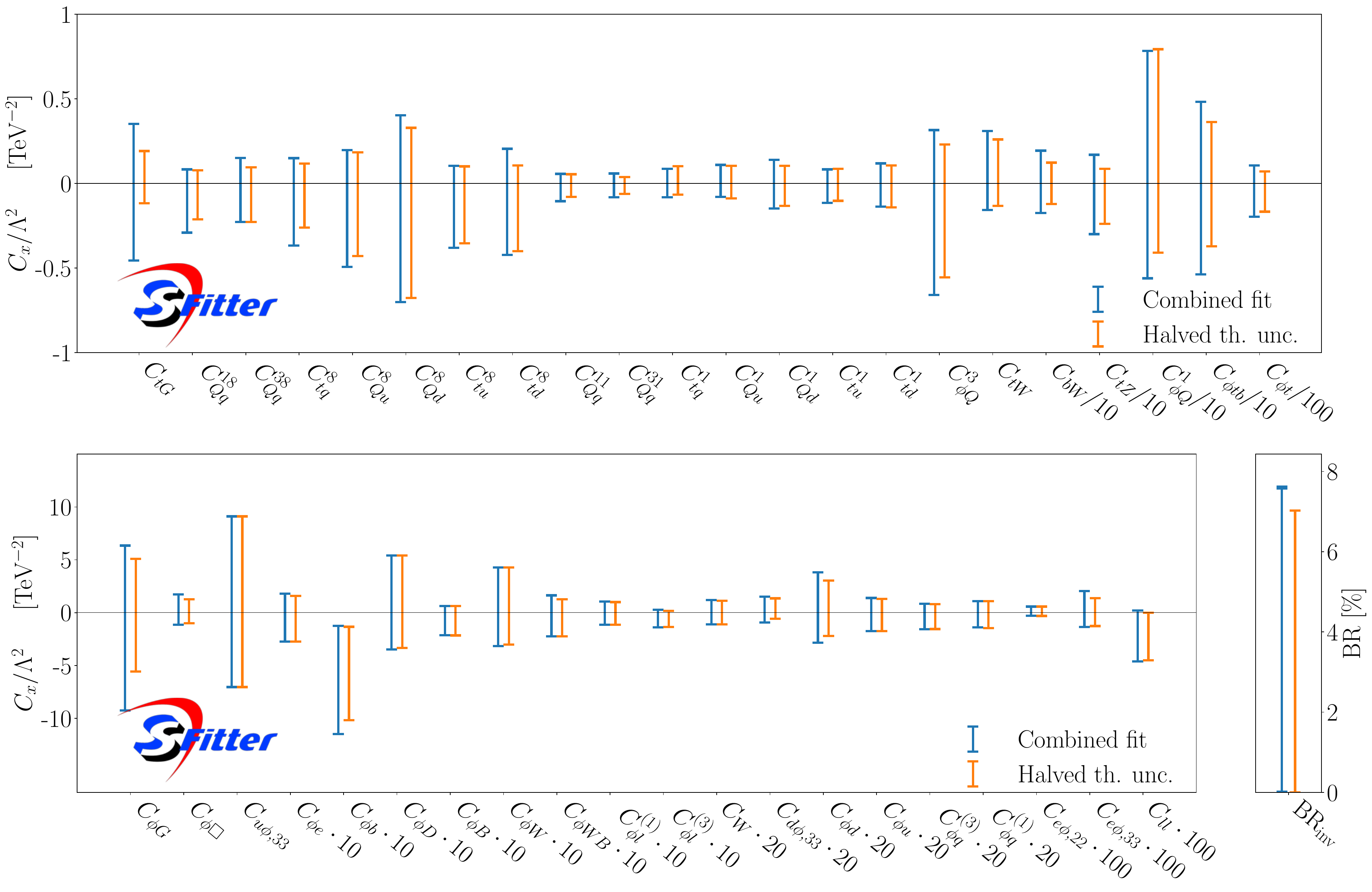}
    \caption{Results from a combined SMEFT analysis of the top sector and the 
    Higgs-electroweak sector, showing the constraints at 95\% CL on 43 degrees 
    of freedom, resulting from a profiled likelihood.}
    \label{fig:combo}
\end{figure}

Finally, making use of the numerical improvements in the \sfitter implementation, 
we can combine the top-sector SMEFT analysis from this paper with the \sfitter 
analysis of the Higgs, di-boson, and electroweak precision observables, 
Ref.~\cite{Brivio:2022hrb}. This combination has been studied in the literature in detail,
showing that the two sectors are linked, for instance, through $\ope_{tG}$~\cite{Ellis:2020unq,Ethier:2021bye}.

We confirm this state of the art and show the combined \sfitter profile likelihood
of the two sectors in Fig.~\ref{fig:combo}.  In total, 43 degrees of freedom are 
constrained: the 22 coefficients constrained by the top sector and discussed in 
Sec.~\ref{sec:eft}, and 21 additional operators relevant to the Higgs, di-boson 
and electroweak observables.  The notation and conventions for these 21 operators 
are provided in App.~\ref{app:combined}.
From the detailed discussion above and in Ref.~\cite{Brivio:2022hrb}, it is 
clear that the challenges and limitations of the global analyses in the 
two sectors are not the same. We show the limits at 95\% CL from one-dimensional profile 
likelihoods of the combined fit (blue) and under the assumption of theory uncertainties 
reduced by a factor of 2 (orange).  The numerical values of the constraints shown in 
Fig.~\ref{fig:combo} are provided in Tab.~\ref{tab:numerical_res}.

In the top sector, we find strong constraints on the four-fermion operators.  
The constraints on their Wilson coefficients are driven by kinematic distributions such 
as the ATLAS measurement of boosted top discussed in Sec.~\ref{subsec:boost}, and therefore 
theory uncertainties do not play an important role in their constraints. Conversely, the 
constraint on $C_{tG}$ improves significantly when theory uncertainties are halved, 
indicating that theory uncertainties dominate constraints obtained from top quark pair 
production total cross sections.  Similarly, this hypothetical reduction of theory uncertainties 
has an effect on the constraints obtained from single top, $t \bar{t} W$ and $t \bar{t} Z$ on 
coefficients such $C_{tW}$, $C_{bW}$, and $C_{tZ}$.

On the other hand, we observe no significant changes in the constraints on the operators relevant 
to the Higgs, di-boson and electroweak sectors, shown in the lower half of Fig.~\ref{fig:combo}, 
when theory uncertainties are reduced.  The exception is $C_{\phi G}$, which also benefits from 
the top quark data through its correlation with $C_{tG}$ and $C_{t \phi}$.  This is in agreement 
with Ref.~\cite{Brivio:2022hrb}, where it was found that in the Higgs-gauge sector, systematic 
uncertainties are the dominant source of uncertainty for many of the observables in this sector.

\section{Outlook}
\label{sec:outlook}

Global SMEFT analyses are an exciting development at the LHC, as
they combine their role as a precision hadron collider with the goal 
of interpreting all measurements in terms of precision quantum field 
theory. This precision theme implies that even if we know that the 
current measurements do not rule out the Standard Model, limits on 
SMEFT Wilson coefficients tell us important information about fundamental 
physics. 

To extract limits on fundamental physics parameters, we need a 
comprehensive uncertainty treatment covering experimental statistical 
uncertainties, experimental systematics, and theory uncertainties. For 
the latter two, it is crucial that we include correlations.
Public likelihoods are the state of the art in communicating 
such experimental results to a broader community. 
We include, for the first time, public ATLAS likelihoods 
for cross section measurements in a global analysis.
These public likelihoods allow us to systematically evaluate the effects
of correlations of systematic and theory uncertainties on a global analysis.

The basis of the global \sfitter analysis is a fully exclusive likelihood.
It includes a large set of rate and kinematic measurements, either 
pre-processed by ATLAS or CMS, unfolded, or extracted and backward-engineered
from experimental publications. The uncertainty treatment is especially
flexible, including a choice of flat nuisance parameters for
correlated theory uncertainties. Starting from the fully exclusive 
likelihood, we can employ a profile likelihood or a Bayesian marginalization
to extract limits on individual Wilson coefficients. In the top sector, 
we find no significant difference between the two statistical approaches.

The focus of this paper was on the role of different uncertainties,
their correlations and the role of public likelihoods in this context. 
In a similar analysis, albeit without public likelihoods,
we found that in the electroweak sector, the correlations were crucial, 
whereas the theory uncertainties were not (yet) a limiting factor~\cite{Brivio:2022hrb}.
Intriguingly, the situation in the top sector is the opposite: theory uncertainties are 
crucial, while the correlations of experimental systematics have a limited 
impact on the SMEFT limits. This reflects the QCD nature and the vast 
statistics of top pair production.

We have demonstrated that public likelihoods provide a much more flexible
approach to handling nuisance parameters. However, fully leveraging their potential currently proves difficult due to the large number of measurements included in our global analysis.
We emphasize that this is not a final statement about 
public likelihoods in SMEFT analyses. 
The reason is that we find kinematic measurements of boosted top 
pair production to be the driver behind improved SMEFT limits. 
For unfolded kinematic measurements, there are no public likelihoods 
available yet, but we are looking forward to implementing them in \sfitter
in the future.

We finished this study of the impact of theory uncertainties in a consistent theory 
framework of LHC data by performing the first combined \sfitter analysis of the Higgs, electroweak, and top sectors. 
This further displayed the limiting effect of theory uncertainties on the constraining power of modern top measurements compared to those in the Higgs sector.

\subsection*{Acknowledgements}

We would like to thank Dirk Zerwas, James Moore, and Luca Mantani for discussions on the
challenges of Monte Carlo toys, as well as Sabine Kraml and  
Lukas Heinrich for the encouragement to use public likelihoods.
We are also grateful to Luca Mantani for his help with the calculation of SMEFT predictions 
and Tomas Dado for his hands-on help with public likelihoods.
This research is
supported by the Deutsche Forschungsgemeinschaft (DFG, German Research
Foundation) under grant 396021762 – TRR 257 \textsl{Particle Physics
Phenomenology after the Higgs Discovery}. 
NE is funded by the Heidelberg IMPRS \textsl{Precision Tests of 
Fundamental Symmetries}.  MM acknowledges support from the Alexander von Humboldt Foundation.
We acknowledge support by the
state of Baden-W\"urttemberg through bwHPC and the German Research
Foundation (DFG) through grant no INST 39/963-1 FUGG (bwForCluster
NEMO).

\clearpage
\appendix
\section{Toys vs Markov chain}
\label{app:toys}

The current \sfitter methodology relies on a Markov chain to encode
the fully exclusive likelihood given in Eq.\eqref{eq:sfitter_like}. It then 
allows for a profile likelihood or 
marginalization to remove nuisance parameters and extract limits on individual Wilson coefficients~\cite{Lafaye:2007vs,Brivio:2022hrb}.  
Past \sfitter analyses used an alternative methodology known as
Monte Carlo toys~\cite{Lafaye:2007vs,Lafaye:2009vr,Biekotter:2018rhp} or Monte Carlo
replicas~\cite{NNPDF:2021njg,Hartland:2019bjb,Giani:2023gfq,Kassabov:2023hbm}. 
This method has not been used in \sfitter since the
SMEFT analysis of the top sector~\cite{Brivio:2019ius}.  In
Ref.~\cite{Kassabov:2023hbm}, its shortcomings for a SMEFT
analysis of the top sector are discussed in detail. In this appendix,
we provide an additional discussion of Monte Carlo toys in the \sfitter context.

\subsubsection*{Likelihood}

The basis of, essentially, all LHC analyses is the likelihood of a
given measurement, $d$, compared to a model or theory
prediction $m(c)$, which depends on parameters or Wilson coefficients
$c$~\cite{Brivio:2022hrb}. To simplify this discussion, we
approximate it as a Gaussian,
\begin{align}
    \log p(d|c)
    = - \frac{\left[ d - m(c) \right]^2}{2 \delta^2}
    = - \frac{\chi^2(c)}{2} \; ,
    \label{eq:def_like}
\end{align}
with an uncertainty $\delta$ assigned to the measurement. 

The toys method describes possible outcomes of a measurement, given its uncertainty,
with a nuisance parameter. In the Gaussian limit, the outcome of a
measurement is sampled around the mean $\bar{d}$,
\begin{align}
\label{eq:appdk}
     d_k \sim \mathcal{N}(\bar{d}, \delta) \; .
\end{align}
To extract the likelihood in Eq.\eqref{eq:def_like} we generate a set
of toy-measurements $d_k$, mimicking the outcomes of 
actual measurements. The basic frequentist
assumption is that we can maximize the likelihood for each
toy-measurement,
\begin{align}
    c_k = \text{argmax} \; p(d_k|c) \; ,
\label{eq:toys_cond}
\end{align}
to extract the maximum-likelihood parameters for each outcome, and
infer the likelihood over model space from the density of these points $\{c_k\}$. 

We can compare the toys to a simple Markov chain, where we are only interested
in a sample of points representing a given likelihood distribution, without 
any downstream task. At each step, 
the Markov chain proposes a new parameter point and keeps it with 
the probability defined by the likelihood in 
Eq.\eqref{eq:def_like}. As for the toys, the Markov chain 
encodes the likelihood through a density of points. The difference 
is that the toys start from the 
distribution of the experimental measurements and extract the 
likelihood from maximum-likelihood points, while the Markov 
chain collects points proportionally to the likelihood. Both 
methods give the same results, provided their algorithms respect 
the assumption that the distribution of maximum-likelihood points 
reproduces the underlying probability or likelihood.

An interesting challenge is the description of correlations between measurements 
in the likelihood. For the Gaussian case, we 
can describe them with a correlation matrix if the 
correlation is less than 100\%.
To describe correlations for non-Gaussian likelihoods, we
introduce nuisance parameters, which increase
the effective dimensionality of the likelihood.

We have introduced toys and the Markov chains as ways 
to construct a likelihood $p(d|c)$ for a given dataset $d$. 
We can as well introduce them as tools to construct 
the probability map $p(c|d)$. 
If the prior $p(c)$ entering the two algorithms is constant over a wide enough range,
the distribution of points will describe the likelihood as well as the
probability $p(c|d)$, the only difference being a 
normalization constant.

\subsubsection*{Circular flat direction}

\begin{figure}[t]
    \centering
    \includegraphics[width=0.495\textwidth]{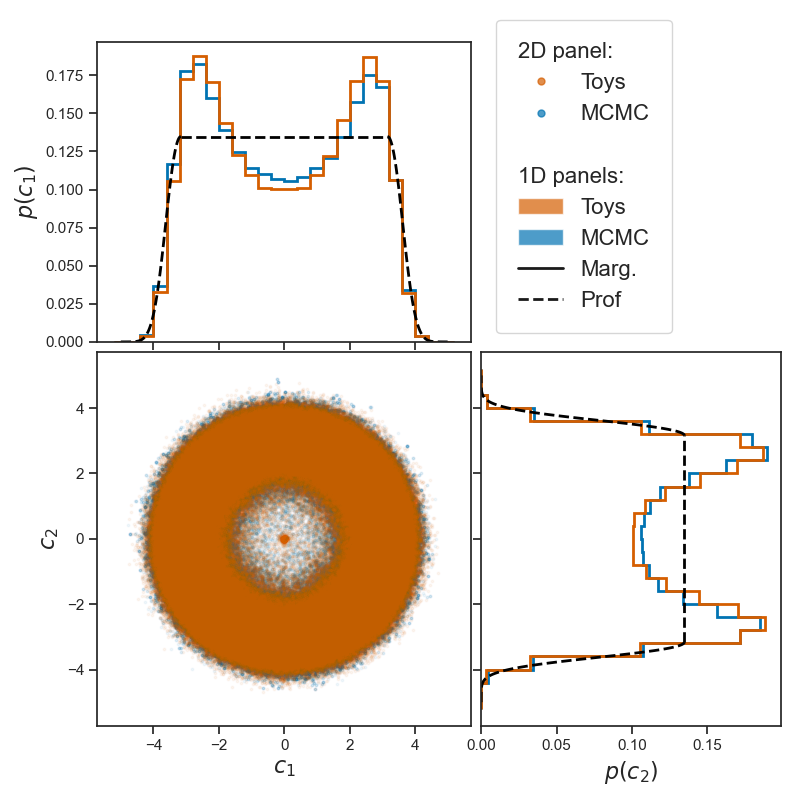}
    \caption{Comparison between toys and MCMC, constraining
      $c_1$ and $c_2$ using the measurement $d_1$ from
      Eq.\eqref{eq:meas1}. The 1-dimensional distributions show the
      profile likelihoods, as well as the marginalized probabilities. The profiling is 
      obtained by analytically maximizing the log-likelihood of Eq.\eqref{eq:def_like} along each axis.}
    \label{fig:toymccircle1}
\end{figure}

A key feature of the SMEFT analysis in the top sector is that, typically, 
4-fermion operators have extremely small interferences with the SM. The likelihood as a function
of the Wilson coefficients is dominated by the squares of these
coefficients. A measurement constraining two different Wilson
coefficients could then read
\begin{align}
    d \sim m(c) = m_\text{SM} + 0.1 c_1^2 + 0.1 c_2^2 \; .
\end{align}
If we consider $c_i^2$ the model parameters, we can always find two squared Wilson
coefficients which solve this relation. In contrast, if we consider $c_i$ as 
model parameters, we can only solve it if the measurement is an upward
fluctuation, $d > m_\text{SM}$. An example would be
\begin{align}
    d_1 = 6 > m_\text{SM,1}= 5
    \qquad \text{solved by} \qquad 
    0.1 c_1^2 + 0.1 c_2^2 = 1 \; .
    \label{eq:meas1}
\end{align}
We show the correlated values of $c_1$ and $c_2$ in 
Fig.~\ref{fig:toymccircle1}. The Markov chain and the toys form
the same circle. The width of this circular flat direction is given by
$\delta_{1,2} = 0.3$. The 2-dimensional likelihood is extracted by binning, with Markov chains
and toys producing the same result. 

The 2-dimensional likelihood can now be reduced in 
dimensionality. Here, profiling and marginalization 
lead to differences through volume
effects~\cite{Brivio:2022hrb}, as seen in the 
1-dimensional distributions 
in Fig.~\ref{fig:toymccircle1}. This difference is independent 
of the toys and Markov chain, which completely agree.

\subsubsection*{Unexplainable underfluctuation}

\begin{figure}[t]
    \includegraphics[width=0.495\textwidth]{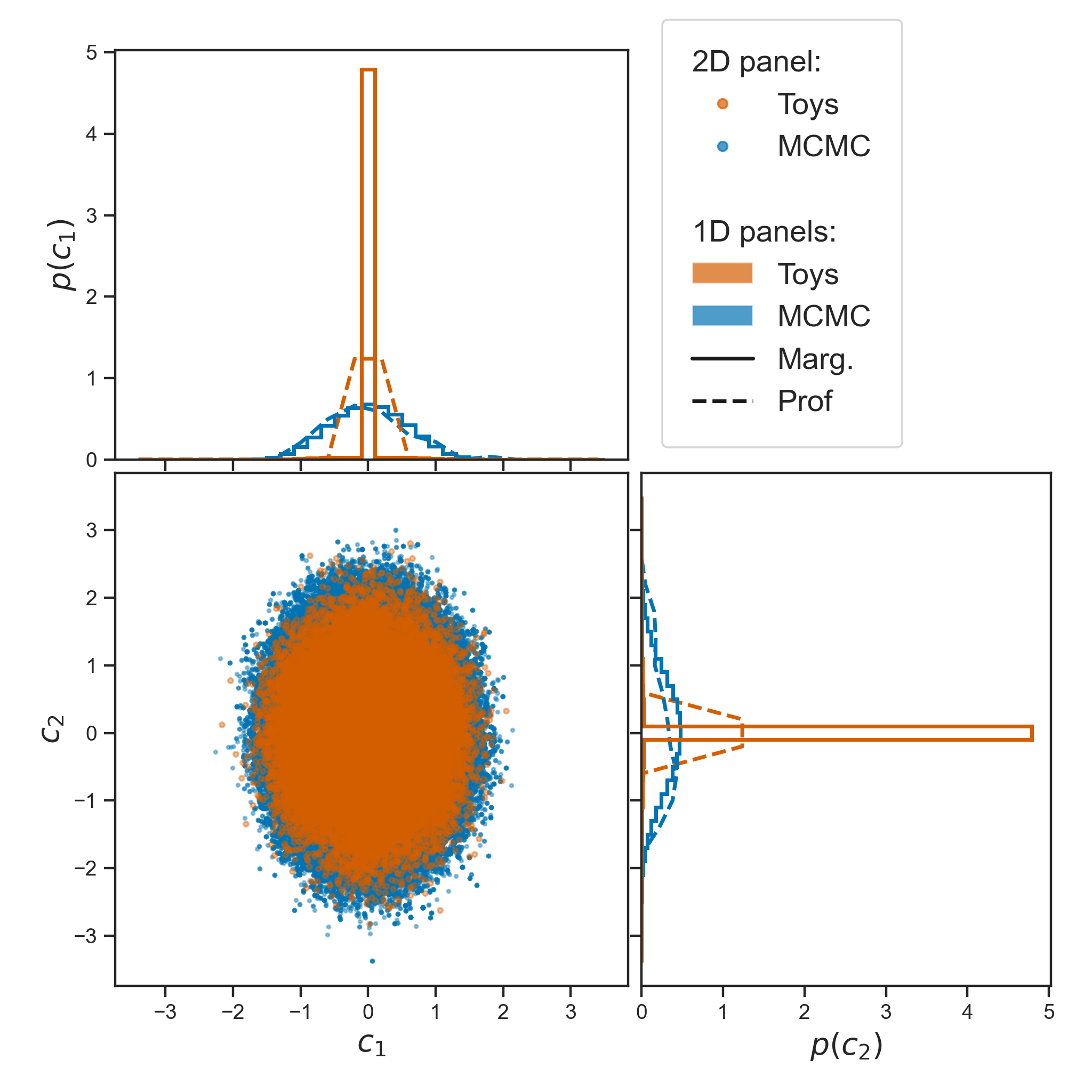}
    \includegraphics[width=0.495\textwidth]{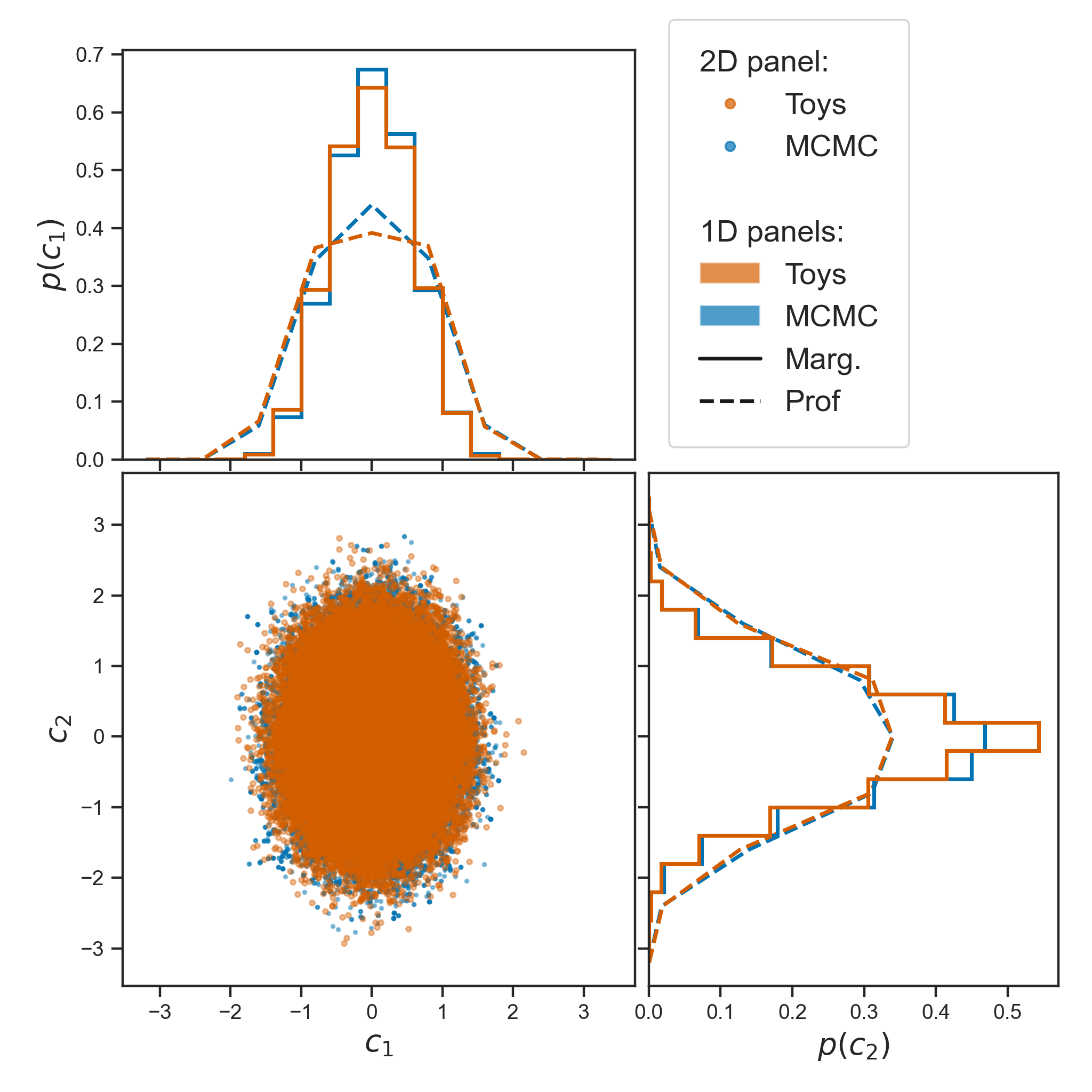}
    \caption{Left: comparison between toys and MCMC, constraining
      $c_1$ and $c_2$ using the measurement $d_2$ from
      Eq.\eqref{eq:meas2}. Right: the same comparison, but removing the
      peak of toys at the boundary $c_{1,2} \approx 0$. The 1-dimensional distributions show the
      profile likelihoods as well as the marginalized probabilities.}
    \label{fig:toymccircle2}
\end{figure}

A problem occurs when we encounter a negative fluctuation in the measurement,
which cannot be mapped to the usual minimum in model space. For instance, 
\begin{align}
    d_2 = 6 < m_\text{SM,2}= 7
    \qquad \text{requiring} \qquad 
    0.2 c_1^2 + 0.1 c_2^2 \stackrel{!}{=} - 1 \; .
    \label{eq:meas2}
\end{align}
This relation cannot be solved for real $c_i$, and the spread from
$\delta_{1,2} = 0.3$ is too small to cure this problem.  The
likelihoods obtained from toys and from the Markov chain are both given in the 
left panel of Fig.~\ref{fig:toymccircle2}, and they differ. 
The Markov chain includes points with a finite
likelihood offset below the theoretical maximum. 
The toys are derived from
Eq.\eqref{eq:toys_cond}, returning $c_1
= c_2 = 0$ if they cannot reach the true maximum. For the right panel of Fig.~\ref{fig:toymccircle2}, we modify 
the toys algorithm to remove the maximum-likelihood peak at the parameter
space boundary.  This is done by retaining only the samples $d_k$ in Eq.\eqref{eq:appdk} which satisfy $d_k > m_\text{SM,2}$. With this modification, the Markov chain and the toys have the 
same likelihood.

Summarizing Fig.~\ref{fig:toymccircle2}, the 
toys and the Markov chain treat 
the unwanted parameter space $c_i^2 < 0$ differently. While the toys provide a
maximum-likelihood parameter point for each assumed measurement, the
Markov chain removes the unwanted points. 
In terms of a prior on the numerical implementation of the scanning, we can 
understand the two methods as 
\begin{align}
  &\text{toys} \quad & p(d|c) \Bigg|_\text{phys} &\sim p(c|d) \Bigg|_\text{phys} = \max (\Delta, p(c|d)) \notag \\
  &\text{Markov chain} \quad & p(c) &= \Theta(c^2) \; ,
\end{align}
where $\Delta$ is chosen to remove the numerically broadened peak at $c =0$.
In the right panel of Fig.~\ref{fig:toymccircle2} we see that for allowed $c_i$ 
the two methods give the same result but with a different 
normalization of the probability $p(c|d)$~\cite{Kassabov:2023hbm}. If we
work with the likelihood $p(d|c)$, the normalization does not matter.
If we include all toy experiments, the two distributions in 
Fig.~\ref{fig:toymccircle2} reflect the fact that the two methods are asking 
different statistical questions. 

\subsubsection*{Spreading out the peak}

The situation given by Eq.\eqref{eq:meas2} is just an extremely
unlucky outcome. Of very few measurements, one happens to be many standard deviations away
from the prediction, while we expect such an outlier only
once in many more measurements. 

\begin{figure}[t]
    \includegraphics[width=0.495\textwidth]{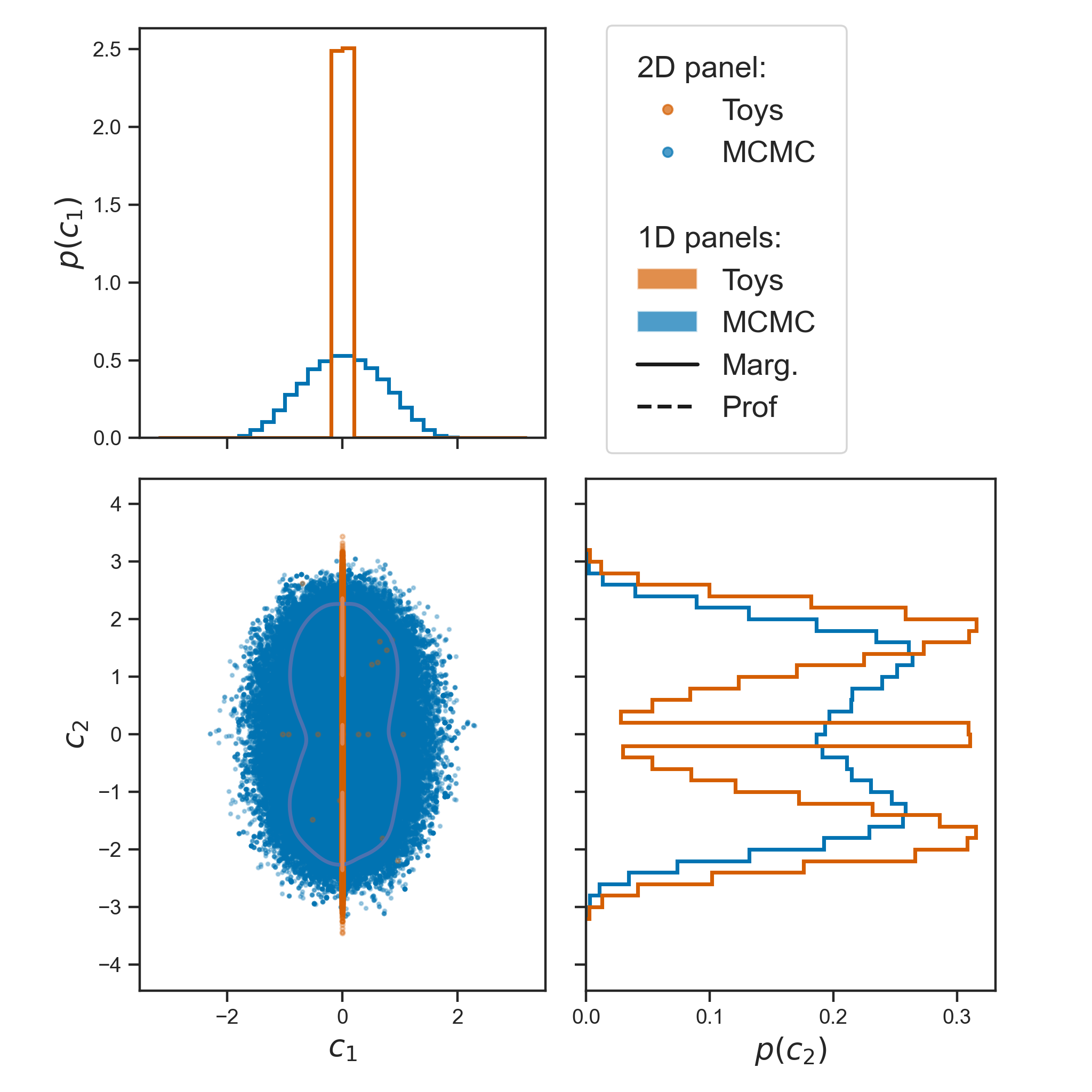}  
    \includegraphics[width=0.495\textwidth]{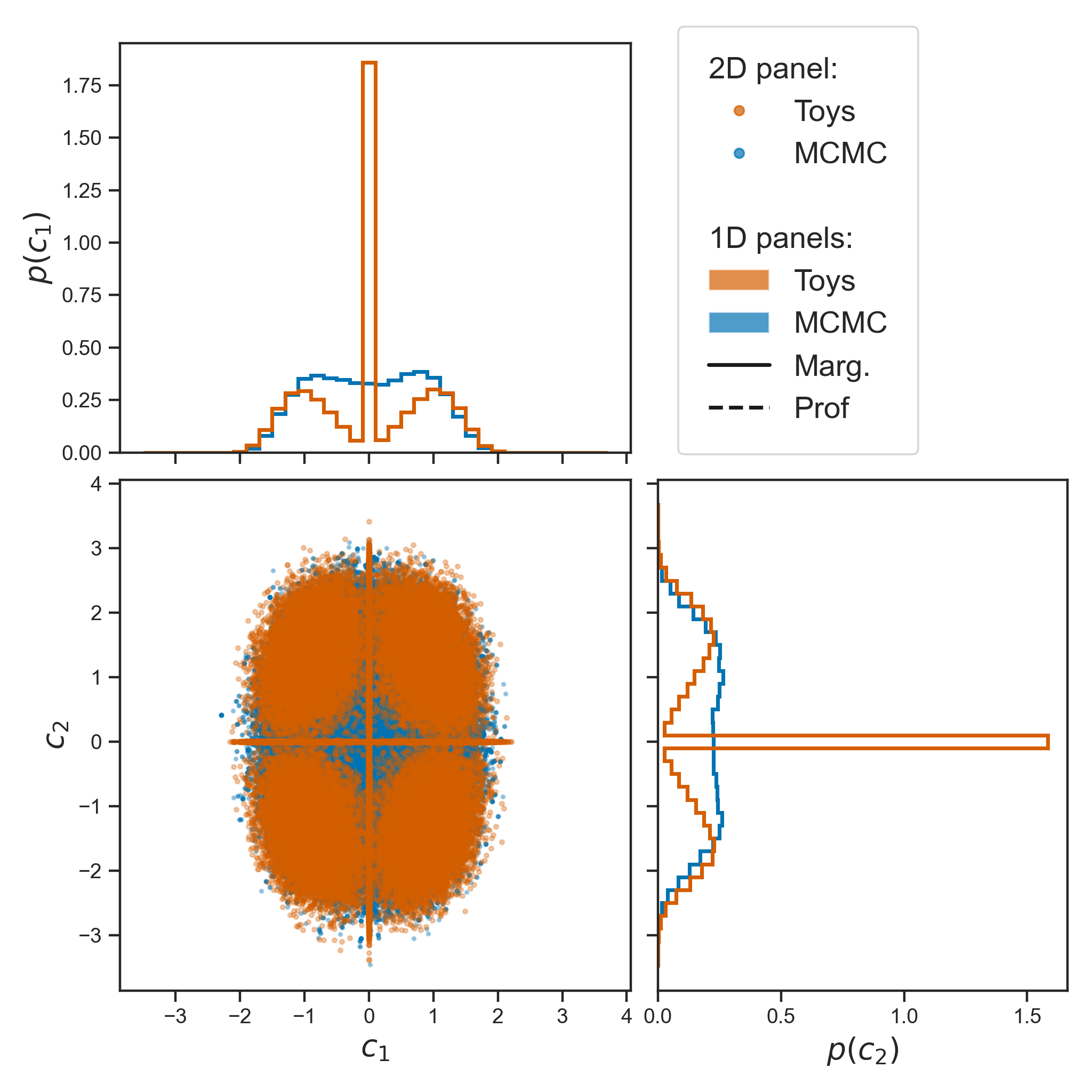}    
    \caption{Left: comparison between toys and MCMC for the combination of $d_1$ 
    and $d_2$ in terms of the $c_i$. Right: the same for 
    a combination of $d_1$, $d_2$, and $d_3$.}
    \label{fig:toymccombo}
\end{figure}

Moreover, in a realistic global analysis,
many measurements together constrain a given model parameter.
To see what happens then, we first combine $d_1$ and $d_2$ in
the left panel of Fig.~\ref{fig:toymccombo}.
The Markov chain generates a valid point
distribution, symmetric in $\pm |c_2|$.  For the toys, the situation is different. On the underfluctuation 
in $d_2$, $c_1$ and
$c_2$ have the same impact, but for the overfluctuation in $d_1$, a
shift in $c_2$ gains more. This is why a peak is observed at $c_1 = 0$ and $c_2$ is adjusted.

Next, we add a third measurement, such that the effect of our
underfluctuation will be compensated by another pull on $c_1$,
\begin{align}
    d_3 = m_\text{SM,3} + 0.2 c_1^2 
    \qquad \text{with} \qquad 
    d_3 = 6, \; m_\text{SM,3}= 5.5 \; .
\end{align}
The central maximum of the corresponding likelihood will be at $c_1 \approx 1.6$. 
In the right panel of Fig.~\ref{fig:toymccombo}, we see that the peak at $c_1=0$ is now accompanied 
by a distribution of finite $c_1$ symmetric around zero. This way, the importance of the 
peak is reduced. For more measurements, this will continue until the underfluctuation will just be 
an expected statistical outlier in a large set of measurements, with little effect on the global likelihood.

\subsubsection*{Determining squared coefficients}

\begin{figure}[t]
    \includegraphics[width=0.495\textwidth]{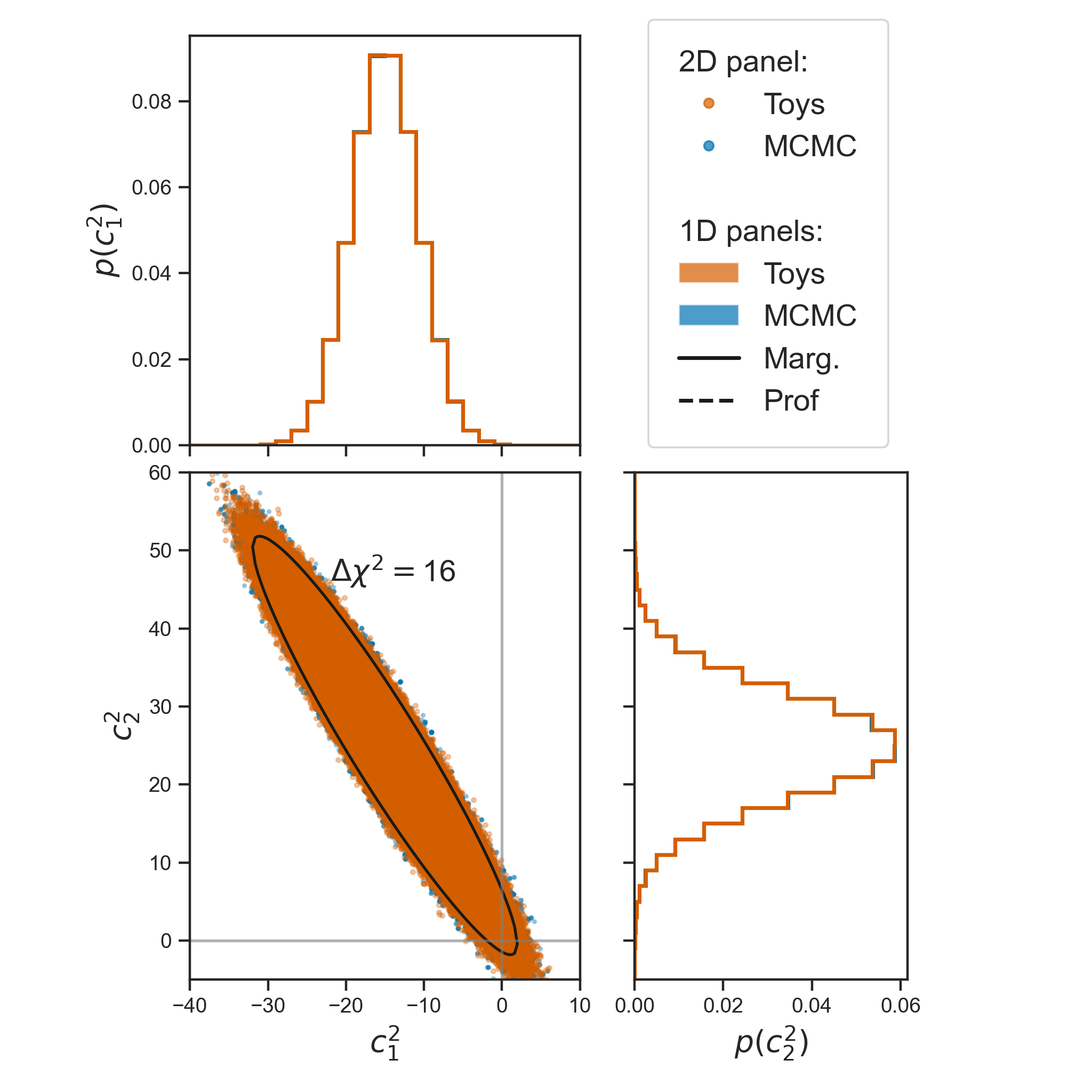}
    \includegraphics[width=0.495\textwidth]{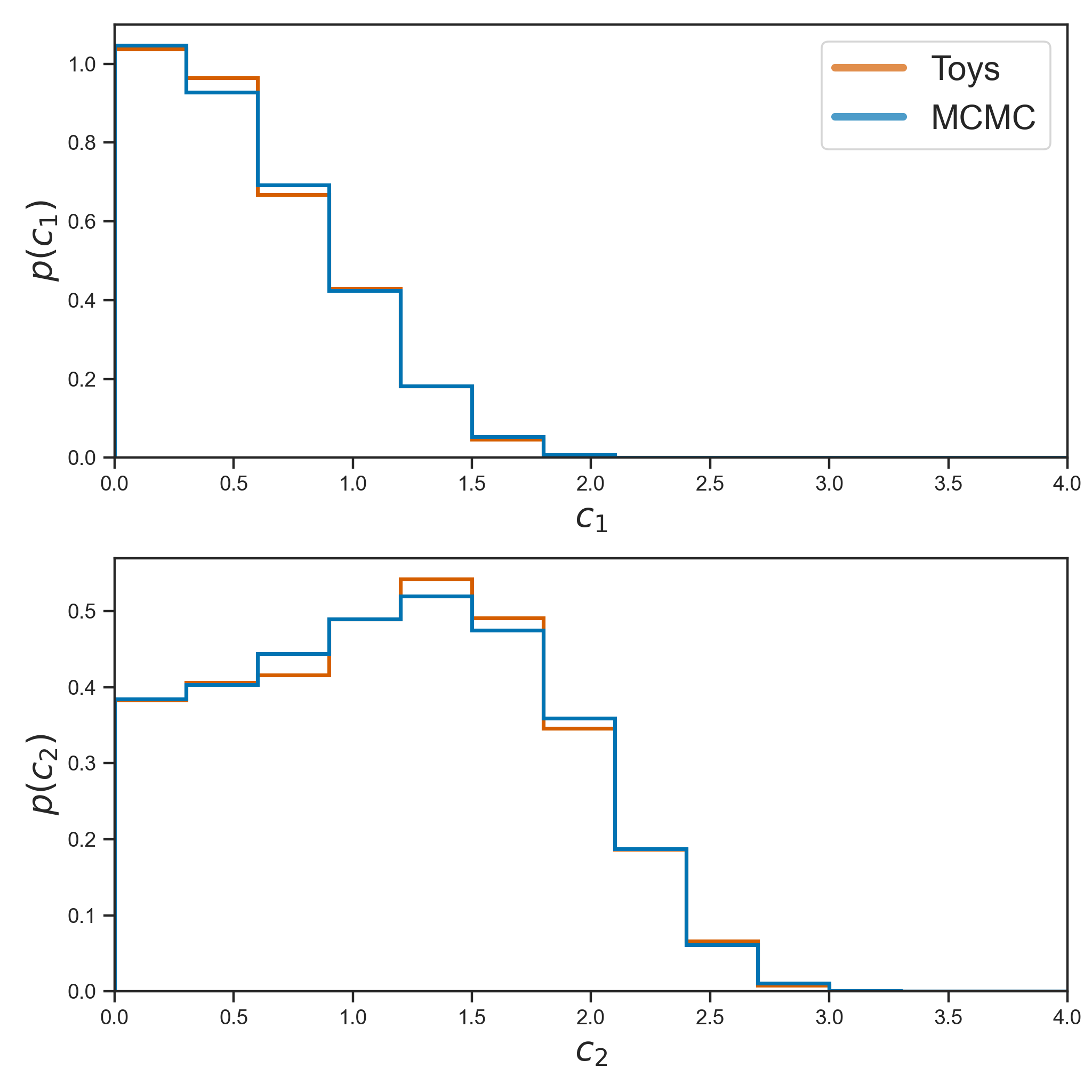}
    \caption{Comparison between toys and MCMC for the combination of
      $d_1$ and $d_2$ in terms of the $c_i^2$, and for the valid
      points only in terms of the $c_i$.}
    \label{fig:toymcsquare}
\end{figure}

Because our measurements show a purely quadratic dependence, we can
try to circumvent our problem by extracting the likelihood 
over a model space defined by $c_1^2$ and $c_2^2$. Instead of first
requiring $c_i^2$ to be positive and then maximizing the
log-likelihood, we first maximize the log-likelihood in terms of the $c_i^2$. In the left panel of
Fig.~\ref{fig:toymcsquare}, we show the results in terms of the $c_i^2$,
with toys and Markov chain in perfect agreement.  
However, almost the entire preferred range gives 
unwanted values for $c_2^2$. The curve defined by $\Delta \chi^2 = 16$   highlights how unlikely it is to obtain points with $c_i^2 > 0$.

In the second step, 
we now select valid parameter points. This 
avoids a hard boundary when optimizing the toys. In the right panel of
Fig.~\ref{fig:toymcsquare}, the blue curves show the  
points in the Markov chain, sampling directly in $c_1$ vs $c_2$. 
These are the same points as those in the upper-right quadrant of 
the left panel of Fig.~\ref{fig:toymccombo}.  In orange, we show the 
one-dimensional distributions of only the allowed parameter points
out of all those plotted in the left panel of Fig.~\ref{fig:toymcsquare}.
Neither the likelihood nor the probability is
parametrization invariant, so we need to apply the
Jacobian
\begin{align}
    p(c) 
    = p(c^2) \; \frac{d c^2}{d c} 
\end{align}
when extracting the shown likelihoods as a function of 
$c_i$ rather than $c_i^2$.  For comparison with the blue curves, the 
orange distribution is reweighted by this Jacobian. Again, 
we find complete agreement between the toys and the Markov chain.

\section{Higgs, top, di-boson and electroweak combination}
\label{app:combined}

In producing the global analysis of Sec.~\ref{sec:global_combo} and
Fig.~\ref{fig:combo}, we have combined the top sector from this paper
with the previous \sfitter analysis of the Higgs, di-boson, and electroweak
sectors of Ref.~\cite{Brivio:2022hrb}, taking all data from within this 
reference. Note that while Ref.~\cite{Brivio:2022hrb} provides constraints on
Wilson coefficients in the HISZ basis; here, we provide all constraints in the
Warsaw basis.

In addition to the 22 operators introduced in Sec.~\ref{sec:eft} and constrained
by the top sector observables, a further 21 Wilson coefficients can be 
constrained by the addition of data from the Higgs, di-boson, and electroweak
sectors. These operators are assumed to follow the same flavor symmetry 
conventions as introduced in Eq.\eqref{eq:symm}, \emph{i.e.} flavor universality
applied to the first two quark generations. 
The notation and conventions for these 21 operators are provided in 
Tab.~\ref{tab:operators}.

\begin{table}[t!]\centering
\renewcommand{\arraystretch}{1.5}
\begin{small} \begin{tabular}{lc|lc}
 \toprule
Coefficient & Operator & Coefficient & Operator\\
\midrule
 \text{$C_{\phi G}$} & \, $\phi^{\dagger} \phi G_{\mu \nu}^{A} G^{A \mu \nu}$ \, & \text{$C_{W}$} & \, $\varepsilon^{IJK} W_{\mu}^{I \nu} W_{\nu}^{J \rho} W_{\rho}^{K \mu}$ \,\\
 \text{$C_{\phi \Box}$} & \, $(\phi^{\dagger} \phi ) \Box (\phi^{\dagger} \phi )$ \, & \text{$C_{d \phi, 33}$} & \, $(\phi^{\dagger} \phi)(\bar{Q}_{3} b \phi)$ \, \\
 \text{$C_{u \phi, 33}$} & \, $(\phi^{\dagger} \phi)(\bar{Q}_{3} t \phi)$ \, & \text{$C_{\phi d}$} & \, $\sum_{i=1}^{2}(\phi^{\dagger} i \overleftrightarrow{D_{\mu}} \phi) (\bar{d}_i \gamma^{\mu} d_i) $ \,\\
 \text{$C_{\phi e}$} & \, $(\phi^{\dagger} i \overleftrightarrow{D_{\mu}} \phi) (\bar{e} \gamma^{\mu} e) $ \, & \text{$C_{\phi u}$} & \, $\sum_{i=1}^{2}(\phi^{\dagger} i \overleftrightarrow{D_{\mu}} \phi) (\bar{u}_i \gamma^{\mu} u_i) $ \,\\
 \text{$C_{\phi b}$} & \, $(\phi^{\dagger} i \overleftrightarrow{D}_{\mu} \phi) (\bar{b} \gamma^{\mu} b)$ \, & \text{$C_{\phi q}^{(3)}$} & \, $\sum_{i=1}^{2}(\phi^{\dagger} i \overleftrightarrow{D_{\mu}} \phi) (\bar{q}_i \gamma^{\mu} q_i) $ \,\\
 \text{$C_{\phi D}$} & \, $(\phi^{\dagger} D^{\mu} \phi ) ^{*} (\phi^{\dagger} D^{\mu} \phi )$ \, & \text{$C_{\phi q}^{(1)}$} & \, $\sum_{i=1}^{2}(\phi^{\dagger} i \overleftrightarrow{D_{\mu}} \phi) (\bar{q}_{i} \tau^{I} \gamma^{\mu} q_{i}) $ \,\\
 \text{$C_{\phi B}$} & \, $\phi^{\dagger} \phi B_{\mu \nu} B^{\mu \nu}$ \, & \text{$C_{e \phi, 22}$} & \, $(\phi^{\dagger} \phi)(\bar{l}_{2} \mu \phi)$ \, \\
 \text{$C_{\phi W}$} & \, $\phi^{\dagger} \phi W_{\mu \nu}^{I} W^{I \mu \nu}$ \, & \text{$C_{e \phi, 33}$} & \, $(\phi^{\dagger} \phi)(\bar{l}_{3} \tau \phi)$ \,\\
 \text{$C_{\phi WB}$} & \, $\phi^{\dagger} \tau^{I} \phi W_{\mu \nu}^{I} B^{\mu \nu}$ \, & \text{$C_{ll}$} & \, $(\bar{l} \gamma_{\mu} l)(\bar{l} \gamma^{\mu} l)$ \, \\
 \text{$C_{\phi l}^{(1)}$} & \, $(\phi^{\dagger} i \overleftrightarrow{D_{\mu}} \phi) (\bar{l} \gamma^{\mu} l) $ \, & $\text{BR}_\text{inv}$ & invisible Higgs decays \\
 \text{$C_{\phi l}^{(3)}$} & $(\phi^{\dagger} i \overleftrightarrow{D}^{I}_{\mu} \phi) (\bar{l} \tau^{I} \gamma^{\mu} l)$  & & \\
 \bottomrule
 \end{tabular} \end{small}
\caption{Additional Wilson coefficients of the Warsaw basis entering the combined
analysis of the Higgs, di-boson, top, and electroweak sectors. The 21 degrees of
freedom shown here are included in the global analysis of 
Sec.~\ref{sec:global_combo} alongside the 22 operators already constrained by top
sector observables.  In total, 43 coefficients are constrained in the global
analysis.}\label{tab:operators}
\end{table}

Table~\ref{tab:numerical_res} reports the numerical values of the boundaries of the 95\% CL intervals shown
in Fig.~\ref{fig:combo}.

\begin{table}[t!]\centering
\renewcommand{\arraystretch}{1.5}
\begin{small}  \begin{tabular}{>{$}l<{$}cc|>{$}l<{$}cc}
 \toprule
\text{Coefficient} & Full analysis & Halved theory unc. & \text{Coefficient} & Full analysis & Halved theory unc.\\
 \midrule
 
C_{\phi G} & [-9.25, 6.35] & [-5.56, 5.1] & 
C_{tG} & [-0.46, 0.35] & [-0.12, 0.19]\\
C_{\phi \Box} & [-1.14, 1.72] & [-1.0, 1.27] &
C_{Qq}^{(18)} & [-0.29, 0.08] & [-0.21, 0.08]\\
C_{u \phi, 33 } & [-7.03, 9.11] & [-7.03, 9.11] &
C_{Qq}^{(38)} & [-0.23, 0.15] & [-0.23, 0.09]\\
C_{\phi e} \times 10 & [-2.73, 1.79] & [-2.73, 1.59] &
C_{tq}^{(8)} & [-0.37, 0.15] & [-0.26, 0.12]\\
C_{\phi b} \times 10 & [-11.46, -1.25] & [-10.17, -1.33]  &
C_{Qu}^{(8)} & [-0.49, 0.2] & [-0.43, 0.18]\\
C_{\phi D} \times 10 & [-3.48, 5.4] & [-3.33, 5.4] &
C_{Qd}^{(8)} & [-0.7, 0.4] & [-0.68, 0.33]\\
C_{\phi B} \times 10 & [-2.12, 0.63] & [-2.15, 0.63] &
C_{tu}^{(8)} & [-0.38, 0.1] & [-0.35, 0.1]\\
C_{\phi W} \times 10 & [-3.15, 4.29] & [-3.0, 4.28] &
C_{td}^{(8)} & [-0.42, 0.2] & [-0.4, 0.11]\\
C_{\phi WB} \times 10 & [-2.24, 1.64] & [-2.24, 1.26] &
C_{Qq}^{(11)} & [-0.1, 0.06] & [-0.08, 0.05]\\
C_{\phi l}^{(1)} \times 10 & [-1.14, 1.07] & [-1.14, 1.0] &
C_{Qq}^{(31)} & [-0.08, 0.06] & [-0.06, 0.04]\\
C_{\phi l}^{(3)} \times 10 & [-1.38, 0.27] & [-1.36, 0.17] &
C_{tq}^{(1)} & [-0.08, 0.09] & [-0.07, 0.1]\\
C_W \times 20 & [-1.1, 1.2] & [-1.1, 1.12] &
C_{Qu}^{(1)} & [-0.08, 0.11] & [-0.09, 0.1]\\
C_{d \phi, 33}\times 20 & [-0.94, 1.51] & [-0.58, 1.36] &
C_{Qd}^{(1)} & [-0.15, 0.14] & [-0.13, 0.1]\\
C_{\phi d}\times 20 & [-2.83, 3.81] & [-2.19, 3.05] &
C_{tu}^{(1)} & [-0.11, 0.08] & [-0.1, 0.09]\\
C_{\phi u}\times 20 & [-1.75, 1.39] & [-1.75, 1.31] &
C_{td}^{(1)} & [-0.14, 0.12] & [-0.14, 0.11]\\
C_{\phi q}^{(3)}\times 20 & [-1.56, 0.84] & [-1.54, 0.8] &
C_{\phi Q}^{(3)} & [-0.66, 0.32] & [-0.56, 0.23]\\
C_{\phi q}^{(1)}\times 20 &  [-1.39, 1.08] & [-1.46, 1.08] &
C_{tW} & [-0.16, 0.31] & [-0.13, 0.26]\\
C_{e \phi, 22}\times 100 & [-0.29, 0.58] & [-0.3, 0.58] &
C_{bW} /10 & [-0.17, 0.19] & [-0.12, 0.12]\\
C_{e \phi, 33}\times 100 & [-1.35, 2.06] & [-1.26, 1.37] &
C_{tZ} /10 & [-0.3, 0.17] & [-0.24, 0.09]\\
C_{ll}\times 100 & [-4.61, 0.21] & [-4.51, 0.0] &
C_{\phi Q}^{(1)} /10 & [-0.56, 0.78] & [-0.41, 0.79]\\
\text{BR}_\text{inv} & [0, 7.6] & [0, 7.03] &
C_{\phi tb} /10 & [-0.54, 0.48] & [-0.37, 0.36]\\
& & & C_{\phi t} /100 & [-0.2, 0.11] & [-0.17, 0.07]\\
\bottomrule
\end{tabular} \end{small}
\caption{Numerical values for the 95\% CL limits shown in Fig.~\ref{fig:combo}. We emphasize that the 
reduction of the theory uncertainties by a factor two is entirely hypothetical.}
\label{tab:numerical_res}
\end{table}

\clearpage
\bibliography{tilman,generative,literature,toplikelihoods,top-eft,citing_us} 

\end{document}